\documentclass[review,authoryear,3p,times,10pt]{elsarticle}

\usepackage{multirow,setspace,times,amssymb,amsmath,graphicx,color,rotating,subfigure,url}
\usepackage{lineno,color}
\usepackage{natbib}
\usepackage{booktabs}%
\usepackage{longtable}%
\usepackage{rotating}
\usepackage{lscape}
\usepackage{ulem}
\usepackage{threeparttable}
\graphicspath{{Figures/}}
\usepackage[table]{xcolor}
\usepackage{tabularx}
\usepackage{graphicx} %use graph format
\usepackage{epstopdf}
\usepackage{mathrsfs}
\usepackage{makecell}
\usepackage[bookmarks=true,colorlinks,linkcolor=blue,anchorcolor=blue,citecolor=blue,unicode]{hyperref}
\usepackage{bookmark}

\usepackage{todonotes}
\usepackage{ragged2e}
\usepackage[font=small]{caption}
\hypersetup{CJKbookmarks=true}%
\makeatletter
\def\ps@pprintTitle{%
	\let\@oddhead\@empty
	\let\@evenhead\@empty
	\let\@oddfoot\@empty
	\let\@evenfoot\@oddfoot
}
\makeatother

\begin{document}
% \begin{CJK*}{GBK}{Song} % Use default fonts from CJK (see below)

\begin{frontmatter}
%\newpage
\title{Tail dependence structure and extreme risk spillover effects between the international agricultural futures and spot markets}

\author[SB]{Yun-Shi Dai}
\author[SB,RCE]{Peng-Fei Dai}
\author[SB,RCE,DM]{Wei-Xing Zhou\corref{WXZ}}
\ead{wxzhou@ecust.edu.cn}
\cortext[WXZ]{Corresponding author.} %Corresponding to: 130 Meilong Road, P.O. Box 114, School of Business, East China University of Science and Technology, Shanghai 200237, China.}

\address[SB]{School of Business, East China University of Science and Technology, Shanghai, China}
\address[RCE]{Research Center for Econophysics, East China University of Science and Technology, Shanghai, China}
\address[DM]{School of Mathematics, East China University of Science and Technology, Shanghai, China}

\begin{abstract}
This paper combines the Copula-CoVaR approach with the ARMA-GARCH-skewed Student-t model to investigate the tail dependence structure and extreme risk spillover effects between the international agricultural futures and spot markets, taking four main agricultural commodities, namely soybean, maize, wheat, and rice as examples. The empirical results indicate that the tail dependence structures for the four futures-spot pairs are quite different, and each of them exhibits a certain degree of asymmetry. In addition, the futures market for each agricultural commodity has significant and robust extreme downside and upside risk spillover effects on the spot market, and the downside risk spillover effects for both soybeans and maize are significantly stronger than their corresponding upside risk spillover effects, while there is no significant strength difference between the two risk spillover effects for wheat, and rice. This study provides a theoretical basis for strengthening global food cooperation and maintaining global food security, and has practical significance for investors to use agricultural commodities for risk management and portfolio optimization.
\end{abstract}


\end{frontmatter}

%\tableofcontents

\section{Introduction}

Food security is a fundamental issue concerning human survival. Currently, the global food security situation remains grim and complex. According to {\textit{The State of Food Security and Nutrition in the World 2022}} released by the Food and Agriculture Organization of the United Nations (FAO)\footnote{\url{https://www.fao.org/publications/sofi/en/}}, the number of people affected by hunger worldwide reached 828 million in 2021, a marked increase of about 150 million since the COVID-19 pandemic broke out, and almost 670 million people are predicted to still be facing hunger in 2030. {\textit{The Global Report on Food Crises 2022}} published by the Global Network Against Food Crises (GNAFC)\footnote{\url{http://www.fightfoodcrises.net/resources/all-publications/en/}} suggests that around 193 million people across 53 countries or regions were in acute food insecurity in 2021, representing a noticeable rise of about 40 million people compared with 2020. In fact, the current global food crisis is the result of several mutually reinforcing drivers, including geopolitical conflicts, economic shocks, weather extremes, and the COVID-19 pandemic. Particularly, the Russia-Ukraine conflict has further exposed the inherent characteristics of the interconnected world food system, intensified the vulnerability of global and local food systems, and caused serious consequences for global food security. With the increasing prominence of food security, the dependence structure of the global food market and its internal risk transmission have gradually become a hot topic in recent years.

The global food market can be segmented into the food spot market and the food futures market. The food futures market developed on the basis of the food spot market has two primary functions of price discovery and risk aversion, which is of great significance for stabilizing price fluctuations and regulating supply and demand in the food spot market \citep{Yang-Bessler-Leatham-2001-JFuturesMark, Joseph-Sisodia-Tiwari-2014-EconModel, Adammer-Bohl-2018-JFuturesMark, Arzandeh-Frank-2019-AmJAgrEcon, Li-Chavas-2023-AmJAgrEcon}. By playing their respective market roles, the food spot and futures markets complement each other and thus form a relatively complete modern food market system. However, with the international situation fraught with tension and turbulence, the frequency of sharp rises and falls in food futures prices has increased in recent years. Violent and frequent price fluctuations exacerbate the risks of the food futures market, which has become a potential threat to the global food security and economic stability. In addition, the current global food crisis has led to growing concerns about global food shortage, and a new round of food trade protectionism has risen, further aggravating the tension between food supply and demand. Fears of food shortages have also heightened the speculative sentiment of global capital on the food market, which in turn has driven food spot prices more volatile. Therefore, it is critical to clarify the tail dependence structure and assess the extreme risk spillovers between the international agricultural futures and spot markets, which is of great reference value for enhancing global food cooperation, strengthening global food system governance, and maintaining global food security.

Most of the traditional methods applied in the studies on financial risk measurement assume the returns of financial assets being normally distributed. However, in real economic activities, financial asset returns tend to exhibit the distribution characteristics of leptokurtosis, fat tail and skewness, as well as the volatility characteristics of asymmetry and heteroscedasticity, so the assumption of normal distribution is no longer applicable. For the sake of better capturing the actual distribution of financial asset returns, the Student t distribution with a longer tail than the normal distribution is widely used for financial risk management, but the symmetric Student t distribution is still unable to solve the asymmetric distribution problem, which may lead to the underestimation of financial risks \citep{Wagner-Marsh-2005-JEmpirFinanc, Eling-2012-InsurMathEcon}. Considering that the skewed Student-t distribution is capable of accurately describing the asymmetry of financial asset returns, we introduce it in this paper to analyze the empirical data and construct various risk measures for the agricultural futures and spot markets.

In addition, the traditional correlation coefficients can only measure the degree of linear correlation between variables. However, with the development of economic globalization, the dependence structure among financial markets has become more and more complex, gradually showing the characteristics of nonlinearity and asymmetry. Thus, measuring the dependence between assets with traditional correlation coefficients is generally considered to be not accurate enough for risk management. Due to its significant advantages in describing nonlinearity and tail dependence, the copula approach is regarded as a useful tool for depicting the dependence structure between various assets or markets, especially under extreme price movements \citep{Patton-2012-JMultivarAnal, Liu-Ji-Fan-2017-QuantFinanc}. In the modeling process, the copula functions have no restriction on the selection of marginal distributions, which makes the descriptions of marginal distributions heterogeneous. Moreover, different single copula functions or even mixed copula functions can be applied to depict diverse dependence structures when modelling the joint distribution. Hence, the copula method is conducive to constructing more flexible and robust models to explore the dependence structures between various markets, which is exactly the method adopted in this paper.

We select the futures and spot of four main agricultural commodities, namely soybean, maize, wheat, and rice, and apply the ARMA-GARCH-skewed Student-t approach to specify the marginal distribution for each return series. We then assess the tail dependence between the agricultural futures returns and their corresponding spot returns based on eight different single copula models. To take the possible asymmetric tail dependence into consideration, we further construct mixed copula models combining the Gumbel copula model and the survival Gumbel copula model, so as to analyze the dependence structure for each agricultural pair in a more comprehensive way. The results show that the tail dependence structures of various agricultural pairs are quite different, and each of them exhibits a certain degree of asymmetry. In addition, we calculate the downside and upside $CoVaR$ and $\Delta CoVaR$ dynamics for agricultural spot returns based on the estimates of the single copula and mixed copula models, respectively, and adopt the Kolmogorov-Smirnov (K-S) test to examine the extreme downside and upside risk spillover effects from the agricultural futures markets to the agricultural spot markets, as well as the possible asymmetry between the two risk spillover effects.

To our knowledge, this paper is the first to investigate the tail dependence structure and the extreme risk spillover effects between the global agricultural futures and spot markets, combining the Copula-CoVaR approach with the ARMA-GARCH-skewed Student-t model. The method we adopt is capable of fully considering the data characteristics of the agricultural return series, such as autocorrelation and heteroscedasticity, and thus clarifying the asymmetric tail dependence structure and evaluating the extreme risk spillover effects more accurately and comprehensively. Since the modeling of dependence structures in finance contributes to derivative pricing, asset allocation, and portfolio optimization, the empirical findings of our paper provide practical implications for using agricultural commodities to manage risks, especially tail risk interdependence and extreme risk spillovers.

The rest of this paper is organized as follows. Section~\ref{S1:LitRev} provides the literature review. Section~\ref{S1:Methodology} details the methodology for model construction and estimation. Section~\ref{S1:Data} introduces the data source and presents the basic statistical description. Section~\ref{S1:EmpAnal} presents and discusses the empirical results on the tail dependence structure and extreme risk spillover effects between the international agricultural futures and spot markets. Section~\ref{S1:Conclude} draws conclusions and gives some implications.

\section{Literature review}
\label{S1:LitRev}

Value-at-Risk ($VaR$) is one of the most commonly used risk measurement methods in finance, which plays an important role in risk management, performance assessment, and risk-based capital calculation. A flood of literature has adopted various approaches to compute $VaR$, including the historical simulation method \citep{Cabedo-Moya-2003-EnergyEcon, Perignon-Smith-2010-JBankFinanc, Aramonte-Rodriguez-Wu-2013-JBankFinanc}, the extreme value theory \citep{Longin-2000-JBankFinanc, Marimoutou-Raggad-Trabelsi-2009-EnergyEcon, Cifter-2011-PhysicaA}, and GARCH-type models \citep{Yu-Li-Jin-2010-EconomRev, Predescu-Stancu-2011-EconComputEconCybernStud, Laporta-Merlo-Petrella-2018-EnergyEcon}. The significant advantages of $VaR$ lie in its concise interpretation, simple calculation, broad comparability and wide applicability \citep{Krause-2003-JRiskFinanc,Jorion-2006}. 

However, under the framework of coherent risk measures, $VaR$ has been criticized for not satisfying the conditions of the sub-additivity axiom and failing in considering extreme events \citep{Danielsson-2002-JBankFinanc, Acerbi-Tasche-2010-EconNotes}. \cite{Boucher-Danielsson-Kouontchou-Maillet-2014-JBankFinanc} and \cite{Danielsson-James-Valenzuela-Zer-2016-JFinancStabil} believe that a key reason for the lack of accuracy of $VaR$ prediction in practical application is the uncertainty of model specification and estimation, which also supports the view of \cite{Jorion-2009-EurFinancManag} that $VaR$ prediction is subject to model risks. In addition, the subprime mortgage crisis that broke out in the United States in 2007 quickly spread to other countries, resulting in a global financial crisis and economic recession. As a mainstream method of risk measurement, $VaR$ pays too much attention to the risk of a single institution or a single market, and ignores the risk transmission between financial institutions and markets, thus leading to serious underestimations of the potential occurrence of risk events and the risk magnitude posed by subprime mortgages \citep{Chen-Wang-Zhang-2019-JEconometrics}. After the financial crisis, $VaR$ has been widely questioned, and the methods of financial risk measurement were in urgent need of innovation.

% Adrian-Brunnermeier-2008-FEDNSR, Adrian-Brunnermeier-2011-NBER, 

Against this background, \cite{Adrian-Brunnermeier-2016-AmEconRev} propose risk measures of Conditional Value-at-Risk ($CoVaR$) and delta Conditional Value-at-Risk ($\Delta CoVaR$), providing new ideas for risk management practice. Subsequently, \cite{Girardi-Ergun-2013-JBankFinanc} redefine $CoVaR$ and interpret the systemic risk contribution of an individual institution as the difference between its $CoVaR$ in the normal state and its $CoVaR$ in the crisis state. $CoVaR$ is a more effective and comprehensive risk measurement method, which has been widely recognized by academics, taking into account the complex interdependent structure among different financial markets and being able to measure the ``extra'' risks arising from the existence of other markets and the interaction between various markets.

Numerous scholars have employed $CoVaR$ and $\Delta CoVaR$ to measure the risk spillovers between different financial markets such as the exchange rate, securities, banking, and insurance markets. \cite{Hardie-Wang-Yu-2016-JEconom} adopt $CoVaR$ in conjunction with network and tail event approaches to study the interdependence and risk transmission between the financial institutions in the United States, which can be divided into broker-dealers, insurance companies, depository companies, and others. \cite{Boako-Alagidede-2017-JMultinatlFinancM} use $CoVaR$ to analyze the risk spillover effects of extreme downside currency prices on African stock markets, finding that exchange rate risk may spill over to several African stock markets, especially during periods of market turbulence, and the rise of stock prices is often together with the depreciation of domestic currencies. \cite{Jian-Wu-Zhu-2018-EmergMarkRev} compute the intraday dynamic $CoVaR$ based on an extended MV-CAViaR model, and explore the asynchronous risk spillovers between the Chinese equity and futures markets. \cite{Chen-Chen-Jin-Wei-Yu-2019-EmergMarkFinancTr} empirically investigate the connection between Internet finance and traditional financial markets by using $CoVaR$ to quantify the degree of risk spillovers, and conclude that the risks of Internet finance are more likely to be transmitted to the banking market, followed by the insurance market and finally the securities market. \cite{Hanif-Mensi-Vo-2021-FinancResLett} apply $CoVaR$ approach to examine the impact of the COVID-19 on the spillover effects across ten stock sectors in China and the United States. \cite{Abuzayed-Bouri-AlFayoumi-Jalkh-2021-EconAnalPolicy} combine $CoVaR$ and $\Delta CoVaR$ calculated by the DCC-GARCH model to analyze the systemic risk spillovers between the international equity market and domestic equity markets in the countries that are most impacted by the COVID-19 pandemic. \cite{Duarte-Eisenbach-2021-JFinanc} and \cite{Wang-Gao-Huang-Sun-Chen-Tang-Di-2022-IntRevFinancAnal} employ $\Delta CoVaR$ to investigate the systemic risk and risk contagion of the U.S. banking industry and global equity markets, respectively. \cite{Aloui-BenJabeur-MeftehWali-2022-ResIntBusFinanc} calculate $CoVaR$ to explore the risk spillover effects from the Chinese equity market to the G7 equity markets before and during the COVID-19 pandemic.

With the deepening of the financialization process of commodity markets, the research on the extreme risk spillover effects is no longer limited to the traditional financial markets, but is expanding to the broader financial system covering commodity markets. \cite{Mensi-Hammoudeh-Shahzad-AlYahyaee-Shahbaz-2017-EnergyEcon, Mensi-Hammoudeh-Shahzad-Shahbaz-2017-JBankFinanc} analyze the dependence between crude oil and monetary markets in MENA and several other countries, as well as the dependence between crude oil and stock markets in developed regions, with the combination of the variational mode decomposition approach and diverse copula functions, and then evaluate their respective risk spillover effects by calculating $CoVaR$. \cite{Li-Wei-2018-EnergyEcon} investigate the dependence structure between the crude oil and the Chinese stock market in raw, short- and long-term investment horizons before and after the financial crisis, and quantify $VaR$, $CoVaR$ and $\Delta CoVaR$ to assess the downside and upside risk spillover effects of the crude oil market on the Chinese stock market. \cite{Ji-Liu-Fan-2019-EnergyEcon, Ji-Liu-Zhao-Fan-2020-IntRevFinancAnal} examine the dynamic dependence and risk spillovers between WTI crude oil and the exchange rate markets in China and the United States, different types of oil shocks and BRICS stock returns, respectively. \cite{Meng-Nie-Mo-Jiang-2020-Energy} assess the spillover effects of the extreme downward and upward movements of the global crude oil prices on the commodity sectors in China based on the $CoVaR$ risk measure, illustrating that the downward risk spillover effect shows a greater impact on the commodity sectors in China. \cite{Sun-Liu-Wang-Li-2020-IntRevFinancAnal} discuss the risk spillover effects from global commodity markets to both international and Chinese maritime markets using the GARCH-Copula-CoVaR model, and find evidence that the extreme risk spillovers vary with time points. \cite{Tiwari-Trabelsi-Alqahtani-Raheem-2020-EnergyEcon}, \cite{Chen-Wen-Li-Yin-Zhao-2022-EnergyEcon} and \cite{Tian-Alshater-Yoon-2022-EnergyEcon} all explore the dependence structures and risk spillovers between oil and equity markets of different economies, and consistently verify the existence of significant risk spillover effects from the oil market to the equity markets.

As one of the main categories of commodity markets, the risk spillover between the agricultural commodity markets and other financial markets has begun to receive some attention. \cite{Ji-Bouri-Roubaud-Shahzad-2018-EnergyEcon} utilize a time-varying CoVaR-Copula model with a switching dependence to explore the tail dependence structure and risk spillover effects between the international energy and agricultural markets, finding that the information spillover from the oil and natural gas markets may aggravate the risk exposure of the agricultural markets. \cite{Kumar-Tiwari-Raheem-Hille-2021-ResourPolicy} investigate the dependence of oil and agricultural commodities in four different market states including rising oil prices-rising agricultural prices, falling oil prices-falling agricultural prices, rising oil prices-falling agricultural prices, and falling oil prices-rising agricultural prices, the results of which support the opinion that the collapses of the oil market and the agricultural market occur synchronously, and the quantified $CoVaR$ and $\Delta CoVaR$ provide convincing evidence of the significant risk spillover effects from the oil market to the agricultural market. \cite{Hanif-Hernandez-Shahzad-Yoon-2021-QRevEconFinanc} combine various copula functions with $CoVaR$ to analyze the nonlinear and dynamic dependence between oil and global food prices, as well as the downward and upward risk spillover effects, and conclude that there exist both tail dependence and asymmetric risk spillover effects between oil and food commodities.

Reviewing tons of relevant literature on extreme risk spillovers, very few academic research involve agricultural commodities, and these existing studies mainly focus on the risk spillovers between agricultural commodity markets and other markets, ignoring the tail dependence and risk spillovers inside the global agricultural market, namely between the agricultural futures and spot markets. Therefore, we combine the copula method with the $CoVaR$ risk measure to investigate the tail dependence structure and extreme risk spillover effects between the international agricultural futures and spot markets, considering that copulas can accurately describe the nonlinear, asymmetrical, and tail-dependent characteristics of the dependence structure among financial variables, and $CoVaR$ is capable of comprehensively measuring the magnitudes and directions of risk spillovers between various markets.

\section{Methodology}
\label{S1:Methodology}

\subsection{Copula modelling}

A copula is defined as a multivariate cumulative distribution function whose margins are uniformly distributed over the interval [0, 1]. Copulas are capable of capturing the dependence structure of multivariate distributions. The theorem proposed by \cite{Sklar-1959-PublInstStatistUnivParis} underlies most applications of copulas. According to Sklar's theorem, given a bivariate joint distribution function and respective univariate marginal distribution functions, there exists a copula distribution function that can describe the dependence structure between these two variables. As a result, the joint distribution function $F$ of the bivariate time series $\mathbf{r}_{t} = \left( r_{1,t},r_{2,t}\right )$ can be decomposed as follows:
\begin{equation}
    F\left( r_{1,t},r_{2,t};\theta \right) = C_{t} \left( F_{1}\left(r_{1,t};\theta_{1} \right),F_{2}\left( r_{2,t};\theta_{2} \right);\theta_{c} \right),
    \label{Eq:Joint_distribution_function}
\end{equation}
where the parameter set $\theta = \left(\theta_{1}^{\prime},\theta_{2}^{\prime},\theta_{c}^{\prime}\right)^{\prime}$, $C_{t}$ denotes the copula distribution function, and $F_{1}$ and $F_{2}$ are the marginal distribution functions of return series $r_{1}$ and $r_{2}$, respectively.

The bivariate joint density function of $r_{1,t}$ and $r_{2,t}$ can be expressed as the product of the copula density function and two marginal density functions under the assumption that all the cumulative distribution functions are differentiable, that is,
\begin{equation}
    f\left( r_{1,t},r_{2,t};\theta \right) = c_{t} \left( F_{1}\left(r_{1,t};\theta_{1} \right),F_{2}\left( r_{2,t};\theta_{2} \right);\theta_{c} \right) \cdot f_{1}\left(r_{1,t};\theta_{1} \right) \cdot f_{2}\left( r_{2,t};\theta_{2}\right),
    \label{Eq:Joint_density_function}
\end{equation}
where $c_ {t} $ is the copula density function, and $f_{1}$ and $f_{2}$ are the marginal density functions of return series $r_{1}$ and $r_{2}$, respectively.

\subsubsection{Single copula}

There are various families of bivariate copulas which differ in the detail of the dependence they represent, such as the elliptical copulas and Archimedean copulas. The elliptical copula families are symmetric about the center, and Normal copula and Student-t copula are the most typically recommended members. Archimedean copula families are defined by their generator functions, among which Clayton copula and Gumbel copula are commonly used in practice.

Normal copula assumes tail independence. The distribution and density functions of the bivariate Normal copula can be given by 
\begin{equation}
    C_{N}\left( u_{1},u_{2}; \rho \right) = \int\nolimits_{-\infty}^{\phi^{-1} \left( u_{1} \right) } \int\nolimits_{-\infty}^{\phi^{-1} \left( u_{2} \right) } \frac{1}{2\pi\sqrt{1-\rho^{2}}} \exp \left\{-\frac{s_{1}^{2}+s_{2}^{2}-2\rho s_{1} s_{2}}{2 \left(1-\rho^{2} \right)} \right\} \mathrm{d}s_1\mathrm{d}s_2
    \label{Eq:Normal_copula}
\end{equation}
and
\begin{equation}
    c_{N}\left( u_{1},u_{2}; \rho \right) = \frac{1}{\sqrt{1-\rho^{2}}} \exp \left\{ \frac{2\rho \phi^{-1} \left( u_{1} \right) \phi^{-1} \left( u_{2} \right) - \phi^{-1} \left( u_{1} \right)^{2} - \phi^{-1} \left( u_{2} \right)^{2}  }{2 \left(1-\rho^{2} \right)} + \frac{\phi^{-1} \left( u_{1} \right)^{2} + \phi^{-1} \left( u_{2} \right)^{2}}{2}  \right\},
    \label{Eq:Normal_copula_density}
\end{equation}
where $u_{1}$, $u_{2}$ are the probability integral transforms of $r_{1}$, $r_{2}$ based on their respective marginal distribution functions, $\phi^{-1} \left( \cdot \right)$ denotes the inverse of the normal cumulative distribution function, and $\rho$ is the parameter of Normal copula whose value range is $(-1, 1)$. The Kendall rank correlation coefficient $\tau_{N}$ corresponding to the bivariate Normal copula can be calculated by $2\arcsin(\rho)/\pi$, while the lower tail dependence $\lambda_{N} ^ {\mathrm{low}} $ and the upper tail dependence $\lambda_ {N} ^ {\mathrm{up}} $ are both 0.

Student-t copula assumes symmetric tail dependence. The distribution and density functions of the bivariate Student-t copula can be given by 
\begin{equation}
    C_{S}\left( u_{1},u_{2}; \rho, \nu \right) = \int\nolimits_{-\infty}^{T_{\nu}^{-1} \left( u_{1} \right) } \int\nolimits_{-\infty}^{T_{\nu}^{-1} \left( u_{2} \right) } \frac{1}{2\pi\sqrt{1-\rho^{2}}} \left[ 1 + \frac{s_{1}^{2}+s_{2}^{2}-2\rho s_{1}s_{2}}{\nu\left( 1-\rho^{2} \right)} \right]^{-\frac{\nu+2}{2}} \mathrm{d}s_1\mathrm{d}s_2
    \label{Eq:Student_copula}
\end{equation}
and
\begin{equation}
    c_{S}\left( u_{1},u_{2}; \rho, \nu \right) = \rho^{-\frac{1}{2}}\frac{ \Gamma \left( \frac{\nu+2}{2} \right) \Gamma\left( \frac{\nu}{2} \right) }{ \left[ \Gamma\left( \frac{\nu+1}{2} \right) \right]^{2} } \frac{ \left[ 1+\frac{ T_{\nu}^{-1} \left( u_{1} \right)^{2} + T_{\nu}^{-1} \left( u_{2} \right)^{2} - 2\rho T_{\nu}^{-1} \left( u_{1} \right) T_{\nu}^{-1} \left( u_{2} \right)}{\nu\left( 1-\rho^{2} \right)} \right]^{-\frac{\nu+2}{2}} }{ \left[ \left( 1+\frac{T_{\nu}^{-1} \left( u_{1} \right)^{2}}{\nu} \right) \left( 1+\frac{T_{\nu}^{-1} \left( u_{2} \right)^{2}}{\nu} \right) \right]^{-\frac{\nu+2}{2}} },
    \label{Eq:Student_copula_density}
\end{equation}
where $T_{\nu}^{-1} \left( \cdot \right)$ denotes the inverse of the Student't cumulative distribution function with $\nu$ degrees of freedom, and $\rho$ is the parameter of Student-t copula whose value is restricted to the interval $(-1, 1)$. The Kendall rank correlation coefficient $\tau_{S}$ corresponding to the bivariate Student-t copula can be expressed as $2\arcsin\left( {\rho} \right)/\pi$, while the lower tail dependence $\lambda_{S}^{\mathrm{low}}$ and the upper tail dependence $\lambda_{S}^{\mathrm{up}}$ are both $2T_{\nu+1}\left( -\sqrt{\nu+1}\sqrt{\left( 1-\rho \right)/\left( 1+\rho \right)} \right)$ where $T_{\nu+1}$ is the Student't cumulative distribution function with $\nu+1$ degrees of freedom.

Clayton copula is sensitive to changes at the lower tail, so it can be used to describe the lower tail dependence between variables. The distribution and density functions of the bivariate Clayton copula can be expressed as
\begin{equation}
    C_{C}\left( u_{1},u_{2}; \alpha \right) = \left( u_{1}^{-\alpha} + u_{2}^{-\alpha} - 1 \right)^{-\frac{1}{\alpha}}
    \label{Eq:Clayton_copula}
\end{equation}
and
\begin{equation}
    c_{C}\left( u_{1},u_{2}; \alpha \right) = \frac{\left( 1+\alpha \right)  u_{1} ^{-\alpha-1} u_{2} ^{-\alpha-1}}{\left( u_{1}^{-\alpha} + u_{2}^{-\alpha} - 1 \right)^{2+\frac{1}{\alpha}}},
    \label{Eq:Clayton_copula_density}
\end{equation}
where $\alpha$ is the parameter of Clayton copula whose value range is $(0, +\infty)$. The Kendall rank correlation coefficient $\tau_{C}$ corresponding to the bivariate Clayton copula can be computed by $\alpha /\left( 2+\alpha \right)$. The lower tail dependence $\lambda_{C}^{\mathrm{low}}$ equals $2^{-1/ \alpha}$, and the upper tail dependence $\lambda_{C}^{\mathrm{up}}$ equals 0.

Gumbel copula is appropriate to depict the upper tail dependence between variables because it is sensitive to changes at the upper tail. The distribution and density functions of the bivariate Gumbel copula can be expressed as
\begin{equation}
    C_{G}\left( u_{1},u_{2}; \alpha \right) = \exp \left\{ -\left[ \left( -\ln u_{1} \right)^{\alpha} + \left( -\ln u_{2} \right)^{\alpha} \right]^{\frac{1}{\alpha}}  \right\}
    \label{Eq:Gumbel_copula}
\end{equation}
and
\begin{equation}
    c_{G}\left( u_{1},u_{2}; \alpha \right) = \frac{ C_{G}\left( u_{1},u_{2} \right) \left( \ln u_{1} \cdot \ln u_{2} \right)^{\alpha-1} }{u_{1}u_{2} \left[ \left( -\ln u_{1} \right)^{\alpha} + \left( -\ln u_{2} \right)^{\alpha} \right]^{2-\frac{1}{\alpha}}} \left\{ \left[ \left( -\ln u_{1} \right)^{\alpha}+\left( -\ln u_{2} \right)^{\alpha} \right]^{\frac{1}{\alpha}} +\alpha -1 \right\},
    \label{Eq:Gumbel_copula_density}
\end{equation}
where $\alpha$ is the parameter of Gumbel copula whose value range is $(1, +\infty)$. The Kendall rank correlation coefficient $\tau_{G}$ corresponding to the bivariate Gumbel copula can be calculated by $1-1/\alpha$. The lower tail dependence $\lambda_{G}^{\mathrm{low}}$ equals 0, and the upper tail dependence $\lambda_{G}^{\mathrm{up}}$ equals $2-2^{1/ \alpha}$.

As a survival function of Clayton copula, the survival Clayton copula (180-degree rotated Clayton copula) can describe the upper tail dependence between variables. Similarly, the survival Gumbel copula (180-degree rotated Gumbel copula), as a survival function of Gumbel copula, can describe the lower tail dependence between variables. Moreover, to verify that there is no negative correlation between agricultural futures and spot, we also introduce the 90-degree and 270-degree rotated Clayton copulas, which can capture the lower-upper tail dependence and upper-lower tail dependence, respectively. The distribution functions of the survival Clayton copula, survival Gumbel copula, 90-degree rotated Clayton copula, and 270-degree rotated Clayton copula can be defined as follows:
\begin{equation}
    C_{SC}\left( u_{1}, u_{2}; \alpha \right) = u_{1} + u_{2} - 1 + \left[ \left(1-u_{1}\right)^{-\alpha} + \left( 1-u_{2}\right)^{-\alpha} - 1 \right]^{-\frac{1}{\alpha}},\ \alpha \in (0, +\infty),
    \label{Eq:Survival_Clayton_copula}
\end{equation}

\begin{equation}
    C_{SG}\left( u_{1}, u_{2}; \alpha \right) = u_{1} + u_{2} - 1 + \exp \left\{ -\left[ \left( -\ln (1-u_{1}) \right)^{\alpha} + \left( -\ln (1-u_{2}) \right)^{\alpha} \right]^{\frac{1}{\alpha}}  \right\},\ \alpha \in (1, +\infty),
    \label{Eq:Survival_Gumbel_copula}
\end{equation}

\begin{equation}
    C_{R_{90}C}\left( u_{1}, u_{2}; \alpha \right) = u_{2} - \left[ \left(1-u_{1}\right)^{-\alpha} + u_{2}^{-\alpha} - 1 \right]^{-\frac{1}{\alpha}},\ \alpha \in (-\infty, 0),
    \label{Eq:90degree_rotated_Clayton_copula}
\end{equation}

\begin{equation}
    C_{R_{270}C}\left( u_{1}, u_{2}; \alpha \right) = u_{1} - \left[ u_{1}^{-\alpha} + \left( 1-u_{2}\right)^{-\alpha} - 1 \right]^{-\frac{1}{\alpha}},\ \alpha \in (-\infty, 0).
    \label{Eq:270degree_rotated_Clayton_copula}
\end{equation}

\subsubsection{Mixed copula}

After sorting out the different characteristics of the above-mentioned single copula functions, it can be found that a single copula model can only capture symmetric tail dependence or asymmetric upper or lower tail dependence, which makes it difficult to accurately depict the dependence structure between variables. Considering the complexity and heterogeneity of realistic financial markets, we further construct mixed copula models to describe the dependence structure in a more comprehensive way.

Mixed copula functions are convex combinations of finite single copula functions \citep{Nelsen-2006}. Specifically, the distribution function of a mixed copula composed of $N$ single copulas can be expressed as
\begin{equation}
    C_{M}\left( u_{1}, u_{2}; \theta_{M} \right) = \sum\limits_{i=1}^{N} \omega_{i} C_{i} \left( u_{1}, u_{2}; \theta_{c}^{i} \right),
    \label{Eq:Mixed_copula}
\end{equation}
where $\theta_{M} = \left( \left( \theta_{c}^{1} \right)^{\prime},\cdots,\left( \theta_{c}^{N} \right)^{\prime},\omega_{1},\cdots,\omega_{N} \right) ^{\prime}$, $\theta_{c}^{i}$ denotes the parameter set of the $i$-th single copula function, and $\omega_{i}$ denotes the weight parameter of the $i$-th single copula function which satisfies $0 \leq \omega_{i} \leq 1$ and $\sum\nolimits_{i=1}^{N} \omega_{i}=1$.

By incorporating the above mixed copula function into our model, the joint distribution function of the bivariate time series $\mathbf{r}_{t} = \left( r_{1,t},r_{2,t} \right)$ can be converted from Eq. (\ref{Eq:Joint_distribution_function}) to Eq. (\ref{Eq:Joint_distribution_function_mixed_copula}), that is,
\begin{equation}
    F\left( r_{1,t},r_{2,t};\theta \right) = \sum\limits_{i=1}^{N} \omega_{i} C_{i,t} \left( F_{1}\left(r_{1,t};\theta_{1} \right),F_{2}\left( r_{2,t};\theta_{2} \right); \theta_{c}^{i} \right),
    \label{Eq:Joint_distribution_function_mixed_copula}
\end{equation}
where $C_{i,t}(\cdot)$ is the distribution function of the $i$-th single copula.

Similarly, the joint density function can also be converted from Eq. (\ref{Eq:Joint_density_function}) to Eq. (\ref{Eq:Joint_density_function_mixed_copula}), that is,
\begin{equation}
    f\left( r_{1,t},r_{2,t};\theta \right) = f_{1}\left(r_{1,t};\theta_{1} \right) \cdot f_{2}\left( r_{2,t};\theta_{2}\right) \cdot \sum\limits_{i=1}^{N} \omega_{i} c_{i,t} \left( F_{1}\left(r_{1,t};\theta_{1} \right),F_{2}\left( r_{2,t};\theta_{2} \right);\theta_{c}^{i} \right),
    \label{Eq:Joint_density_function_mixed_copula}
\end{equation}
where $c_{i,t}(\cdot)$ is the density function of the $i$-th single copula.

The logarithmic likelihood function of Eq. (\ref{Eq:Joint_density_function_mixed_copula}) can be expressed as
\begin{equation}
    L\left( \Theta \right) = L_{c} \left( \psi_{1} \right) + L_{1}\left( \psi_{2,1} \right) + L_{2}\left( \psi_{2,2} \right),
    \label{Eq:Log_likelihood_joint_density_function}
\end{equation}
where $\Theta = \left( \theta_{1}^{\prime},\theta_{2}^{\prime},\left( \theta_{c}^{1} \right)^{\prime},\cdots,\left( \theta_{c}^{N} \right)^{\prime},\omega_{1},\cdots,\omega_{N} \right)^{\prime}$, $L_{c} \left( \psi_{1} \right)$ is the logarithm of the mixed copula density function, and $L_{1}\left( \psi_{2,1} \right)$ and $L_{2}\left( \psi_{2,2} \right)$ are the logarithms of the marginal density functions of $r_{1,t}$ and $r_{2,t}$, which are given by
\begin{equation}
    L_{c} \left( \psi_{1} \right) = \sum\limits_{t=1}^{T} \log \left\{ \sum\limits_{i=1}^{N} \omega_{i} c_{i,t} \left( F_{1}\left(r_{1,t};\theta_{1} \right),F_{2}\left( r_{2,t};\theta_{2} \right);\theta_{c}^{i} \right) \right\},
    \label{Eq:Log_likelihood_copula_density_function}
\end{equation}
\begin{equation}
    L_{1}\left( \psi_{2,1} \right) = \sum\limits_{t=1}^{T} \log \left\{ f_{1} \left(r_{1,t};\theta_{1} \right) \right\},
    \label{Eq:Log_likelihood_marginal_density_r1_function}
\end{equation}
and
\begin{equation}
    L_{2}\left( \psi_{2,2} \right) = \sum\limits_{t=1}^{T} \log \left\{ f_{2} \left(r_{2,t};\theta_{2} \right) \right\},
    \label{Eq:Log_likelihood_marginal_density_r2_function}
\end{equation}
where $\psi_{1} =\theta_M= \left( \left( \theta_{c}^{1} \right)^{\prime},\cdots,\left( \theta_{c}^{N} \right)^{\prime},\omega_{1},\cdots,\omega_{N} \right) ^{\prime}$, $\psi_{2,1} = \theta_{1}^{\prime}$, and $\psi_{2,2} = \theta_{2}^{\prime}$.

\subsection{Marginal distribution modelling}

Considering the autocorrelation and volatility persistence of the return series, we adopt the ARMA$(m, n)$-GARCH$(p, q)$-skewed Student-t model to construct the marginal distributions, which are specified as follows:
\begin{subequations}
  \begin{equation}
    r_{i,t} = \varphi_{0} + \sum\limits_{j=1}^{m} \varphi_{j}r_{i,t-j} + \varepsilon_{i,t} + \sum\limits_{j=1}^{n} \gamma_{j}\varepsilon_{i,t-j},\ i=1,2
    \label{Eq:Marginal_distribution_return},
  \end{equation}
  \begin{equation}
    \varepsilon_{i,t} = \sigma_{i,t}z_{i,t},\ z_{i,t} \sim i.i.d.skst_{v_{i}},
    \label{Eq:Marginal_distribution_error_term}
  \end{equation}
  \begin{equation}
    \sigma_{i,t}^{2} = \alpha_{0} + \sum\limits_{j=1}^{p} \alpha_{j}\varepsilon_{i,t-j}^{2} + \sum\limits_{j=1}^{q} \beta_{j}\sigma_{i,t-j}^{2},
    \label{Eq:Marginal_distribution_conditional_variance}
  \end{equation} 
  \label{Eq:Marginal_distribution}
\end{subequations}
where $\varepsilon_{i,t}$ and $\sigma_{i,t}^{2}$ denote the error term and the conditional variance of the return series, respectively, and $z_{i,t}$ is the standardized residual obeying the skewed Student-t distribution with $v_{i}$ degrees of freedom.

Non-zero skewness and excess kurtosis are allowed for the skewed Student-t distribution, which can present some of the stylized features of financial data. According to \cite{Hansen-1994-IntEconRev}, the density function of the skewed Student-t distribution can be defined as
\begin{equation}
    f\left( z_{t} \mid \nu,\eta \right) = \left\{
    \begin{aligned}
    bc\left[ 1 + \frac{1}{\nu-2}\left( {\frac{bz_{t}+a}{1-\eta}} \right)^{2} \right]^{-(\nu+1)/2}, \ z_{t} < -\frac{a}{b} \\
    bc\left[ 1 + \frac{1}{\nu-2}\left( {\frac{bz_{t}+a}{1+\eta}} \right)^{2} \right]^{-(\nu+1)/2}, \ z_{t} \geq -\frac{a}{b} 
    \end{aligned}
    \right.
    \label{Eq:The_skewed_Student-t_density_distribution}
\end{equation}
where $\nu\in(2,\infty)$ denotes the degrees-of-freedom parameter, and $\eta\in(-1, 1)$ denotes the asymmetric parameter. Moreover, the constants $a$, $b$ and $c$ can be obtained by
\begin{equation}
    \left\{
    \begin{alignedat}{4}
    c &= \frac{\Gamma\left( \frac{\nu+1}{2} \right)} {\Gamma\left( \frac{\nu}{2} \right)\sqrt{\pi(\nu-2)}}\\
    a &= 4\eta c \frac{\nu-2}{\nu-1}\\
    b &= \sqrt{1+3\eta^{2}-a^{2}}
    \end{alignedat}
    \right..
    \label{Eq:The_skewed_Student-t_density_distribution_abc}
\end{equation}

\subsection{Model estimation}

Full information maximum likelihood (FIML) is a complete and systematic method, which can estimate the parameters of the marginal model and the copula model simultaneously. Although FIML can provide the most efficient estimation, the high dimension of the parameter space makes it difficult to maximize the likelihood function. According to \cite{Joe_Xu_1996}, we estimate the mixed copula model by applying the inference for the margins (IFM). Specifically, the IFM method usually consists of two steps: in the first step, the parameters of marginal models are estimated, and in the second step, the parameters of the copula model are estimated given the parameters of marginal models. Particularly, we estimate the marginal ARMA$(m, n)$-GARCH$(p, q)$-skewed Student-t model through different combinations of the lag parameters $m$, $n$, $p$ and $q$ whose values are restricted to the range of 0 to 3. The Akaike information criterion (AIC) is applied as the standard to assess the goodness of fit, and then select the optimal one from a group of candidate models.

In theory, if the exact distribution of the standardized residuals is known, a specific distribution can be used for the transformation of the standardized residuals into a uniform distribution. However, the true distribution of the standardized residuals is empirically unknowable, making it difficult to obtain a uniform distribution by transforming the standardized residuals with a specific distribution. The canonical maximum likelihood (CML) method emphasizes that the transformation of the standardized residuals based on an empirical cumulative distribution function can always contribute to a uniform distribution asymptotically, regardless of the specification of the marginal models. Referring to \cite{Wang-Wu-Lai-2013-JBankFinanc} and \cite{Ji-Bouri-Roubaud-Shahzad-2018-EnergyEcon}, we apply the CML method to obtain a uniform distribution by transforming the standardized residuals according to the empirical marginal cumulative distribution function below:
\begin{equation}
    \hat{F}_{k}(x) = \frac{1}{T+1} \sum\limits_{t=1}^{T} I \left( \hat{\eta}_{k,t} \leq x \right),
    \label{Eq:Empirical_marginal_cumulative_distribution_function}
\end{equation}
where $k=1,2$, and $I(\cdot)$ is the indicator function whose value equals 1 if $\hat{\eta}_{k,t} \leq x$ and 0 otherwise.

For the $j$-th observation of $\hat{\eta}_{k,t}$, its cumulative probability can be obtained by
\begin{equation}
    \hat{u}_{k,j} = \hat{F}_{k} \left( \hat{\eta}_{k,j} \right),
    \label{Eq:Cumulative_probability_for_observation}
\end{equation}
where $j=1,2,...,T$.
Given the parameter estimates of the marginal models, we then estimate the parameters $\psi_{1} = \left( \left( \theta_{c}^{1} \right)^{\prime}, \cdots, \left( \theta_{c}^{N} \right)^{\prime}, \omega_{1}, \cdots, \omega_{N} \right) ^{\prime}$ of the copula model by maximizing the logarithmic likelihood function $L_{c}(\psi_{1})$ as follows:
\begin{equation}
    \psi_{1} = \arg \mathop{ \max}_{\psi_{1}} L_{c}(\psi_{1}).
    \label{Eq:Maximize_the_log-likelihood_function}
\end{equation}

\subsection{Risk spillover measures}

Next, we measure the $VaR$ and $CoVaR$ for the agricultural spot returns based on the estimated results of the marginal and copula models, taking into account both downside and upside risks.

$VaR$ is interpreted as the expected maximum loss of an asset portfolio within a certain period under a given confidence level. Let $r_{1,t}$ and $r_{2,t}$ denote the logarithmic return series of the agricultural futures and spot at time $t=1, \cdots, T$. Given the confidence level $1-\alpha_{i}^{d}$, the downside $VaR$ for $r_{i}$ can be expressed as
\begin{subequations}
  \begin{equation}
    \text{Pr} \left( r_{i,t} \leq VaR_{\alpha_{i}^{d},t}^{r_{i,t}} \right) = \alpha_{i}^{d}, 
    \label{Eq:Downside_VaR}
  \end{equation}
where $i=1,2$, and the value of $\alpha_{i}^{d}$ is set to 0.05 as the measure of the fifth quantile of the return distribution.
Similarly, given the confidence level $1-\alpha_{i}^{u}$, the upside $VaR$ for $r_{i}$ can be expressed as
  \begin{equation}
    \text{Pr} \left( r_{i,t} \geq VaR_{\alpha_{i}^{u},t}^{r_{i,t}} \right) = \alpha_{i}^{u}, 
    \label{Eq:Upside_VaR}
  \end{equation}
\end{subequations}
where the value of $\alpha_{i}^{u}$ is set to 0.95 as the measure of the ninety-fifth quantile of the return distribution.

Based on the constructed ARMA$(m, n)$-GARCH$(p, q)$-skewed Student-t model, $VaR$ for $r_{i}$ can be estimated by
\begin{equation}
    VaR_{\alpha_{i},t}^{r_{i,t}} = \mu_{i,t}+\sigma_{i,t}t_{\nu,\eta}^{-1}(\alpha_{i}),
    \label{Eq:VaR_estimation}
\end{equation}
with
\begin{equation}
    \mu_{i,t} = \varphi_{0} + \sum\limits_{j=1}^{m} \varphi_{j}r_{i,t-j} + \sum\limits_{j=1}^{n} \gamma_{j}\varepsilon_{i,t-j},
    \label{Eq:mu_estimation}
\end{equation}
where $t_{\nu,\eta}^{-1}(\alpha_{i})$ denotes the $\alpha_{i}$ quantile of the skewed Student-t distribution in Eq. (\ref{Eq:The_skewed_Student-t_density_distribution}). The downside $VaR$ values for the agricultural return series can be calculated by the above equations if $\alpha_{i}$ is $\alpha_{i}^{d}$, and the upside $VaR$ values can be computed if $\alpha_{i}$ is $\alpha_{i}^{u}$.

We apply $CoVaR$ to quantify the risk spillover between the agricultural futures and spot returns, because $CoVaR$ has significant advantages in measuring tail dependence and extreme risk spillover between different assets or markets. Specifically, the $CoVaR$ in our paper is interpreted as the $\alpha_{2}$ quantile of the conditional distribution of the agricultural spot returns on the condition that the $\alpha_{1}$ quantile of the conditional distribution of the agricultural futures returns is given. Accordingly, the downside $CoVaR$ for $r_{2}$ can be expressed as
\begin{subequations}
  \begin{equation}
    \text{Pr} \left( r_{2,t} \leq CoVaR_{\alpha_{2}^{d},\alpha_{1}^{d},t}^{r_{2,t} \mid r_{1,t}}  ~\Big|~ r_{1,t} \leq VaR_{\alpha_{1}^{d},t}^{r_{1,t}} \right) = \alpha_{2}^{d} 
    \label{Eq:Downside_CoVaR_estimation},
  \end{equation}
where $\alpha_{1}^{d}$ is set as 0.05.
Similarly, the upside $CoVaR$ for $r_{2}$ can be expressed as
  \begin{equation}
    \text{Pr} \left( r_{2,t} \geq CoVaR_{\alpha_{2}^{u},\alpha_{1}^{u},t}^{r_{2,t} \mid r_{1,t}}  ~\Big|~ r_{1,t} \geq VaR_{\alpha_{1}^{u},t}^{r_{1,t}} \right) = \alpha_{2}^{u}, 
    \label{Eq:Upside_CoVaR_estimation}
  \end{equation}
where $\alpha_{1}^{u}$ is set as 0.95.
\end{subequations}

In order to calculate $CoVaR$ for the agricultural spot returns, the copula method is adopted in this paper. Specifically, we can solve the following equation to obtain the expression of $CoVaR_{\alpha_{2},\alpha_{1},t}^{r_{2,t} \mid r_{1,t}}$ in terms of copulas:
\begin{equation}
    C \left( F_{2,t} \left( CoVaR_{\alpha_{2},\alpha_{1},t}^{r_{2,t} \mid r_{1,t}} \right), F_{1,t} \left( VaR_{\alpha_{1},t}^{r_{1,t}} \right) \right) - \alpha_{2}\alpha_{1}= 0, 
    \label{Eq:CoVaR_estimation_equation}
\end{equation}
where $F_{1,t} \left( VaR_{\alpha_{1},t}^{r_{1,t}} \right) = \alpha_{1}$, $F_{1,t}(\cdot)$ and $F_{2,t}(\cdot)$ denote the marginal distribution functions of the agricultural futures and spot returns, respectively.
Referring to \cite{Reboredo-Ugolini-2015-JIntMoneyFinan, Reboredo-Ugolini-2016-EnergyEcon}, we estimate $CoVaR$ in two steps. Firstly, given the confidence levels for $VaR$ and $CoVaR$ and the exact form of the copula function, the value of $F_{2,t} \left( CoVaR_{\alpha_{2},\alpha_{1},t}^{r_{2,t} \mid r_{1,t}} \right)$ can be obtained by inverting the copula function in Eq. (\ref{Eq:CoVaR_estimation_equation}). Secondly, the value of $CoVaR_{\alpha_{2},\alpha_{1},t}^{r_{2,t} \mid r_{1,t}}$ can be calculated through the inverse of the marginal distribution function of $r_{2,t}$, that is, $F_{2,t}^{-1} \left( F_{2,t} \left( CoVaR_{\alpha_{2},\alpha_{1},t}^{r_{2,t} \mid r_{1,t}} \right) \right)$. Specifically, the downside $CoVaR$ values for $r_{2}$ are computed if $\alpha_{2}=\alpha_{2}^{d}$ and $\alpha_{1}=\alpha_{1}^{d}$, while the upside $CoVaR$ values for $r_{2}$ are evaluated if $\alpha_{2}=\alpha_{2}^{u}$ and $\alpha_{1}=\alpha_{1}^{u}$.

In addition, $\Delta CoVaR$ is further introduced to identify the risk spillover from the agricultural futures market to the agricultural spot market. In our paper, the $\Delta CoVaR$ is interpreted as the change from the $VaR$ for the agricultural spot returns on the condition of an extreme movement of the agricultural futures returns to the $VaR$ for the agricultural spot returns on the condition of a normal state of the agricultural futures returns. Accordingly, the downside and upside $\Delta CoVaR$s are designed as \begin{subequations}
  \begin{equation}
    \Delta CoVaR_{\alpha_{2}^{d},\alpha_{1},t}^{r_{2,t} \mid r_{1,t}} = CoVaR_{\alpha_{2}^{d},\alpha_{1} = 0.05,t}^{r_{2,t} \mid r_{1,t}} - CoVaR_{\alpha_{2}^{d},\alpha_{1} = 0.5,t}^{r_{2,t} \mid r_{1,t}}
    \label{Eq:Downside_delta_CoVaR_estimation}
  \end{equation}
and
  \begin{equation}
    \Delta CoVaR_{\alpha_{2}^{u},\alpha_{1},t}^{r_{2,t} \mid r_{1,t}} = CoVaR_{\alpha_{2}^{u},\alpha_{1} = 0.95,t}^{r_{2,t} \mid r_{1,t}} - CoVaR_{\alpha_{2}^{u},\alpha_{1} = 0.5,t}^{r_{2,t} \mid r_{1,t}},
    \label{Eq:Upside_delta_CoVaR_estimation}
  \end{equation}
\end{subequations}
where the corresponding value of the normal state is set as 0.5, and $CoVaR_{\alpha_{2}^{d},\alpha_{1} = 0.5,t}^{r_{2,t} \mid r_{1,t}}$ and $CoVaR_{\alpha_{2}^{u},\alpha_{1} = 0.5,t}^{r_{2,t} \mid r_{1,t}}$ satisfy
\begin{subequations}
  \begin{equation}
    \text{Pr} \left( r_{2,t} \leq CoVaR_{\alpha_{2}^{d},\alpha_{1} = 0.5,t}^{r_{2,t} \mid r_{1,t}}  ~\Big|~  F_{1,t}(r_{1,t}) = 0.5 \right) = \alpha_{2}^{d}
    \label{Eq:Downside_delta_CoVaR_estimation_requirement}
  \end{equation}
and
  \begin{equation}
    \text{Pr} \left( r_{2,t} \geq CoVaR_{\alpha_{2}^{u},\alpha_{1} = 0.5,t}^{r_{2,t} \mid r_{1,t}}  ~\Big|~  F_{1,t}(r_{1,t}) = 0.5 \right) = \alpha_{2}^{u}.
    \label{Eq:Upside_delta_CoVaR_estimation_requirement}
  \end{equation}
\end{subequations}

On the basis of the above estimation, the Kolmogorov-Smirnov (K-S) test proposed by \cite{Abadie-2002-JAmStatAssoc} is employed to examine the significance of extreme risk spillover effects from the agricultural futures returns to the agricultural spot returns. The K-S test can be given by
\begin{equation}
    KS_{mn} = \left( \frac{mn}{m+n} \right)^{\frac{1}{2}} \mathop{\sup}_{x} \Big| G_{m}(x) - H_{n}(x) \Big|,
    \label{Eq:KS_test}
\end{equation}
where $G_{m}(x)$ and $H_{n}(x)$ denote the distribution functions of $CoVaR$ and $VaR$, whose sample sizes are $m$ and $n$, respectively.

We first examine whether the downside risk spillover of the agricultural futures returns on the agricultural spot returns is statistically significant. According to \cite{Reboredo-Ugolini-2016-EnergyEcon} and \cite{Ji-Bouri-Roubaud-Shahzad-2018-EnergyEcon}, the null hypothesis of the K-S test is designed as $H_{0}: CoVaR_{\alpha_{2}^{d},\alpha_{1}^{d},t}^{r_{2,t} \mid r_{1,t}} = VaR_{\alpha_{2}^{d},t}^{r_{2,t}}$, indicating no significant difference between the downside $CoVaR$ and $VaR$. In contrast, the alternative hypothesis is $H_{1}: CoVaR_{\alpha_{2}^{d},\alpha_{1}^{d},t}^{r_{2,t} \mid r_{1,t}} < VaR_{\alpha_{2}^{d},t}^{r_{2,t}}$, which suggests that there exist significant downside risk spillover effects from the agricultural futures market to the agricultural spot market.

Similarly, the significance of the upside risk spillover effects from the agricultural futures returns to the agricultural spot returns is investigated. The null hypothesis of there being no upside risk spillover is set as $H_{0}: CoVaR_{\alpha_{2}^{u},\alpha_{1}^{u},t}^{r_{2,t} \mid r_{1,t}} = VaR_{\alpha_{2}^{u},t}^{r_{2,t}}$, while the alternative hypothesis is set as $H_{1}: CoVaR_{\alpha_{2}^{u},\alpha_{1}^{u},t}^{r_{2,t} \mid r_{1,t}} > VaR_{\alpha_{2}^{u},t}^{r_{2,t}}$, which implies that the agricultural futures market exhibits significant upside risk spillover effects on the agricultural spot market.

Furthermore, in order to explore the possible asymmetry between the downside and upside risk spillover effects, we apply the K-S test to compare the normalized downside $CoVaR$ and the normalized upside $CoVaR$, in which the null hypothesis of there being no significant strength differences between the downside and upside risk spillover effects is defined as
\begin{equation}
    H_{0}: \frac{CoVaR_{\alpha_{2}^{d},\alpha_{1}^{d},t}^{r_{2,t} \mid r_{1,t}}}{VaR_{\alpha_{2}^{d},t}^{r_{2,t}}} = \frac{CoVaR_{\alpha_{2}^{u},\alpha_{1}^{u},t}^{r_{2,t} \mid r_{1,t}}}{VaR_{\alpha_{2}^{u},t}^{r_{2,t}}}.
    \label{Eq:CoVaR_test}
\end{equation}

\section{Data description}
\label{S1:Data}

\subsection{Data source}

We select four typical agricultural futures listed on the Chicago Board of Trade (CBOT), namely soybean, corn, wheat, and rough rice, as representatives of the international agricultural futures market. The reasons for our data sample selection are as follows. CBOT is the largest and most influential agricultural futures exchange worldwide, and offers a broad range of global benchmarks across major agricultural commodities. In addition, most bulk commodities in the international trade market, including agricultural commodities, are usually denominated and paid in dollars. The agricultural futures prices of CBOT have become the primary reference prices for international pricing of important agricultural commodities. Moreover, soybean, corn, wheat, and rough rice are highly cosmopolitan agricultural commodities, and their futures trading volumes have always been at the forefront of the international commodity futures market. Meanwhile, as the main food crops of the world, the four agricultural commodities also play an essential role in meeting the basic survival needs of human beings and coping with the risks of global food security.

Corresponding to soybean, corn, wheat, and rough rice futures, we include four sub-indexes under the Grains and Oilseeds Index (GOI) issued by the International Grains Council (IGC), namely soybean, maize, wheat, and rice, as the research objects of the international agricultural spot market. The soybean sub-index is derived from the export quotations of soybeans from the United States, Brazil, and Argentina, which are the top three soybean producers in the world, collectively accounting for 81.69\% of the global soybean production in 2021. Similarly, the maize sub-index is calculated by the maize export prices of Brazil, the United States, Argentina, and the Black Sea region. The wheat sub-index consists of ten export quotations of wheat from Argentina, Australia, the Black Sea region, Canada, the European Union, and the United States. The rice sub-index is based on rice export prices from the major rice producers, including India, Pakistan, Thailand, Uruguay, the United States, and Vietnam. Therefore, the four sub-indexes released by IGC can well reflect the international spot-price trends of soybean, maize, wheat, and rice.

Considering that the IGC's agricultural price indexes take January 3, 2000 as the base period, we further select the daily closing prices of continuous contracts of soybean, corn, wheat, and rough rice futures from January 3, 2000 to April 29, 2022, as well as the daily data of soybean, maize, wheat, and rice sub-indexes in the sample period. Our data of agricultural futures are sourced from the Wind database, and the data of agricultural spot are collected from the IGC website.

\subsection{Statistical description}

\begin{figure}[!h]
  \centering
  \includegraphics[width=0.475\linewidth]{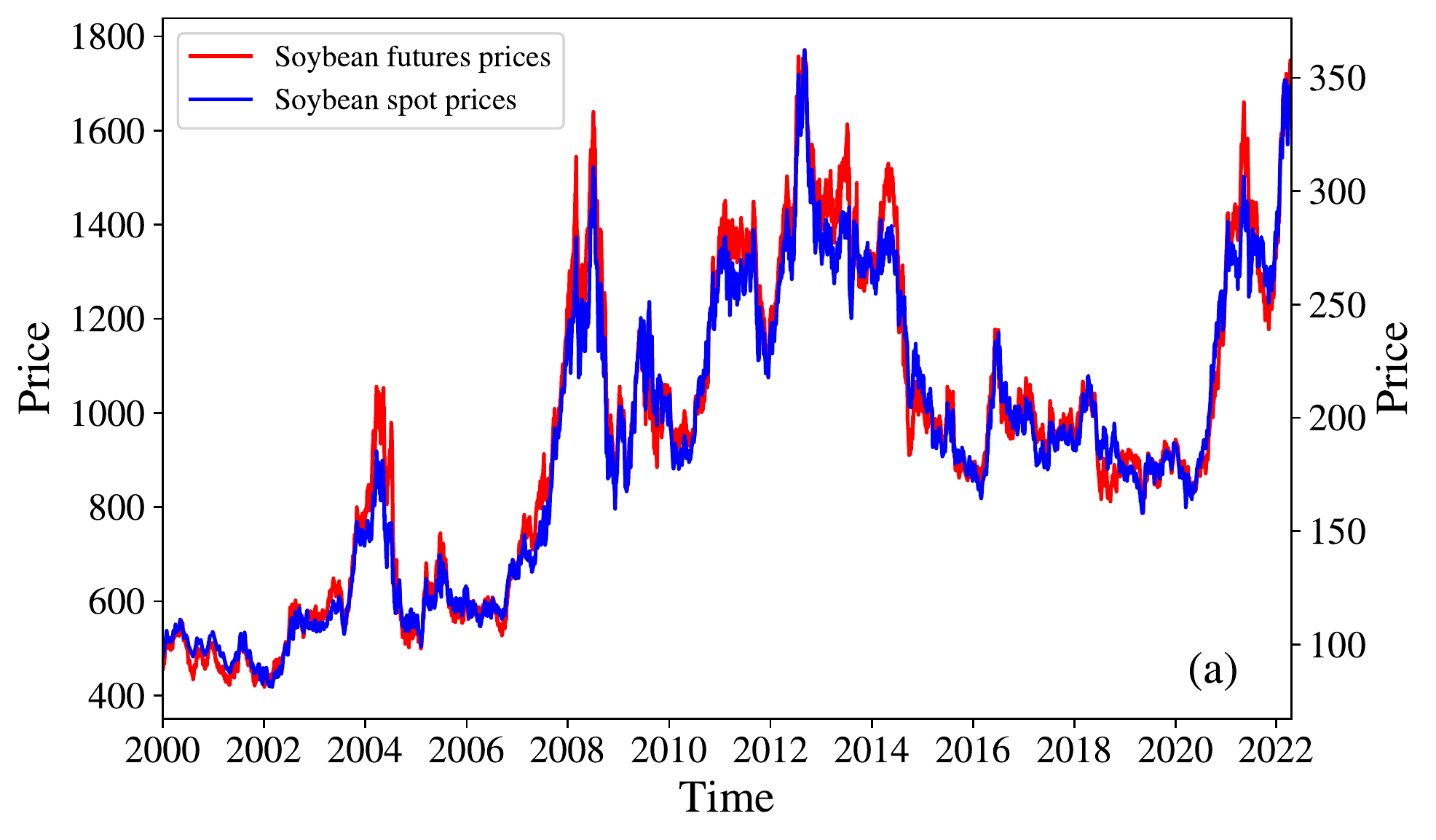}
  \includegraphics[width=0.475\linewidth]{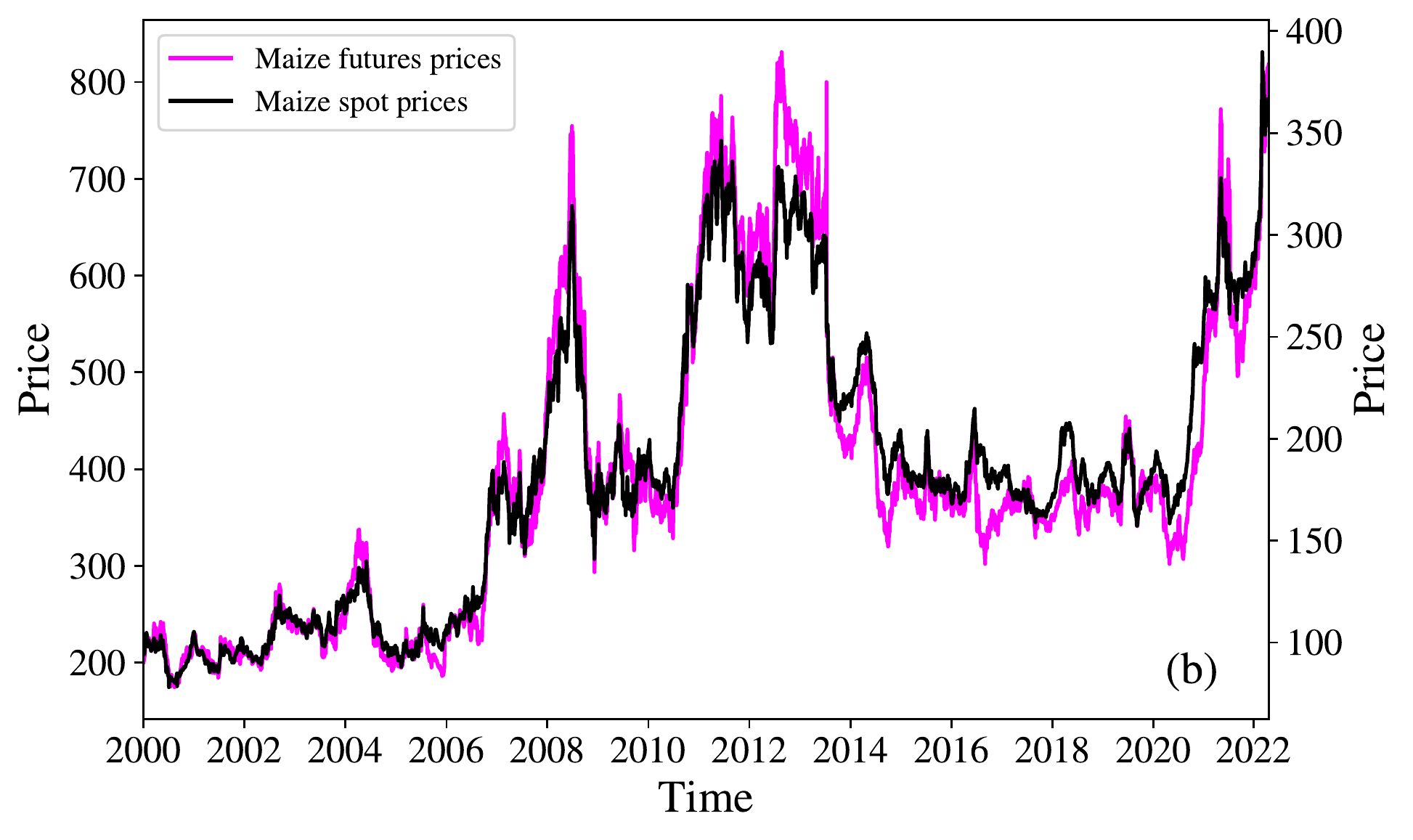}\\
  \includegraphics[width=0.475\linewidth]{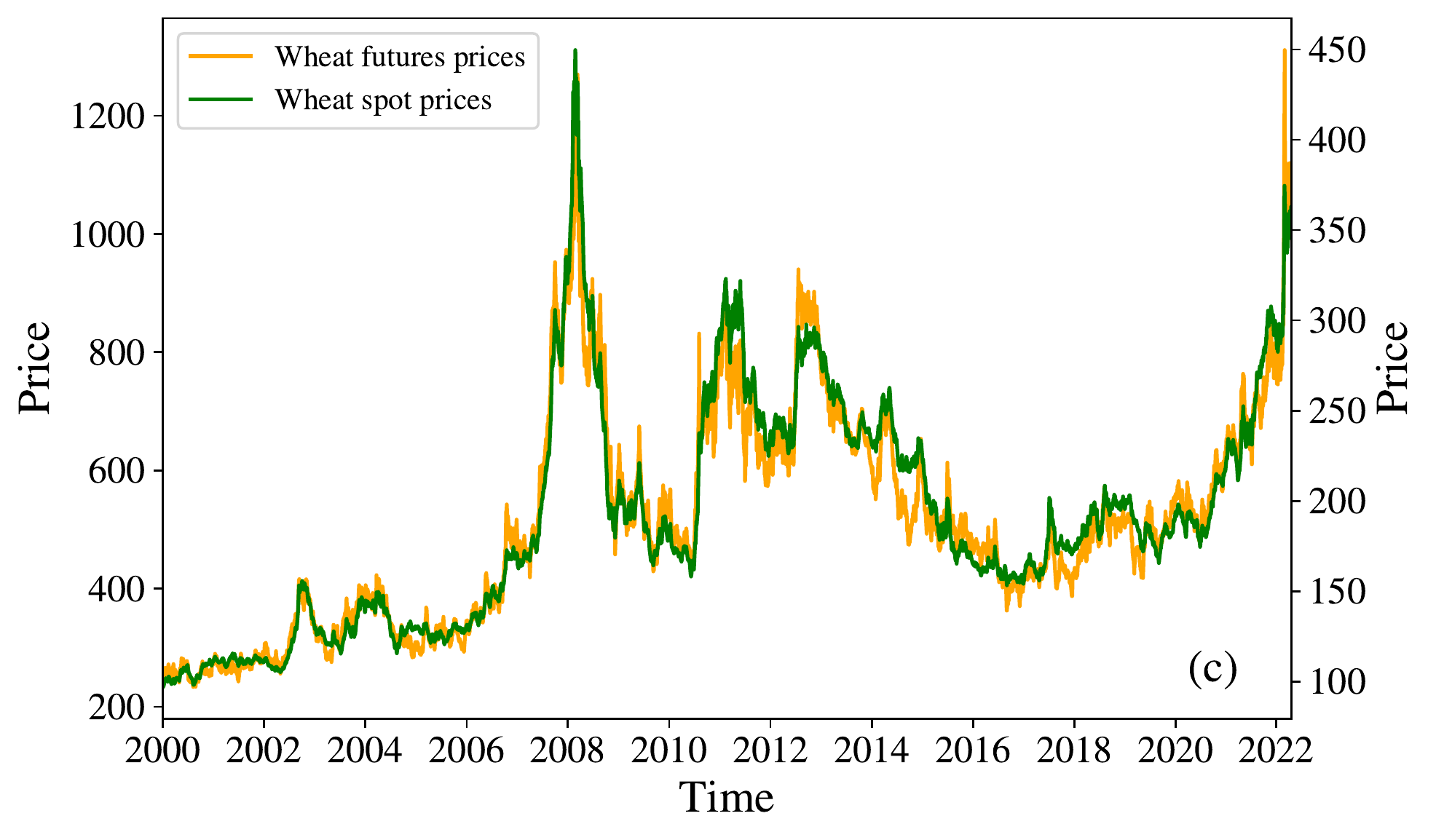}
  \includegraphics[width=0.475\linewidth]{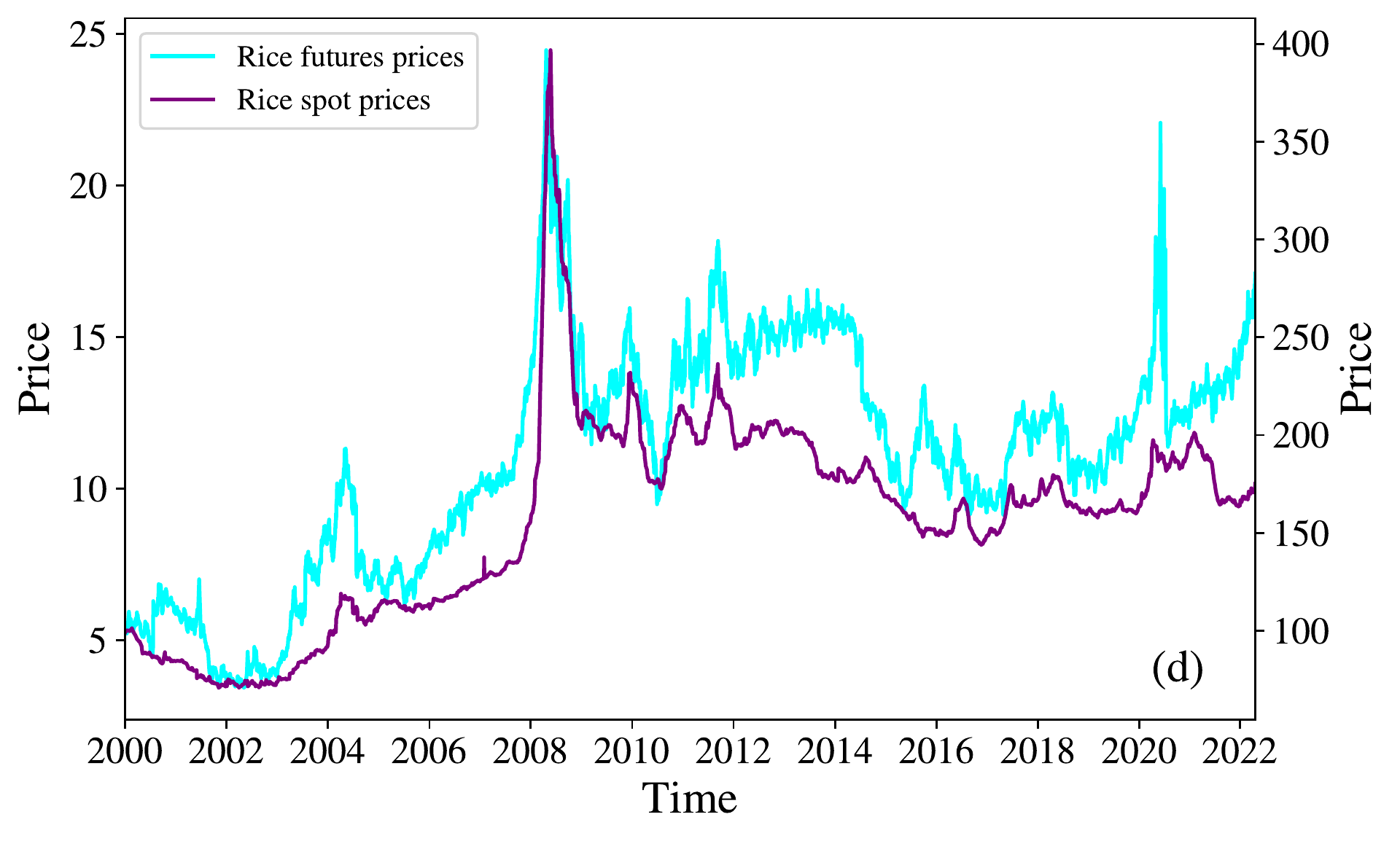}
  \caption{Evolution of futures and spot prices of soybean (a), maize (b), wheat (c), and rice (d).}
\label{Fig:AgroPrice_evolution}
\end{figure}

Figure~\ref{Fig:AgroPrice_evolution}(a)-(d) depict the price evolution of soybean, corn, wheat, and rough rice futures and their corresponding spot. We note that the price trends between futures and spot of soybean, corn, and wheat are highly consistent, while the trends between rough rice futures and rice spot are slightly different. Due to frequent extreme weather from 2006 to 2008, agricultural output declined year after year. The futures and spot prices of all agricultural commodities soared before the global financial crisis in 2008, and then fell sharply, with multiple wave crests during the subsequent period of recovery. Between 2010 and 2012, extreme weather caused widespread production cuts in agricultural commodities, leading many countries to implement protectionist trade policies to restrict food exports, which triggered a resurgence of agricultural price rises around the world. The La Nina events in 2020, as well as the spread of the COVID-19 pandemic, have severely disrupted global food supply and demand. In addition, the gradual recovery of the world economy, combined with frequent geopolitical conflicts, has boosted the prices of the main agricultural commodities since 2021.

Table~\ref{Tab:Correlation_Coefficient} presents the Pearson product-moment correlation coefficients, Kendall's tau coefficients, and Spearman's rank correlation coefficients between futures and spot price series of soybean, maize, wheat, and rice. These three correlation coefficients are the three most important correlation coefficients in statistics, which can be utilized to quantify the correlation between different variables. These correlation coefficients are restricted to the interval $[-1,1]$, where $-$1 indicates a perfect negative correlation, 1 corresponds to a perfect positive correlation, and 0 represents no correlation. From Table~\ref{Tab:Correlation_Coefficient}, we find that the correlation coefficients of the agricultural commodity pairs are large and significant at the 1\% level. These three correlation coefficients obtain a consistent conclusion, that is, there exists a significant positive correlation between the futures and spot price series of each agricultural commodity. Specifically, the positive correlation between soybean futures and spot is the highest, followed by wheat and maize, and finally rice. This result supports the similar trends shown in Figure~\ref{Fig:AgroPrice_evolution}.

\begin{table}[!h]
  \centering
  \setlength{\abovecaptionskip}{0pt}
  \setlength{\belowcaptionskip}{10pt}
  \caption{Correlation coefficients between futures and spot price series of agricultural commodities} %\vspace{-2mm}
  \setlength\tabcolsep{12.3pt}
  \begin{tabular}{lcccc}
    \toprule
        & \multicolumn{1}{c}{Soybean} & \multicolumn{1}{c}{Maize} & \multicolumn{1}{c}{Wheat} & \multicolumn{1}{c}{Rice} \\
    \midrule
    Pearson product-moment correlation coefficient & 0.9893$^{***}$ & 0.9766$^{***}$ & 0.9786$^{***}$ & 0.9200$^{***}$ \\
    Kendall's tau coefficient & 0.9072$^{***}$ & 0.8341$^{***}$ & 0.8855$^{***}$ & 0.7552$^{***}$ \\
    Spearman's rank correlation coefficient & 0.9857$^{***}$ & 0.9545$^{***}$ & 0.9815$^{***}$ & 0.9161$^{***}$ \\
    \bottomrule
  \end{tabular} 
  \begin{flushleft}
    \footnotesize
    \justifying Note: This table presents the results of the three most important correlation coefficients in statistics, which can quantify the correlation between the futures and spot price series of soybean, maize, wheat, and rice. These correlation coefficients are restricted to the interval $[-1,1]$, where 0 represents no correlation, and $-$1 corresponds to a perfect negative correlation while 1 corresponds to a perfect positive correlation. Superscript *** denotes significance at the 1\% level.
 \end{flushleft} 
  \label{Tab:Correlation_Coefficient}%
\end{table}%
% \vspace{-2mm}

%\subsection{Statistical description and tests of the return series}

To calculate the return series of the agricultural futures and spot, we first align the price series by following the steps below. Considering the empty data points before the listing of the agricultural futures, the closing prices of their listing day are used as surrogates. The missing data after the listing of the agricultural futures is supplemented by the prices of the previous day. In the case of the missing data points of the previous day, the prices of the previous two days are used as supplements, and the rest can be done in the same manner. We then match the daily closing price data of continuous contracts of the agricultural futures with the daily price data of the agricultural spot.

Based on the data alignment, we compute the daily logarithmic return of the agricultural futures or spot over a time scale $\Delta t$ and multiply it by 100 as follows:
\begin{equation}\label{Eq:Logarithmic_return}
   r_i(t) = \ln{\frac{P_i(t)}{P_i(t - \Delta t)}} \times 100
\end{equation}
where $i=1,2$, $P_{1}(t)$ and $P_{2}(t)$ denote the daily prices of the agricultural futures and spot, respectively, and the time scale $\Delta t$ equals 1 day.

\begin{figure}[!ht]
\centering
\includegraphics[width=0.246\linewidth]{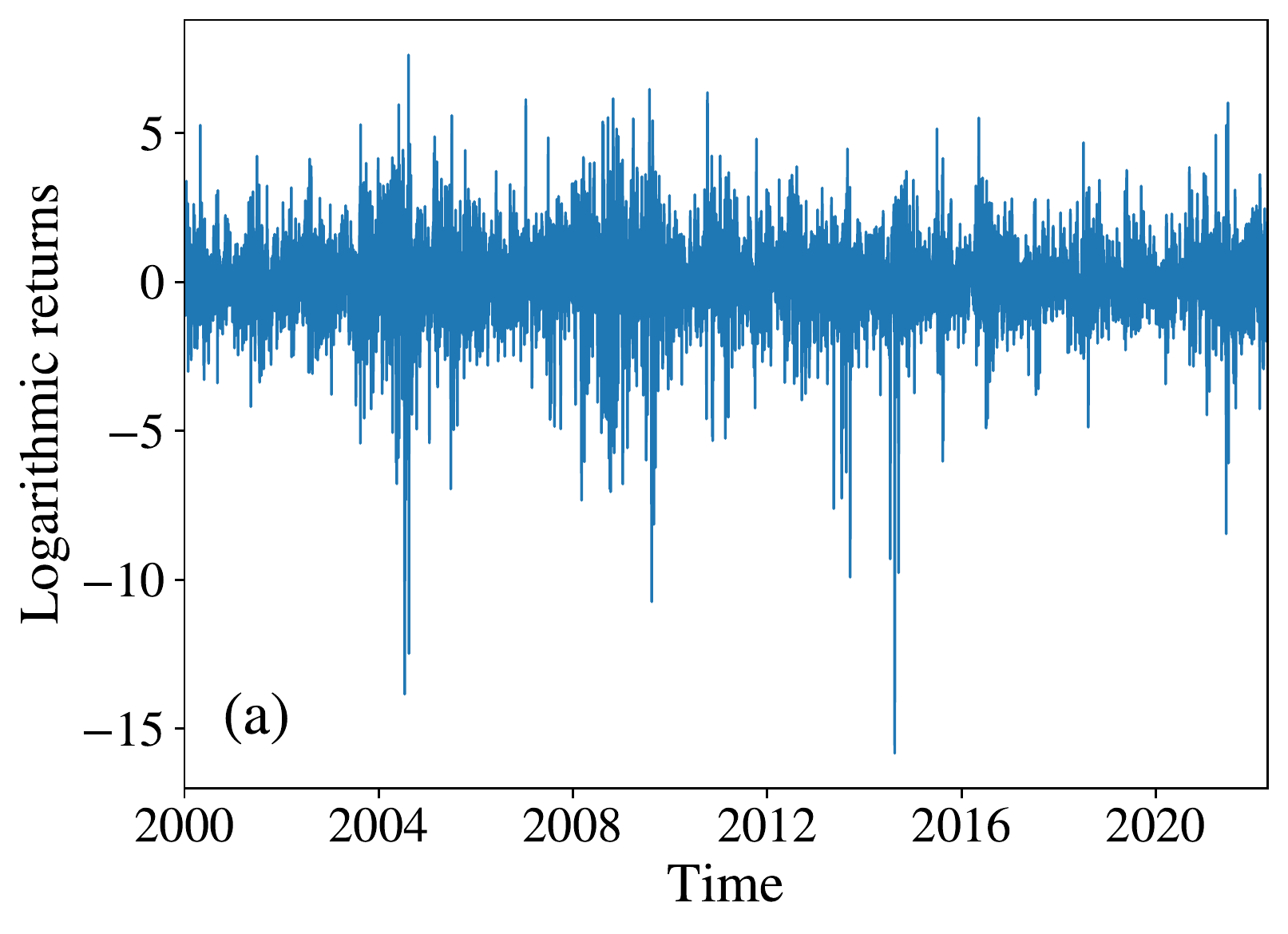}
\includegraphics[width=0.246\linewidth]{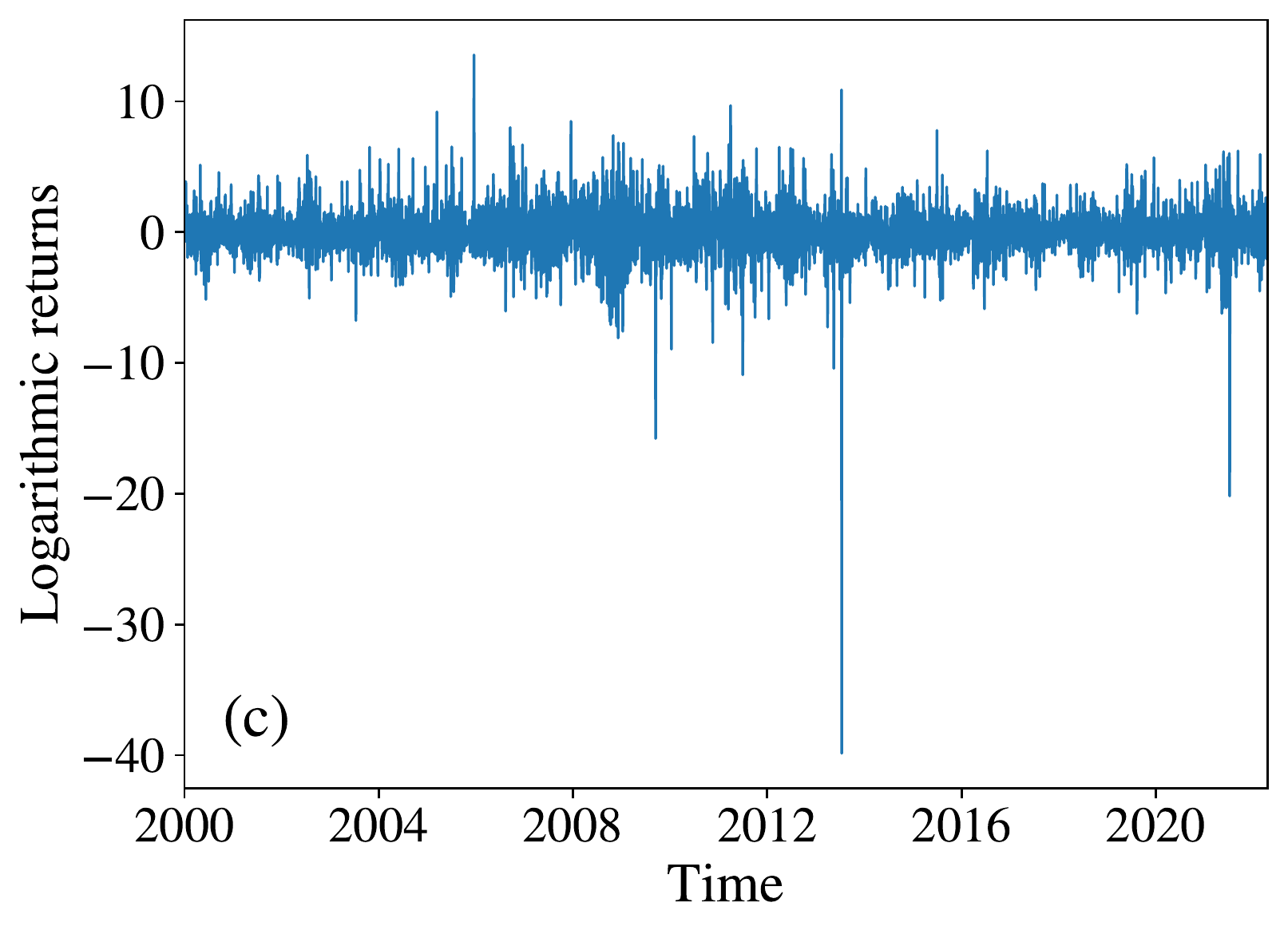}
\includegraphics[width=0.246\linewidth]{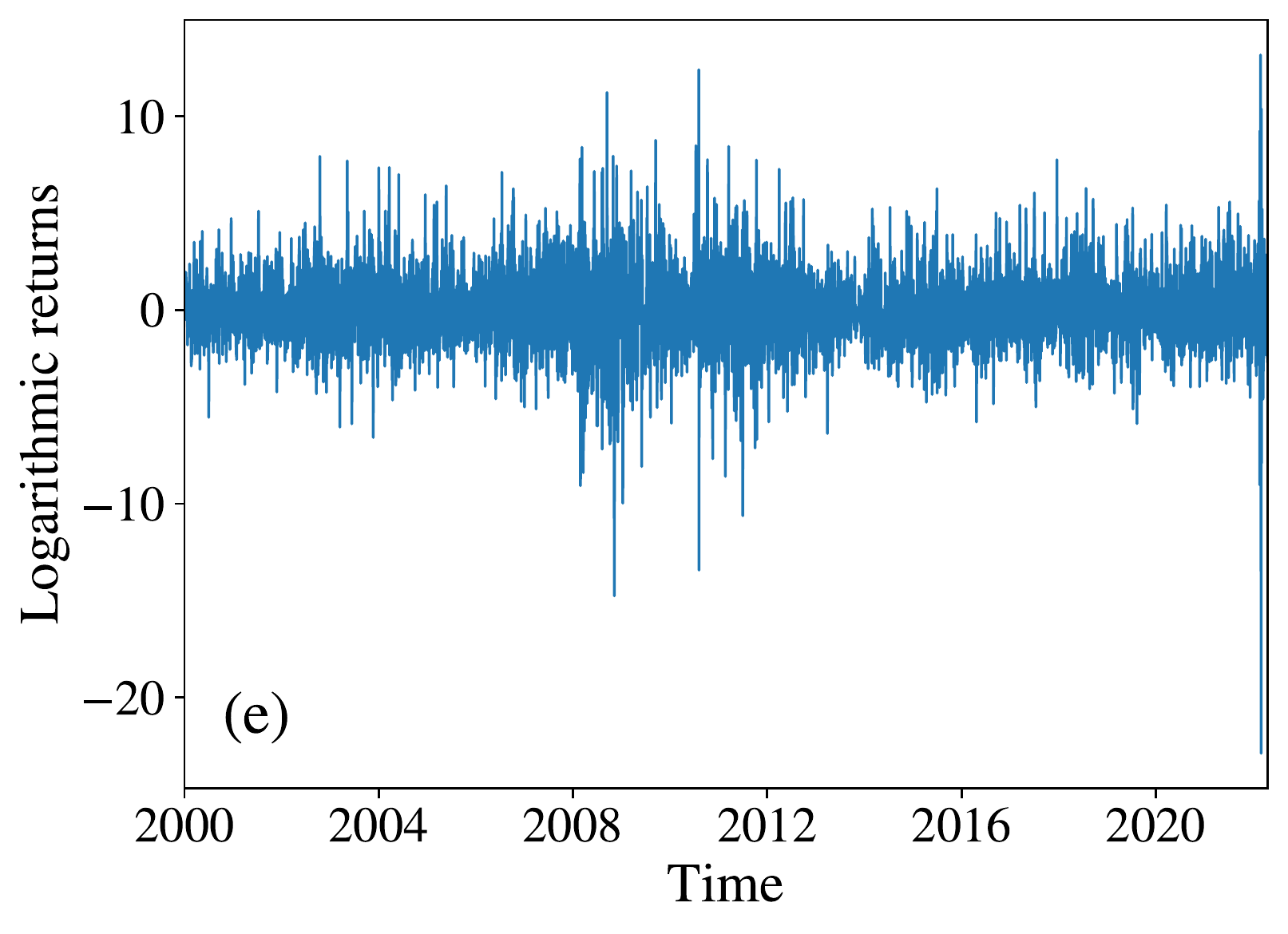}
\includegraphics[width=0.246\linewidth]{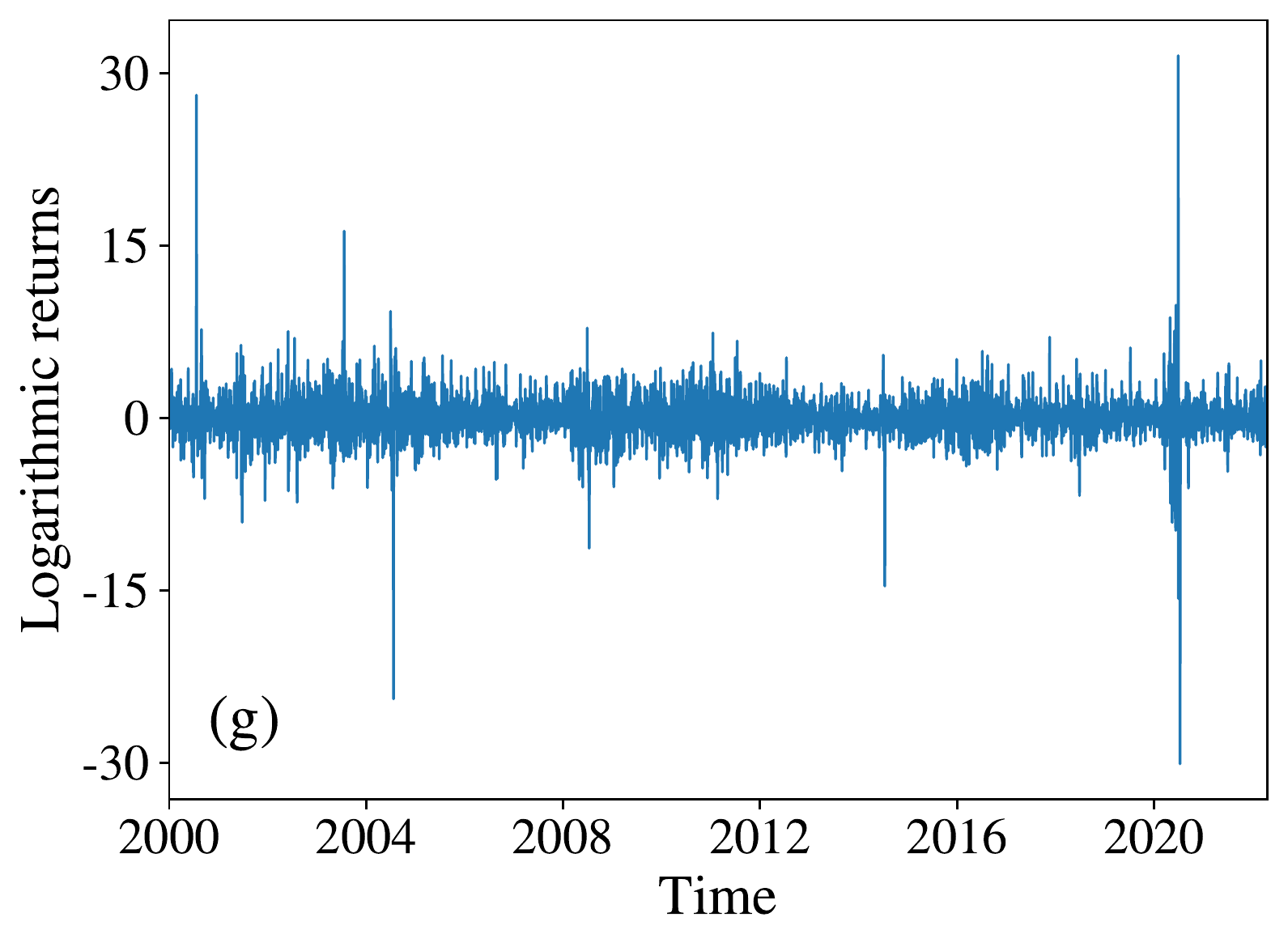}
\includegraphics[width=0.246\linewidth]{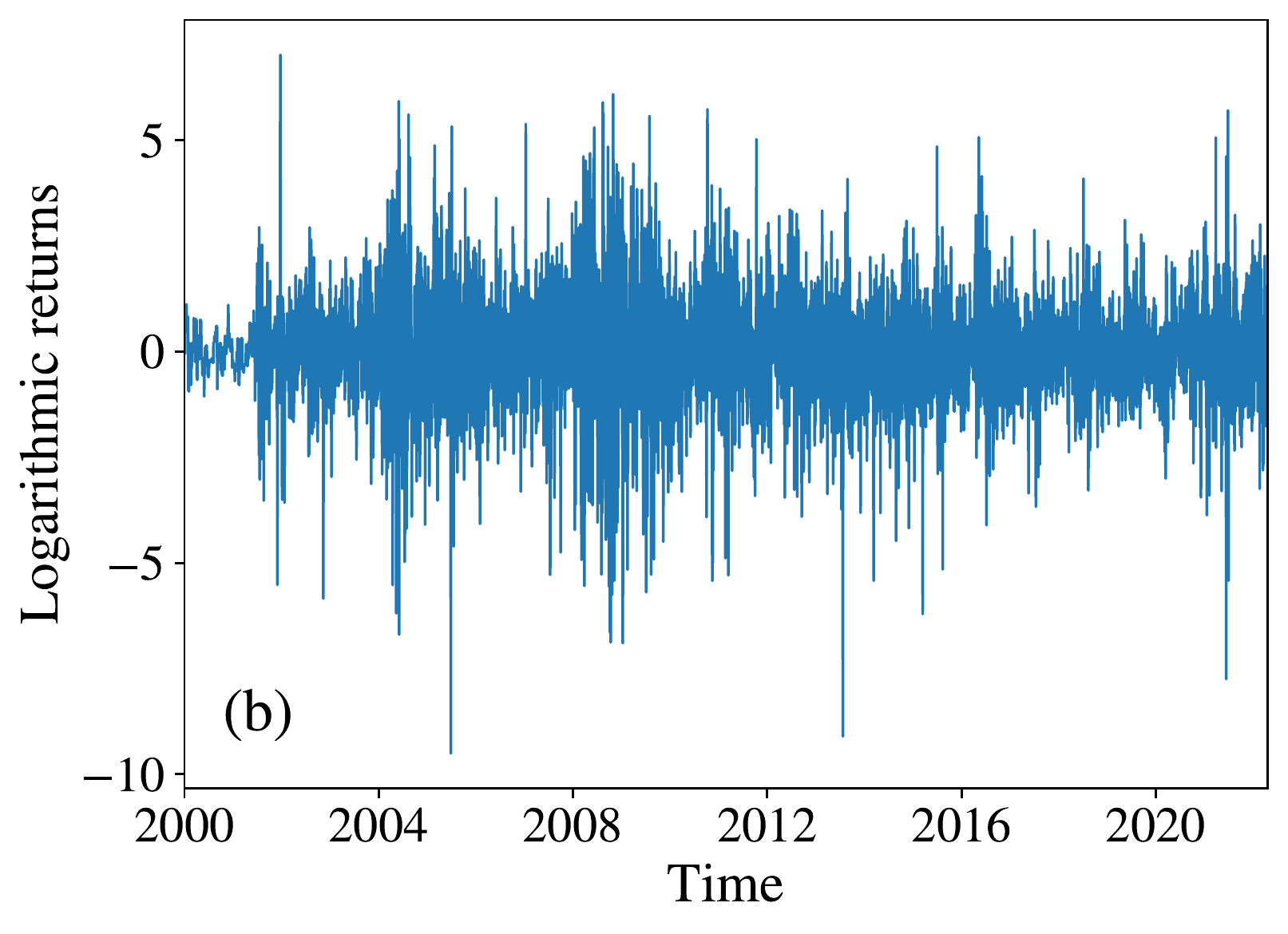}
\includegraphics[width=0.246\linewidth]{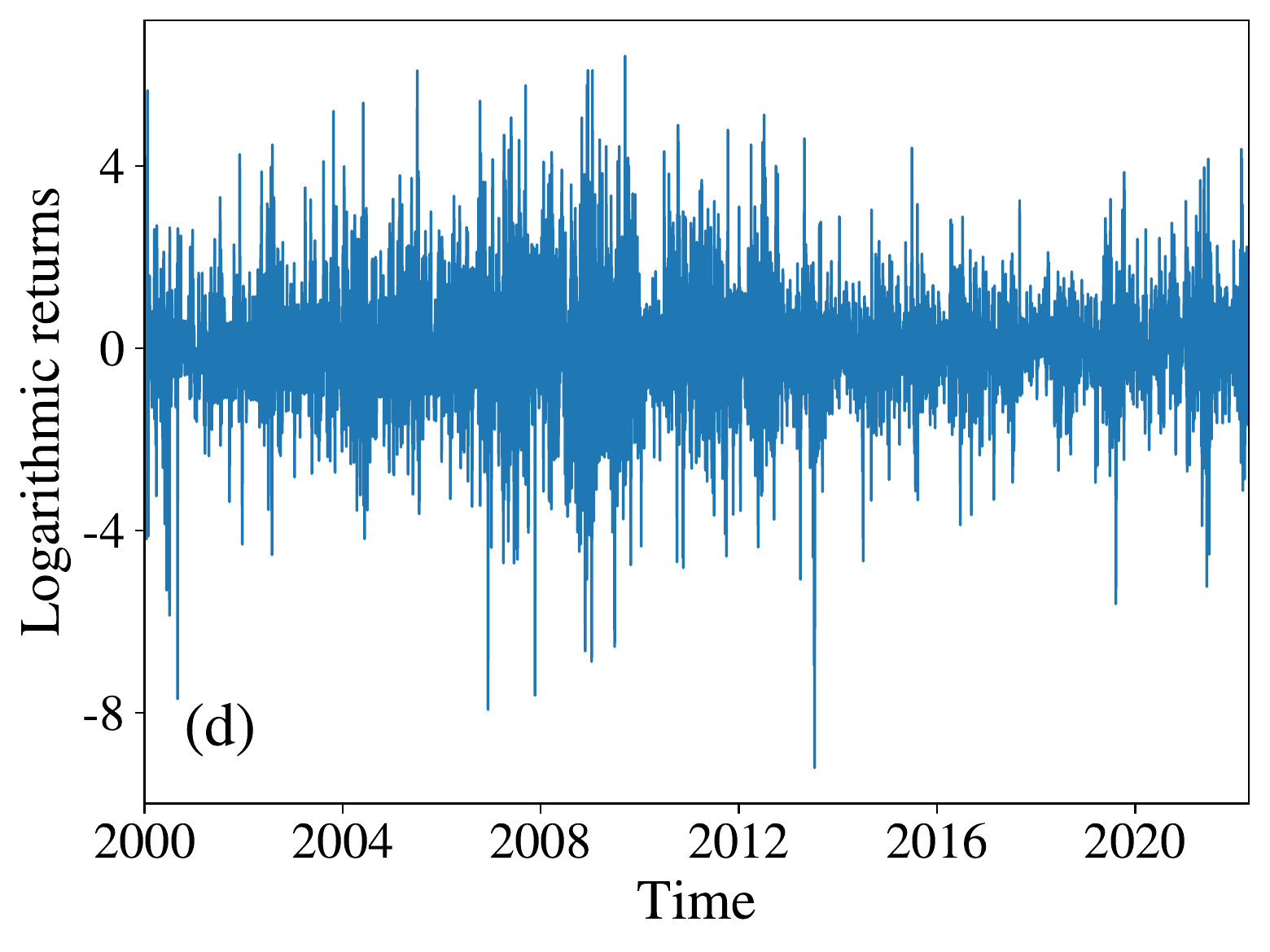}
\includegraphics[width=0.246\linewidth]{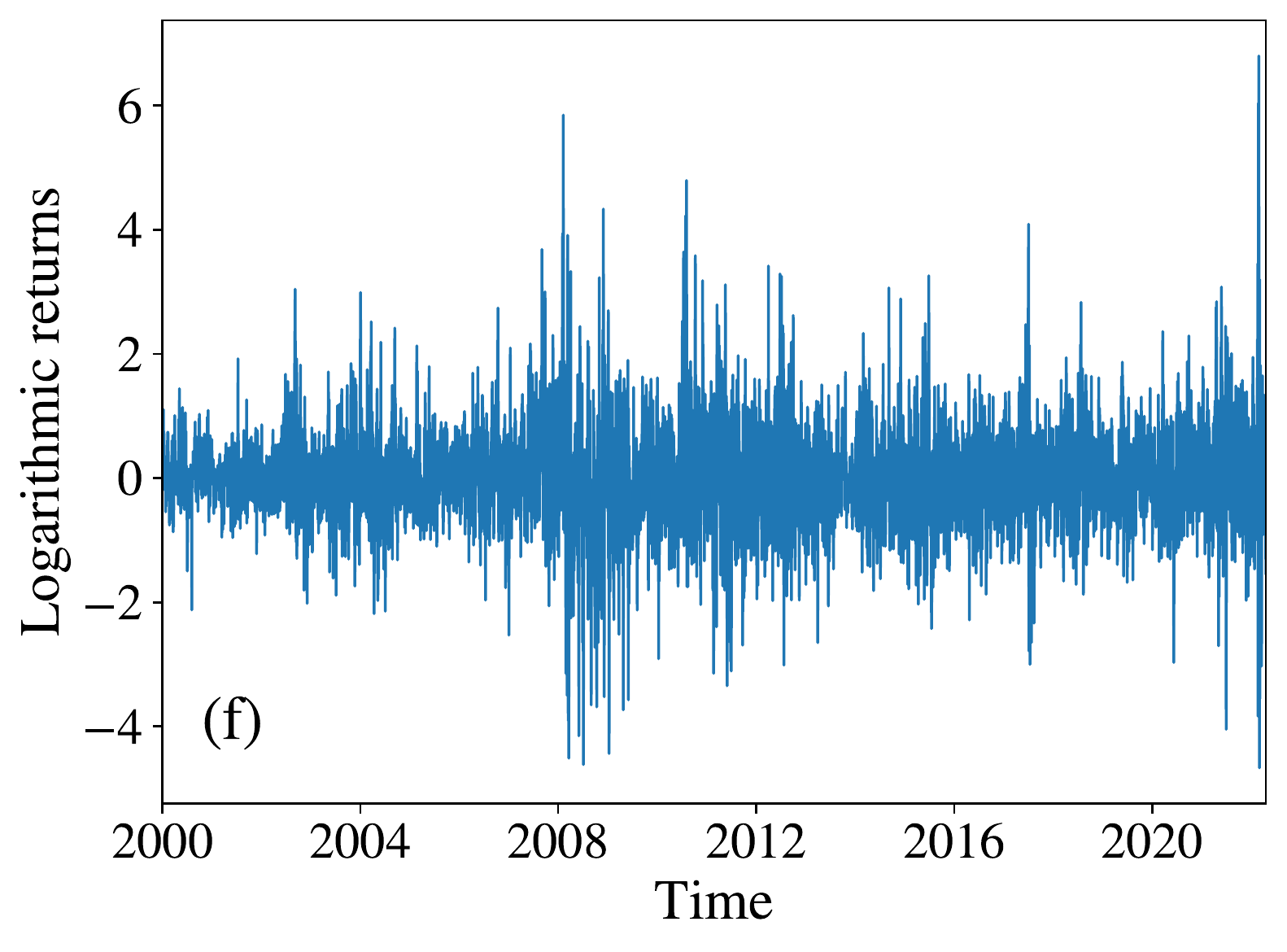}
\includegraphics[width=0.246\linewidth]{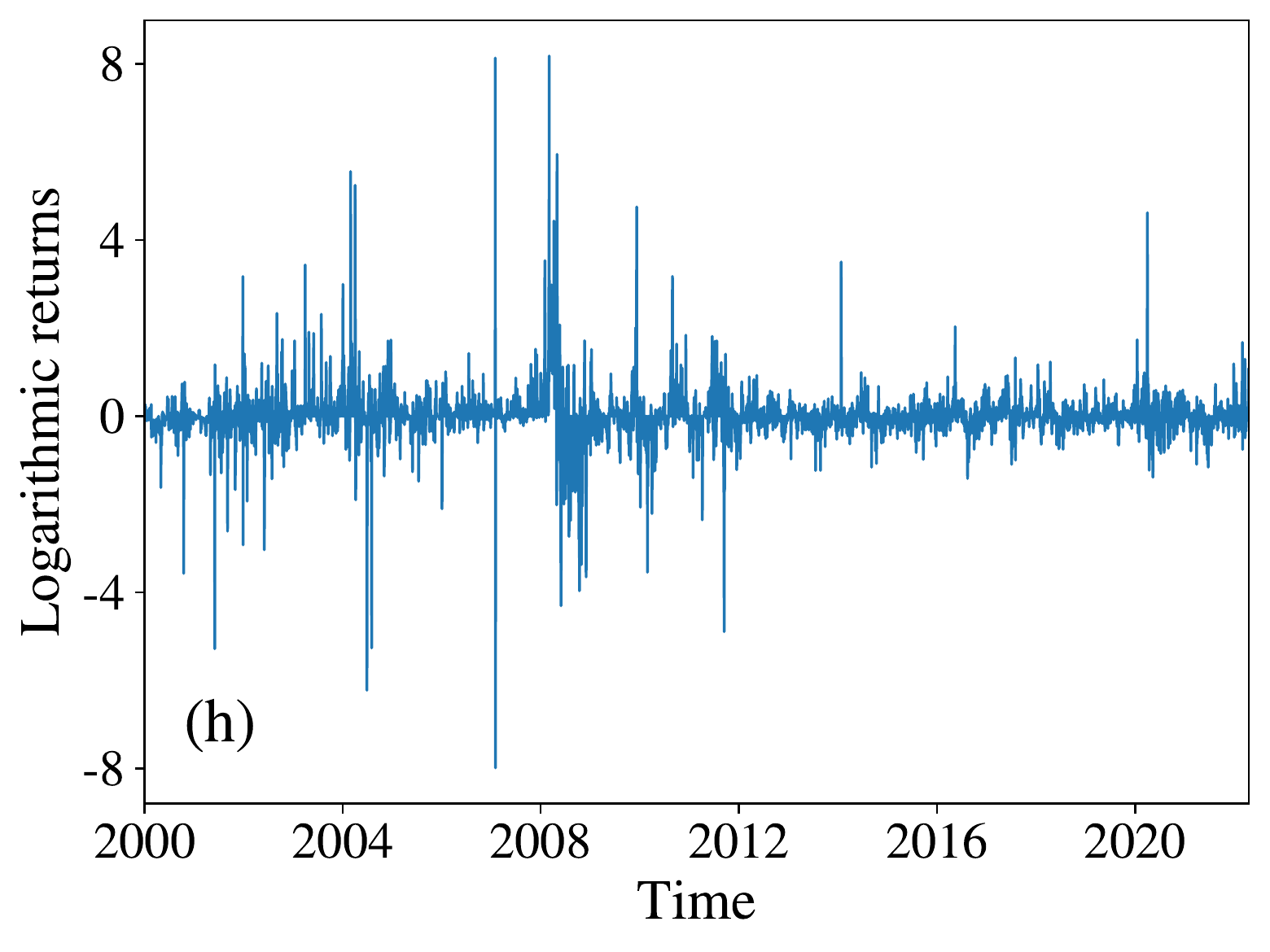}
\caption{Evolution of soybean futures returns (a), soybean spot returns (b), corn futures returns (c), maize spot returns (d), wheat futures returns (e), wheat spot returns (f), rough rice futures returns (g), and rice spot returns (h).}
\label{Fig:AgroReturns_evolution}
\end{figure}

Figure~\ref{Fig:AgroReturns_evolution} describes the dynamic evolution of the return series of soybean, corn, wheat, and rough rice futures and their corresponding spot. We find that the return trends and fluctuation scales vary with different agricultural commodities and different markets, but all return series exhibit the phenomenon of volatility clustering. In addition, the return series of all agricultural commodities fluctuated dramatically during 2004-2006 and 2008-2012, and their fluctuation ranges have increased again since 2020, which is consistent with Figure~\ref{Fig:AgroPrice_evolution}.

Table~\ref{Tab:Agro_Stat_Test} further presents the descriptive statistics of each agricultural return series. After comparing the results, we note that the maximum values (Max), the absolute minimum values ($-$Min) and the standard deviations (Std. Dev.) of the return series for soybean, corn, wheat, and rough rice futures are all larger than those of their respective spot return series, which implies that the agricultural futures markets are more volatile than the agricultural spot markets. The mean value (Mean) of each return series is positive, and smaller than its corresponding standard deviation. From the panel A of Table~\ref{Tab:Agro_Stat_Test}, we also find that the skewness (Skewness) of each agricultural return series is significantly not 0, indicating that the distributions of these return series are all right- or left-skewed. Moreover, the kurtosis (Kurtosis) of each return series is greater than 3, which is the kurtosis of a normal distribution, showing that these agricultural return series have some stylized properties such as leptokurtosis and fat tail.

Table~\ref{Tab:Agro_Stat_Test} also reports several diagnostic test results for the agricultural return series. The Jarque-Bera test is to examine whether the skewness and kurtosis of the sample data conforming to a normal distribution. From the panel B of Table~\ref{Tab:Agro_Stat_Test}, the normality of each agricultural return series is rejected as proved by the Jarque-Bera statistics, whose values are far greater than 0 and significant at the 1\% level. Moreover, the significant ADF and PP statistics, as well as the insignificant KPSS statistics draw a unanimous conclusion that all agricultural return series are stationary. The Ljung-Box test results, including the Q statistics for the return series and the Q$^{2}$ statistics for the squared return series, indicate that all agricultural return series do not follow the white noise process, and confirm the existence of autocorrelation and heteroscedasticity. In addition, the statistics of the ARCH-LM test also verify the presence of ARCH effect in each return series. Hence, GARCH-class models can be constructed in this paper for further study.

\begin{table}[!ht]
  \centering
  \setlength{\abovecaptionskip}{0pt}
  \setlength{\belowcaptionskip}{10pt}
  \caption{Descriptive statistics and diagnostic tests for futures and spot return series of agricultural commodities}
  \setlength\tabcolsep{3pt}   \resizebox{\textwidth}{!}{ 
    \begin{tabular}{l r@{.}l r@{.}l r@{.}l r@{.}l c r@{.}l r@{.}l r@{.}l r@{.}l}
    \toprule
         & \multicolumn{8}{c}{Futures} && \multicolumn{8}{c}{Spots}  \\
    \cline{2-9} \cline{11-18}
         & \multicolumn{2}{c}{Soybean} & \multicolumn{2}{c}{Corn} & \multicolumn{2}{c}{Wheat} & \multicolumn{2}{c}{Rough rice} && \multicolumn{2}{c}{Soybean} & \multicolumn{2}{c}{Maize} & \multicolumn{2}{c}{Wheat} & \multicolumn{2}{c}{Rice}  \\
    \midrule
    \multicolumn{18}{l}{\textit{Panel A: Descriptive statistics}} \\
    Max & 7&6292 & 13&5589 & 13&1690 & 31&5128 && 7&0204 & 6&4144 & 6&7993 & 8&1817 \\
    Min & $-$15&8290 & $-$39&8614 & $-$22&8880 & $-$30&1080 && $-$9&5068 & $-$9&2090 & $-$4&6677 & $-$7&9870 \\
    Mean & 0&0236 & 0&0250 & 0&0256 & 0&0211 && 0&0226 & 0&0236 & 0&0232 & 0&0101 \\
    Std. Dev. & 1&5494 & 1&8517 & 2&0228 & 1&8406 && 1&3461 & 1&3336 & 0&8432 & 0&5000 \\
    Skewness & $-$0&9073 & $-$1&9584 & $-$0&0660 & 0&1904 && $-$0&3875 & $-$0&1437 & 0&2832 & 0&9732 \\
    Kurtosis & 7&1195 & 43&9913 & 5&9528 & 47&3749 && 3&7119 & 3&2484 & 4&4666 & 67&4358     \vspace{2mm}\\
    \multicolumn{18}{l}{\textit{Panel B: Diagnostic tests}} \\
    Jarque-Bera & \multicolumn{2}{r}{12627$^{***}$} & \multicolumn{2}{r}{456272$^{***}$} & \multicolumn{2}{r}{8293$^{***}$} & \multicolumn{2}{r}{525033$^{***}$} && \multicolumn{2}{r}{3363$^{***}$} & \multicolumn{2}{r}{2488$^{***}$} & \multicolumn{2}{r}{4742$^{***}$} & \multicolumn{2}{r}{1064644$^{***}$} \\
    ADF         & $-$23&56$^{***}$ & $-$74&79$^{***}$ & $-$16&36$^{***}$ & $-$21&90$^{***}$ && $-$41&60$^{***}$ & $-$15&36$^{***}$  & $-$16&23$^{***}$  & $-$9&22$^{***}$  \\
    PP          & $-$74&64$^{***}$ & $-$74&79$^{***}$ & $-$75&24$^{***}$  & $-$71&65$^{***}$ && $-$74&14$^{***}$  & $-$71&61$^{***}$  & $-$60&57$^{***}$  & $-$66&05$^{***}$  \\
    KPSS        & 0&07 & 0&07 & 0&06 & 0&05 && 0&07 & 0&08 & 0&11 & 0&12  \\
    Q(10)   & 25&13$^{***}$ & 10&55$^{**}$ & 3&98$^{*}$ & 41&49$^{***}$ && 14&84$^{**}$ & 27&52$^{***}$ & 423&75$^{***}$ & 1087&08$^{***}$  \\
    Q(20)       & 32&63$^{**}$ & 20&18$^{**}$ & 32&36$^{**}$ & 67&60$^{***}$ && 28&13$^{**}$ & 50&48$^{***}$ & 472&13$^{***}$ & 1647&48$^{***}$  \\
    Q$^{2}$(10) & 420&71$^{***}$ & 32&87$^{***}$ & 1161&11$^{***}$ & 656&07$^{***}$ && 1185&88$^{***}$ & 748&44$^{***}$ & 2024&60$^{***}$ & 293&96$^{***}$  \\
    Q$^{2}$(20) & 787&59$^{***}$ & 34&38$^{**}$ & 1276&25$^{***}$ & 717&24$^{***}$ && 1893&55$^{***}$ & 1125&95$^{***}$ & 2459&58$^{***}$ & 350&25$^{***}$  \\
    ARCH-LM(20) & 413&59$^{***}$ & 32&06$^{***}$ & 579&55$^{***}$ & 626&97$^{***}$ && 559&61$^{***}$ & 410&59$^{***}$ & 860&11$^{***}$ & 277&38$^{***}$  \\ 
  \bottomrule
    \end{tabular}
    }%
  %  \begin{tablenotes}
  \begin{flushleft}
    \footnotesize
    \justifying Note: This table reports the descriptive statistics and diagnostic test results for each agricultural return series, where the returns are calculated by multiplying the logarithmic returns by 100. The Jarque-Bera test is a normality test, where a statistic much larger than 0 indicates a significant departure from normality. The ADF, PP and KPSS tests are the most common methods of unit root test, and the null hypothesis in both ADF test and PP test is non-stationarity, while the null hypothesis in KPSS test is stationarity. The Q and Q$^{2}$ statistics for the Ljung-Box test appertain to the results of the white noise test on the return and squared return series, respectively. The ARCH-LM test examines the presence of ARCH effect. ***, ** and * denote significance at the 1\%, 5\% and 10\% level, respectively.
%    \end{tablenotes}
\end{flushleft} 
  \label{Tab:Agro_Stat_Test}%
\end{table}%

% \todo[author=周炜星, inline]{搜索“latex 表格 小数点对齐”，可使表1的对齐方式美化。}

\section{Empirical analysis}
\label{S1:EmpAnal}

\subsection{Estimation of marginal models}

Table~\ref{Tab:Agro_Marginal_estimation} presents the estimated results of the marginal ARMA-GARCH-skewed Student-t models, in which the optimal lag parameters for each return series are determined in the range of 0 to 3 according to the AIC. We find that the mean equations of different agricultural futures and spot returns conform to various types of ARMA$(m, n)$ models, and most of the coefficients are significant at the 1\% level. As for the variance equations in panel B of Table~\ref{Tab:Agro_Marginal_estimation}, all coefficients are significant at the 1\% level, and the sum of the ARCH and GARCH terms of each return series is close to 1, demonstrating a high volatility persistence. Furthermore, the estimated values for the asymmetric parameters (Asymmetry) and the degrees-of-freedom parameters (Tail) indicate that the error terms are not normal and can be well specified by distributions with asymmetries and fat tails. Particularly, the asymmetry coefficients of all return series are significantly positive at the 1\% level, further implying that the fat tails are right-skewed, which means that large positive returns are more likely than large negative returns in the agricultural futures and spot markets.

\begin{table}[!ht]
  \centering
  \setlength{\abovecaptionskip}{0pt}
  \setlength{\belowcaptionskip}{10pt}
  \caption{Marginal estimation using the ARMA-GARCH-skewed Student-t model presented in Eq.~(\ref{Eq:Marginal_distribution})}
    \setlength\tabcolsep{2.9pt}
    \resizebox{\textwidth}{!}{
    \begin{tabular}{l r@{.}l r@{.}l r@{.}l r@{.}l c r@{.}l r@{.}l r@{.}l r@{.}l}
    \toprule
    & \multicolumn{8}{c}{Futures} && \multicolumn{8}{c}{Spots} \\
    \cline{2-9} \cline{11-18}
    & \multicolumn{2}{c}{Soybean} & \multicolumn{2}{c}{Corn} & \multicolumn{2}{c}{Wheat} & \multicolumn{2}{c}{Rough rice} && \multicolumn{2}{c}{Soybean} & \multicolumn{2}{c}{Maize} & \multicolumn{2}{c}{Wheat} & \multicolumn{2}{c}{Rice} \\
    \midrule
    \multicolumn{18}{l}{\textit{Panel A: Mean equation in Eq.~(\ref{Eq:Marginal_distribution_return})}} \\
 Constant, $\varphi_0$ & 0&0287$^{*}$ & 0&0258 & 0&0146 & 0&0277 && 0&0056 & 0&0264 & 0&0186 & $-$0&0036 \\
         & (0&0174) & (0&0201) & (0&0164) & (0&0195) && (0&0166) & (0&0173) & (0&0122) & (0&0047) \\
    AR(1), $\varphi_1$ & 0&9839$^{***}$ & $-$1&3241$^{***}$ & $-$0&4347$^{***}$ & \multicolumn{2}{r}{} && 0&7726$^{***}$ & 0&9208$^{***}$ & 0&1808$^{***}$ & 0&8979$^{***}$ \\
         & (0&0287) & (0&0012) & (0&0102) & \multicolumn{2}{r}{} && (0&0631) & (0&1266) & (0&0136) & (0&0289) \\
    AR(2), $\varphi_2$ & $-$0&9233$^{***}$ & $-$1&4236$^{***}$ & 0&3704$^{***}$ & \multicolumn{2}{r}{} && \multicolumn{2}{r}{} & \multicolumn{2}{r}{} & 0&0624$^{***}$ & $-$0&0248$^{**}$ \\
         & (0&0082) & (0&0009) & (0&0114) & \multicolumn{2}{r}{} && \multicolumn{2}{r}{} & \multicolumn{2}{r}{} & (0&0136) & (0&0109) \\
    AR(3), $\varphi_3$ & \multicolumn{2}{r}{} & $-$0&5546$^{***}$ & 0&9398$^{***}$ & \multicolumn{2}{r}{} && \multicolumn{2}{r}{} & \multicolumn{2}{r}{} & 0&0572$^{***}$ & \multicolumn{2}{r}{} \\
         & \multicolumn{2}{r}{} & (0&0007) & (0&0049) & \multicolumn{2}{r}{} && \multicolumn{2}{r}{} & \multicolumn{2}{r}{} & (0&0134) & \multicolumn{2}{r}{} \\
    MA(1), $\gamma_1$ & $-$0&9878$^{***}$ & 1&3465$^{***}$ & 0&4273$^{***}$ & 0&0515$^{***}$ && $-$0&7049$^{***}$ & $-$0&8598$^{***}$ & \multicolumn{2}{r}{} & $-$0&8121$^{***}$ \\
         & (0&0245) & (0&0001) & (0&0110) & (0&0132) && (0&0707) & (0&1247) & \multicolumn{2}{r}{} & (0&0280) \\
    MA(2), $\gamma_2$ & 0&9403$^{***}$ & 1&4420$^{***}$ & $-$0&3918$^{***}$ & \multicolumn{2}{r}{} && \multicolumn{2}{r}{} & $-$0&0443$^{*}$ & \multicolumn{2}{r}{} & \multicolumn{2}{r}{} \\
         & (0&0085) & (0&0000) & (0&0101) & \multicolumn{2}{r}{} && \multicolumn{2}{r}{} & (0&0237) & \multicolumn{2}{r}{} & \multicolumn{2}{r}{} \\
    MA(3), $\gamma_3$ & \multicolumn{2}{r}{} & 0&5795$^{***}$ & $-$0&9495$^{***}$ & \multicolumn{2}{r}{} && \multicolumn{2}{r}{} & \multicolumn{2}{r}{} & \multicolumn{2}{r}{} & \multicolumn{2}{r}{} \\
        & \multicolumn{2}{r}{} & (0&0000) & (0&0000) & \multicolumn{2}{r}{} && \multicolumn{2}{r}{} & \multicolumn{2}{r}{} & \multicolumn{2}{r}{} & \multicolumn{2}{r}{} 
\vspace{2mm}\\
    \multicolumn{18}{l}{\textit{Panel B: Variance equation in Eq.~(\ref{Eq:Marginal_distribution_conditional_variance})}} \\
    Constant, $\alpha_0$ & 0&0281$^{***}$ & 0&0507$^{***}$ & 0&0413$^{***}$ & 0&0508$^{***}$ && 0&0043$^{***}$ & 0&0178$^{***}$ & 0&0030$^{***}$ & 0&0272$^{***}$ \\
           & (0&0067) & (0&0071) & (0&0117) & (0&0119) && (0&0016) & (0&0055) & (0&0011) & (0&0053) \\
    ARCH, $\alpha_1$ & 0&0512$^{***}$ & 0&0671$^{***}$ & 0&0411$^{***}$ & 0&0795$^{***}$ && 0&0614$^{***}$ & 0&0779$^{***}$ & 0&0586$^{***}$ & 0&1989$^{***}$ \\
         & (0&0067) & (0&0048) & (0&0055) & (0&0098) && (0&0072) & (0&0121) & (0&0073) & (0&0291) \\
    GARCH, $\beta_1$ & 0&9367$^{***}$ & 0&9197$^{***}$ & 0&9488$^{***}$ & 0&9090$^{***}$ && 0&9376$^{***}$ & 0&9164$^{***}$ & 0&9391$^{***}$ & 0&8001$^{***}$ \\
           & (0&0081) & (0&0016) & (0&0072) & (0&0104) && (0&0072) & (0&0130) & (0&0076) & (0&0309) \\
    Asymmetry, $\eta$ & 0&9474$^{***}$ & 1&0449$^{***}$ & 1&1374$^{***}$ & 1&0475$^{***}$ && 0&9551$^{***}$ & 1&0211$^{***}$ & 1&0641$^{***}$ & 1&0121$^{***}$ \\
            & (0&0175) & (0&0191) & (0&0222) & (0&0182) && (0&0180) & (0&0183) & (0&0208) & (0&0140) \\
    Tail, $\nu$ & 6&0595$^{***}$ & 4&7667$^{***}$ & 7&9852$^{***}$ & 4&3545$^{***}$ && 6&7250$^{***}$ & 5&4882$^{***}$ & 8&0763$^{***}$ & 2&1419$^{***}$ \\
        & (0&4740) & (0&2992) & (0&7773) & (0&2675) && (0&5183) & (0&3967) & (0&7914) & (0&0203)   
\vspace{2mm}\\
    \multicolumn{9}{l}{\textit{Panel C: Diagnostic tests}} \\
    LL & \multicolumn{2}{c}{$-$9712.75} & \multicolumn{2}{c}{$-$10532.60} & \multicolumn{2}{c}{$-$11411.45} & \multicolumn{2}{c}{$-$10195.39} && \multicolumn{2}{c}{$-$8751.98} & \multicolumn{2}{c}{$-$8852.04} & \multicolumn{2}{c}{$-$6106.19} & \multicolumn{2}{c}{38.18} \\
    AIC & 3&4638 & 3&7565 & 4&0696 & 3&6346 && 3&1208 & 3&1568 & 2&1785 & $-$0&0104 \\
    Jarque-Bera & [0&0000] & [0&0000] & [0&0000] & [0&0000] && [0&0000] & [0&0000] & [0&0000] & [0&0000] \\
    Q(10) & [0&8499] & [0&3405] & [0&7446] & [0&7630] && [0&7418] & [0&1383] & [0&1583] & [0&3152] \\
    Q(20) & [0&9198] & [0&3891] & [0&4326] & [0&2332] && [0&8667] & [0&1412] & [0&0946] & [0&1474] \\
    Q$^{2}$(10) & [0&9206] & [0&1064] & [0&1135] & [0&9855] && [0&9842] & [0&3901] & [0&4335] & [1&0000] \\
    Q$^{2}$(20) & [0&2657] & [0&2228] & [0&4908] & [0&4474] && [0&1019] & [0&3319] & [0&7032] & [1&0000] \\
    ARCH-LM(10) & [0&9213] & [0&1113] & [0&2485] & [0&9848] && [0&9821] & [0&3446] & [0&4606] & [1&0000] \\
    ARCH-LM(20) & [0&2926] & [0&2639] & [0&1056] & [0&4669] && [0&1132] & [0&2936] & [0&7399] & [1&0000] \\
   \bottomrule
    \end{tabular}
    }%
  \begin{flushleft}
    \footnotesize
    \justifying Note: This table presents the estimated results and diagnostic tests for each marginal ARMA-GARCH-skewed Student-t model, in which the optimal lag parameters are determined in the range of 0 to 3 according to the AIC. The Jarque-Bera test, Ljung-Box test and ARCH-LM test are utilized to examine the existence of normality, serial correlation and ARCH effect in the standardized residual sequence for each marginal model, respectively. The standard errors of parameter estimates are listed in parentheses, and the $p$-values of test statistics are reported in square brackets. ***, ** and * denote significance at the 1\%, 5\% and 10\% level, respectively.
  \end{flushleft}
  \label{Tab:Agro_Marginal_estimation}%
\end{table}%

With the exception of the parameter estimates, Table~\ref{Tab:Agro_Marginal_estimation} further provides information on the goodness-of-fit tests for each estimated marginal model, and lists the corresponding $p$-value of each test statistic in the square brackets. It can be found that both the Q statistics for the standardized residual sequences and the Q$^{2}$ statistics for the standardized squared residual sequences fail to reject the null hypothesis of no autocorrelation at the 5\% level. The ARCH-LM test results show that the null hypothesis cannot be rejected even at the 10\% level, indicating that there is no autoregressive conditional heteroskedasticity remaining in the residual sequences of the estimated marginal models. Therefore, comparing the test results in Table~\ref{Tab:Agro_Marginal_estimation} with those in Table~\ref{Tab:Agro_Stat_Test}, we can conclude that the ARMA-GARCH-skewed Student-t models constructed in this paper can well specify the marginal distributions of the agricultural return series.

\subsection{Empirical results based on single copula models}

\subsubsection{Single copula estimation}

We apply eight different types of single copula models, including the Normal copula, Student-t copula, Clayton copula, survival Clayton copula, 90-degree rotated Clayton copula, 270-degree rotated Clayton copula, Gumbel copula, and survival Gumbel copula to assess the dependence structure between the agricultural futures and spot returns. Table~\ref{Tab:Agro_Single_Copula_Estimation} reports the estimated parameters of the single copula models for each pair of the futures-spot returns. We note that the coefficient estimates of the Normal, Student-t, Clayton, survival Clayton, Gumbel and survival Gumbel copula models for other pairs of agricultural commodities are significant at the 5\% level, except for the pair of rough rice futures and rice spot. Meanwhile, the parameter estimates of the 90-degree rotated Clayton copula and 270-degree rotated Clayton copula models for each pair are quite small and statistically insignificant, which further proves the positive correlation between the agricultural futures and spot markets.

\begin{table}[htp]
  \centering
  \setlength{\abovecaptionskip}{0pt}
  \setlength{\belowcaptionskip}{10pt}
  \caption{Estimation of single copula models between futures and spot return series of agricultural commodities}
  \setlength\tabcolsep{16.5pt}
  \resizebox{\textwidth}{!}{
    \begin{tabular}{l r@{.}l r@{.}l r@{.}l r@{.}l}
    \toprule
        & \multicolumn{2}{c}{Soybean} & \multicolumn{2}{c}{Maize} & \multicolumn{2}{c}{Wheat} & \multicolumn{2}{c}{Rice} \\
    \midrule
    \multicolumn{9}{l}{\textit{Panel A: Normal copula}} \\
    $\rho$ & \multicolumn{2}{c}{\textbf{0.8757} (0.0023)} & \multicolumn{2}{c}{\textbf{0.8383} (0.0030)} & \multicolumn{2}{c}{\textbf{0.8068} (0.0036)} & \multicolumn{2}{c}{0.0284 (0.0134)} \\
    LL & 4081&3610 & 3399&6650 & 2948&4270 & 2&2462 \\
    AIC & $-$8160&7220 & $-$6797&3290 & $-$5894&8540 & $-$2&4924 \\
    BIC & $-$8154&0890 & $-$6790&6960 & $-$5888&2210 & 4&1406 \vspace{1mm}\\
    \multicolumn{9}{l}{\textit{Panel B: Student-t copula}} \\
    $\rho$ & \multicolumn{2}{c}{\textbf{0.9098} (0.0028)} & \multicolumn{2}{c}{\textbf{0.8701} (0.0035)} & \multicolumn{2}{c}{\textbf{0.8259} (0.0042)} & \multicolumn{2}{c}{0.0267 (0.0137)} \\
    Dof $\nu$ & \multicolumn{2}{c}{\textbf{2.0001} (0.0895)} & \multicolumn{2}{c}{\textbf{3.0238} (0.1636)} & \multicolumn{2}{c}{\textbf{4.4245} (0.2975)} & \multicolumn{2}{c}{30.1660 (19.2940)} \\
    LL & 5074&4420 & 4039&5780 & 3229&2720 & 0&8133 \\
    AIC & $-$10144&8800 & $-$8075&1560 & $-$6454&5440 & 2&3734 \\
    BIC & $-$10131&6200 & $-$8061&8900 & $-$6441&2780 & 15&6395 \vspace{1mm}\\
    \multicolumn{9}{l}{\textit{Panel C: Clayton copula}} \\
    $\alpha$ & \multicolumn{2}{c}{\textbf{3.1839} (0.0528)} & \multicolumn{2}{c}{\textbf{2.5704} (0.0445)} & \multicolumn{2}{c}{\textbf{1.9760} (0.0376)} & \multicolumn{2}{c}{0.0204 (0.0141)} \\
    LL & 3499&9630 & 2974&3100 & 2319&5990 & 1&0970 \\
    AIC & $-$6997&9260 & $-$5946&6190 & $-$4637&1990 & $-$0&1940 \\
    BIC & $-$6991&2930 & $-$5939&9860 & $-$4630&5660 & 6&4390 \vspace{1mm}\\
    \multicolumn{9}{l}{\textit{Panel D: Survival Clayton copula}} \\
    $\alpha$ & \multicolumn{2}{c}{\textbf{3.4682} (0.0558)} & \multicolumn{2}{c}{\textbf{2.5694} (0.0448)} & \multicolumn{2}{c}{\textbf{2.1084} (0.0390)} & \multicolumn{2}{c}{0.0328 (0.0141)} \\
    LL & 3850&2190 & 2956&3170 & 2486&2570 & 2&9302 \\
    AIC & $-$7698&4390 & $-$5910&6340 & $-$4970&5140 & $-$3&8603 \\
    BIC & $-$7691&8060 & $-$5904&0010 & $-$4963&8810 & 2&7727 \vspace{1mm}\\
    \multicolumn{9}{l}{\textit{Panel E: R$_{90}$-Clayton copula}} \\
    $\alpha$ & \multicolumn{2}{c}{$-$0.0001 (0.0269)} & \multicolumn{2}{c}{$-$0.0001 (0.0244)} & \multicolumn{2}{c}{$-$0.0001 (0.0302)} & \multicolumn{2}{c}{$-$0.0001 (0.0134)} \\
    LL & $-$0&5268 & $-$0&4937 & $-$0&4875 & $-$0&0204 \\
    AIC & 3&0535 & 2&9874 & 2&9750 & 2&0409 \\
    BIC & 9&6865 & 9&6204 & 9&6080 & 8&6739 \vspace{1mm}\\
    \multicolumn{9}{l}{\textit{Panel F: R$_{270}$-Clayton copula}} \\
    $\alpha$ & \multicolumn{2}{c}{$-$0.0001 (0.0291)} & \multicolumn{2}{c}{$-$0.0001 (0.0215)} & \multicolumn{2}{c}{$-$0.0001 (0.0250)} & \multicolumn{2}{c}{$-$0.0001 (0.0147)} \\
    LL & $-$0&5251 & $-$0&5115 & $-$0&4977 & $-$0&0215 \\
    AIC & 3&0501 & 3&0230 & 2&9954 & 2&0431 \\
    BIC & 9&6831 & 9&6560 & 9&6284 & 8&6761 \vspace{1mm}\\
    \multicolumn{9}{l}{\textit{Panel G: Gumbel copula}}  \\
    $\alpha$ & \multicolumn{2}{c}{\textbf{3.4839} (0.0400)} & \multicolumn{2}{c}{\textbf{2.8639} (0.0323)} & \multicolumn{2}{c}{\textbf{2.4905} (0.0277)} & \multicolumn{2}{c}{1.0131 (0.0071)} \\
    LL & 4588&6490 & 3652&8440 & 3026&1990 & 1&9736 \\
    AIC & $-$9175&2990 & $-$7303&6870 & $-$6050&3990 & $-$1&9472 \\
    BIC & $-$9168&6660 & $-$7297&0540 & $-$6043&7660 & 4&6858 \vspace{1mm}\\
    \multicolumn{9}{l}{\textit{Panel H: Survival Gumbel copula}}  \\
    $\alpha$ & \multicolumn{2}{c}{\textbf{3.3819} (0.0388)} & \multicolumn{2}{c}{\textbf{2.8605} (0.0322)} & \multicolumn{2}{c}{\textbf{2.4485} (0.0272)} & \multicolumn{2}{c}{1.0069 (0.0078)} \\
    LL & 4401&2660 & 3657&4610 & 2923&9450 & 0&4054 \\
    AIC & $-$8800&5320 & $-$7312&9230 & $-$5845&8890 & 1&1892 \\
    BIC & $-$8793&8990 & $-$7306&2900 & $-$5839&2560 & 7&8223 \\
    \bottomrule
    \end{tabular}}%
  \begin{flushleft}
    \footnotesize
    \justifying Note: This table reports the estimated parameters and goodness-of-fit measures of eight different single copula models for each pair of the futures-spot returns. $\rho$ and $\alpha$ denote the copula parameters, and Dof is the degrees-of-freedom parameter of the Student-t copula model. LL, AIC and BIC represent the values of the logarithmic likelihood, Akaike information criterion and Bayesian information criterion, respectively. Bold numbers refer to significance at the 5\% level, and the standard errors of parameter estimates are presented in parentheses.
  \end{flushleft}
  \label{Tab:Agro_Single_Copula_Estimation}%
\end{table}%

In addition, Table~\ref{Tab:Agro_Single_Copula_Estimation} also reports the values of the logarithmic likelihood (LL), Akaike information criterion (AIC) and Bayesian information criterion (BIC) of various single copula models for each pair to evaluate the goodness of fit. For rice, the survival Clayton copula model yields the largest LL and the smallest AIC and BIC, followed by the Normal copula and Gumbel copula models. Since the assumption of no tail dependence in the Normal copula may deviate from the reality to some extent, we will not pay much attention to it in the following sections. Specially, both the survival Clayton copula and the Gumbel copula are able to describe the upper tail dependence between variables, indicating that the upper-tail correlation may be relatively common between the rice futures and spot markets. In contrast, among the eight different single copula models, the Student-t copula model results in the largest LL and the smallest AIC and BIC in the futures-spot pairs of soybean, maize and wheat. However, the assumption of symmetric tail dependence in the Student-t copula suggests that the tail dependence structure between the agricultural futures and spot markets when both markets are crashing is the same as that when both markets are booming, which may be too restrictive in empirical analysis.

Further comparison between the results in Table~\ref{Tab:Agro_Single_Copula_Estimation} shows that for the three pairs of soybean, maize and wheat, the Gumbel copula and survival Gumbel copula models can also yield large LL and small AIC and BIC except for the Student-t copula model, which means that the two copula functions are capable of describing the dependence structure between the agricultural futures and spot well. Particularly, the Gumbel copula can capture the upper tail dependence between assets, while the survival Gumbel copula can depict the lower tail dependence. In order to take the asymmetric tail dependence into consideration, we therefore construct mixed copula models combining the Gumbel copula and the survival Gumbel copula to further investigate the dependence structure between the agricultural futures and spot markets.

\subsubsection{Risk spillover measure}

Before the construction of the mixed copula models, we first calculate the measures of the downside and upside risk spillover effects of the agricultural futures returns on the agricultural spot returns, including $VaR$, $CoVaR$ and $\Delta CoVaR$, based on the estimated results of the single copula models. Considering that there is no negative correlation between the agricultural futures and spot returns, we eliminate the 90-degree rotated Clayton and 270-degree rotated Clayton copulas, and then quantify the risk spillovers according to the other six types of single copula models in Table~\ref{Tab:Agro_Single_Copula_Estimation}, respectively.

\begin{figure}[!t]
\centering
\includegraphics[width=0.45\linewidth]{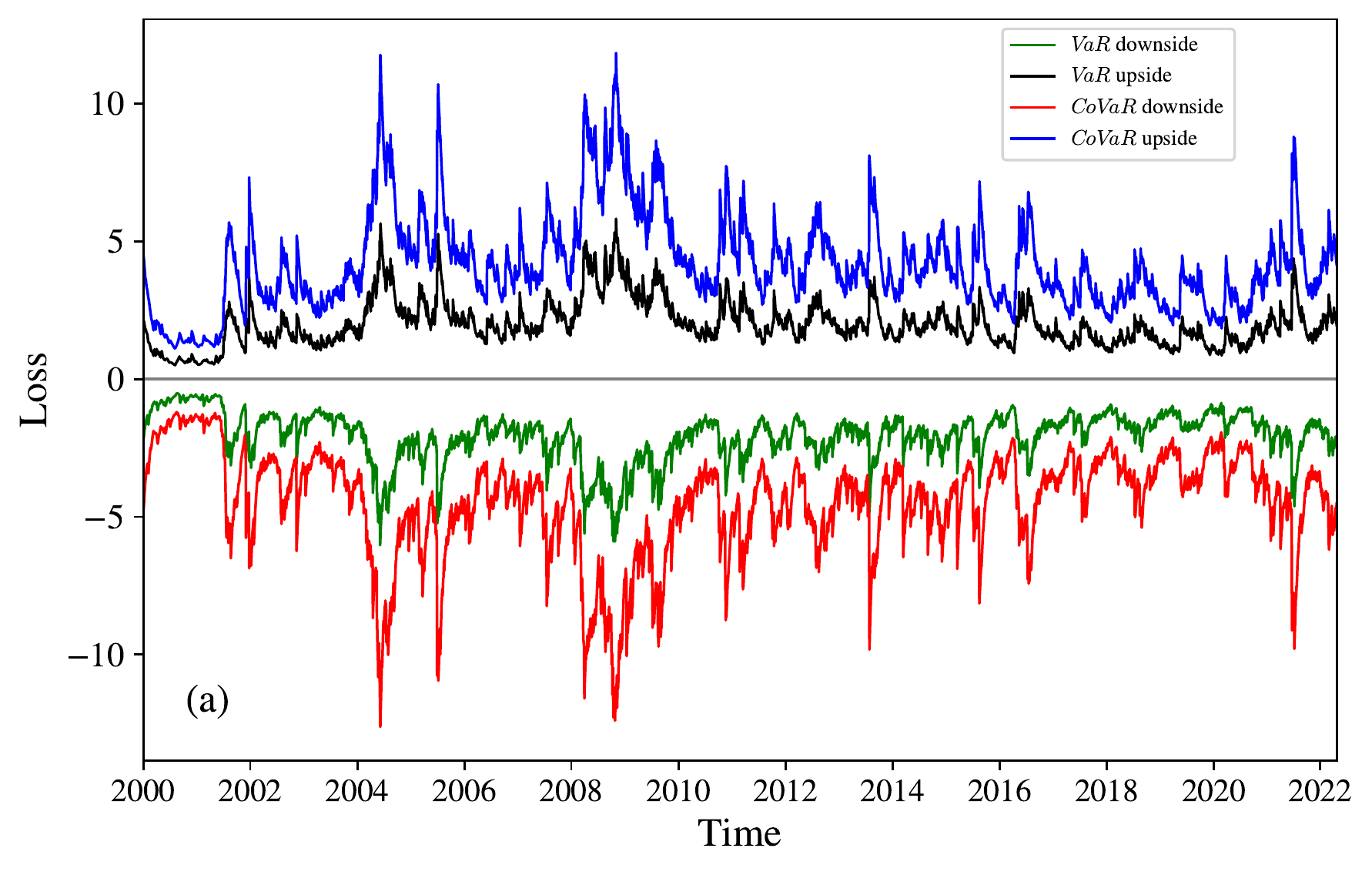}
\includegraphics[width=0.45\linewidth]{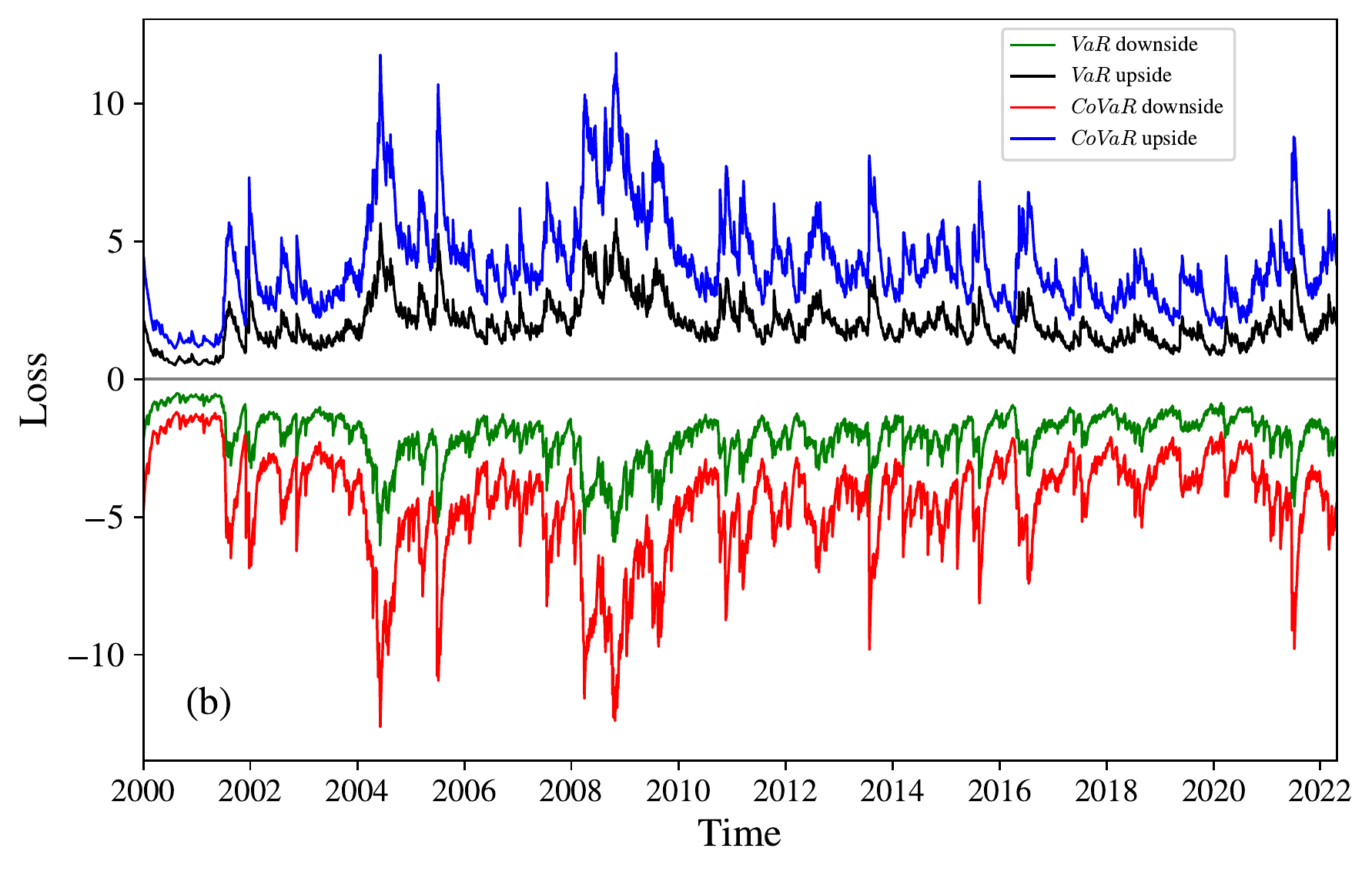}
\includegraphics[width=0.45\linewidth]{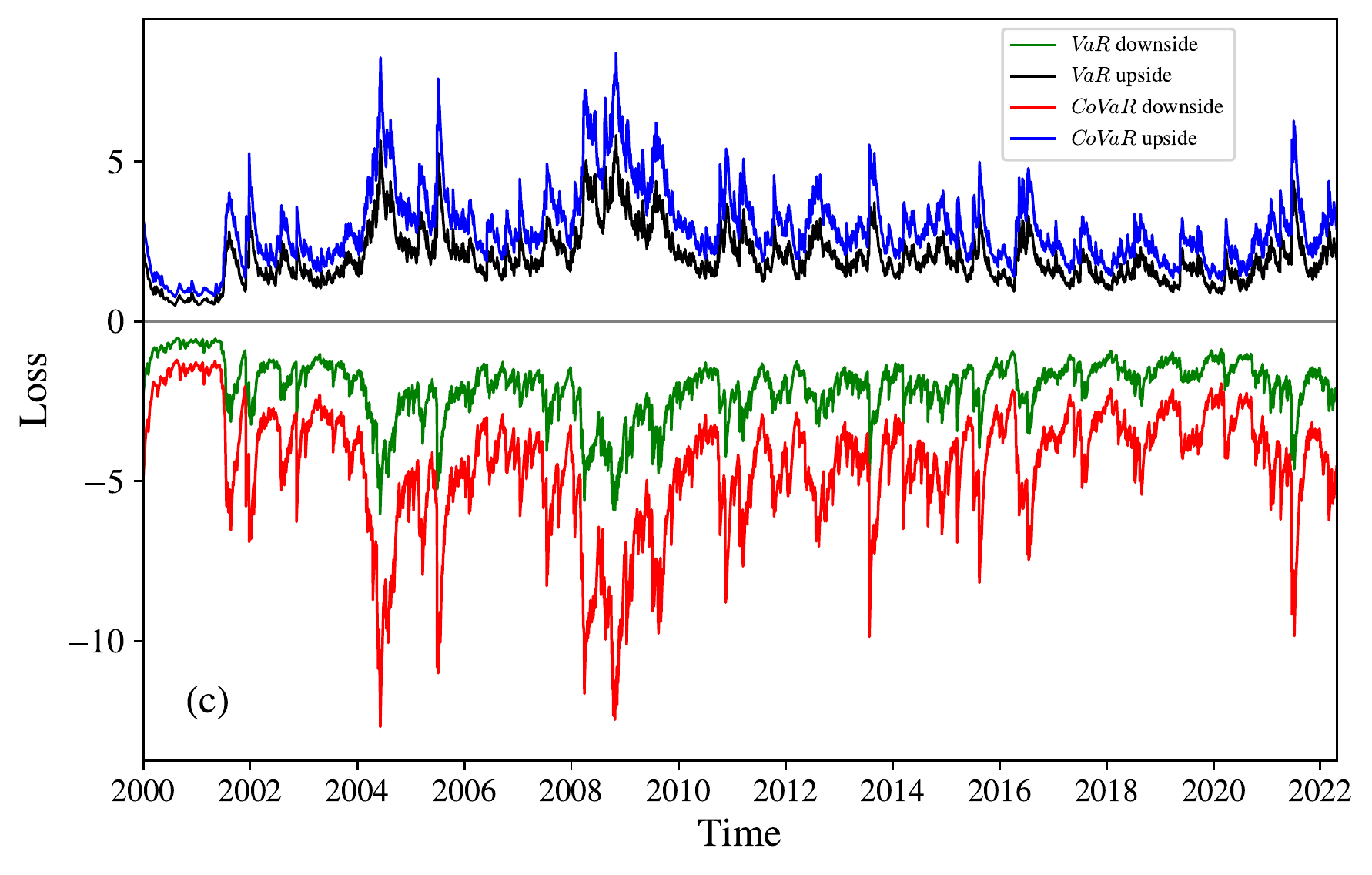}
\includegraphics[width=0.45\linewidth]{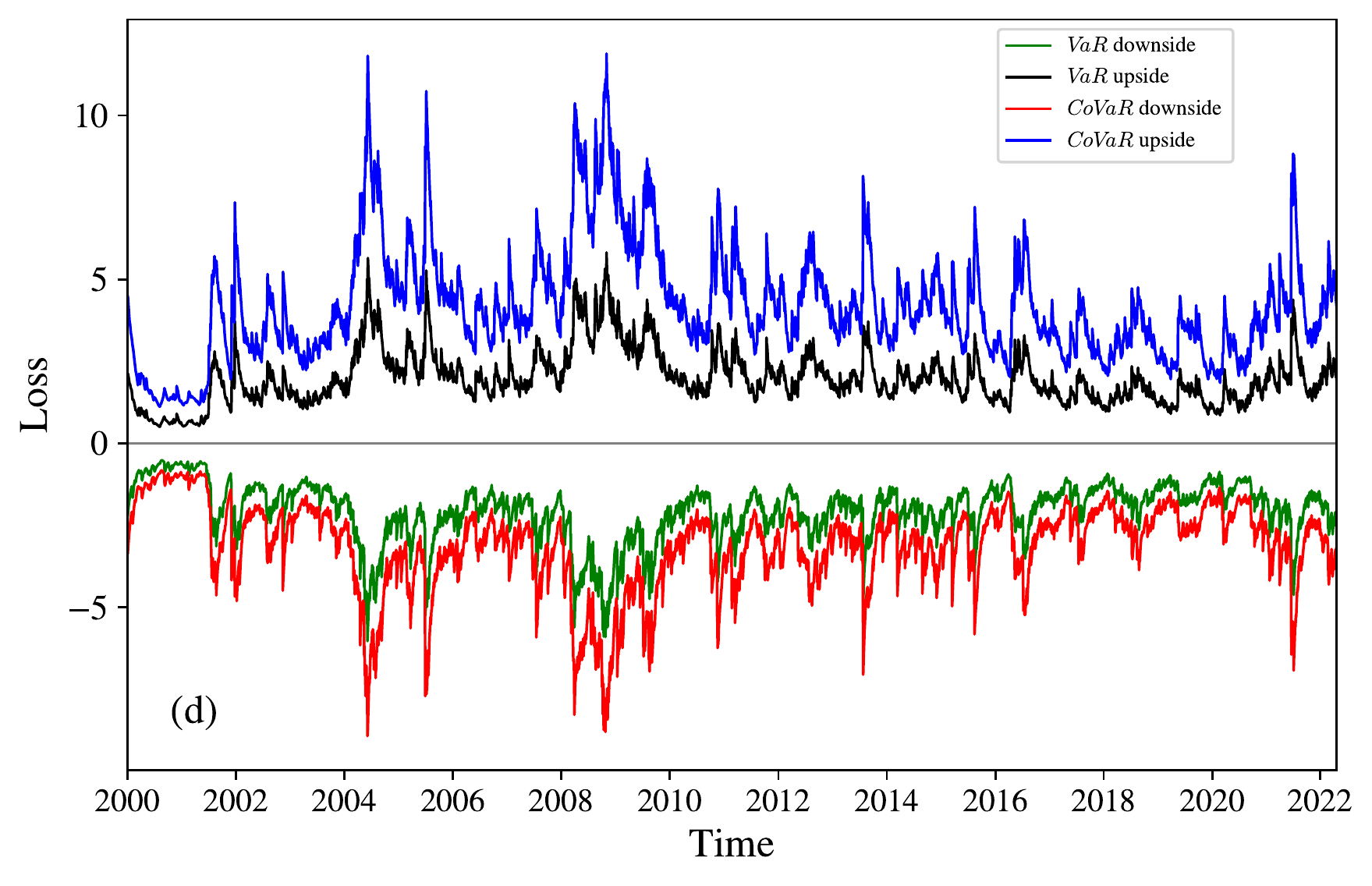}
\includegraphics[width=0.45\linewidth]{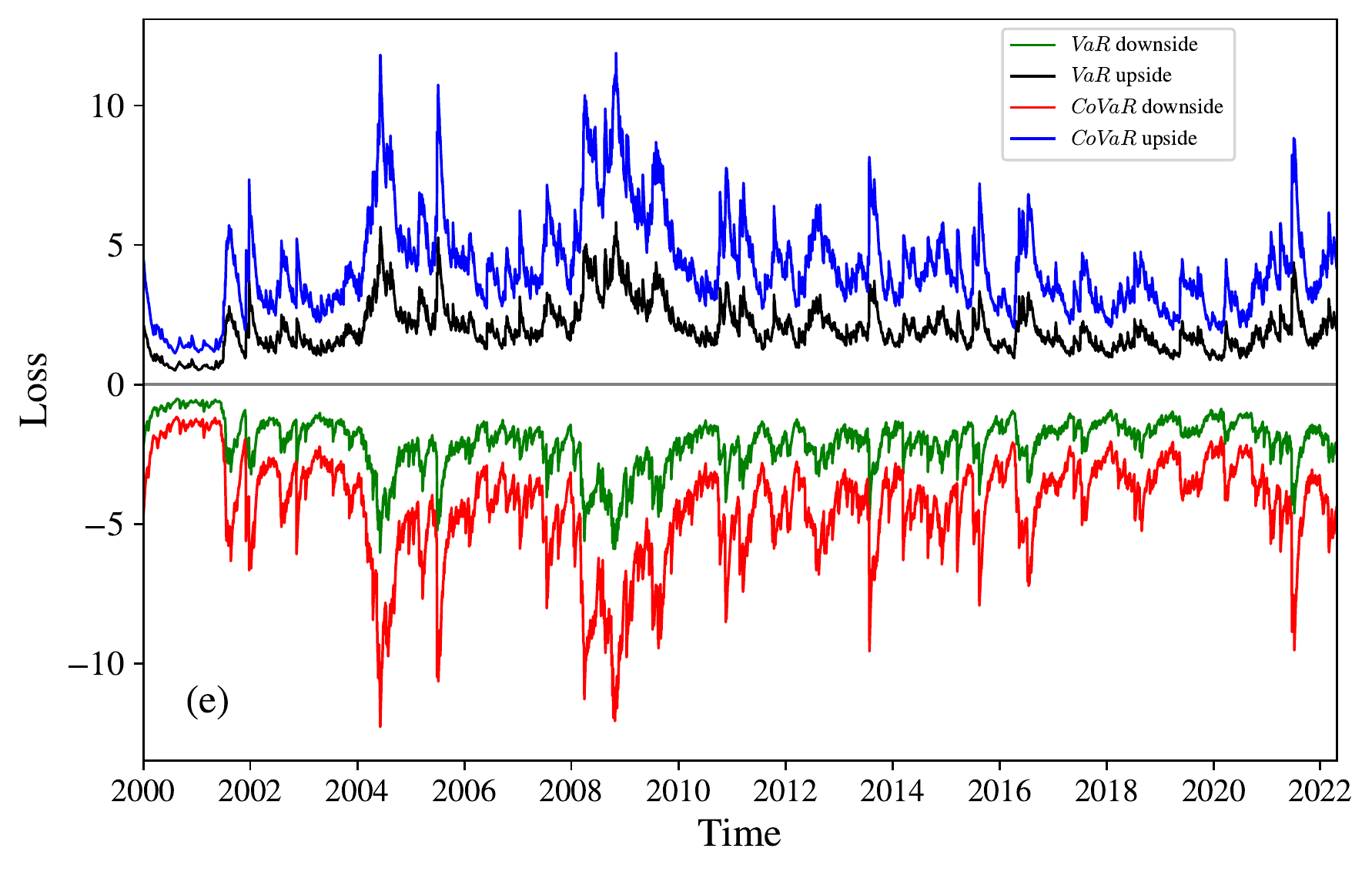}
\includegraphics[width=0.45\linewidth]{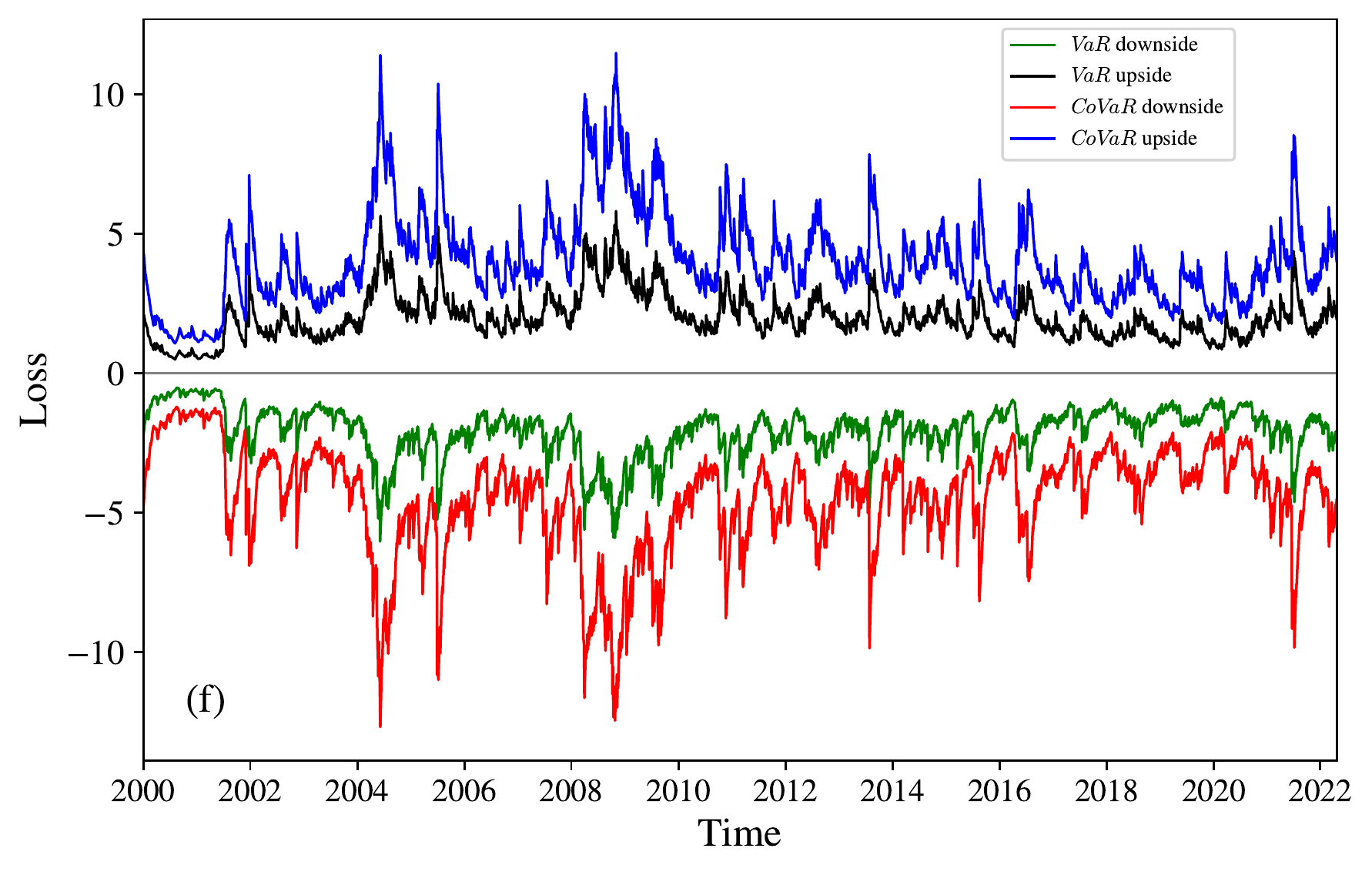}
\caption{$VaR$s and $CoVaR$s for soybean based on Normal copula (a), Student-t copula (b), Clayton copula (c), survival Clayton copula (d), Gumbel copula (e), and survival Gumbel copula (f).}
\label{Fig:VaRs_Soybean_singlecopula}
\end{figure}

\begin{figure}[!t]
\centering
\includegraphics[width=0.45\linewidth]{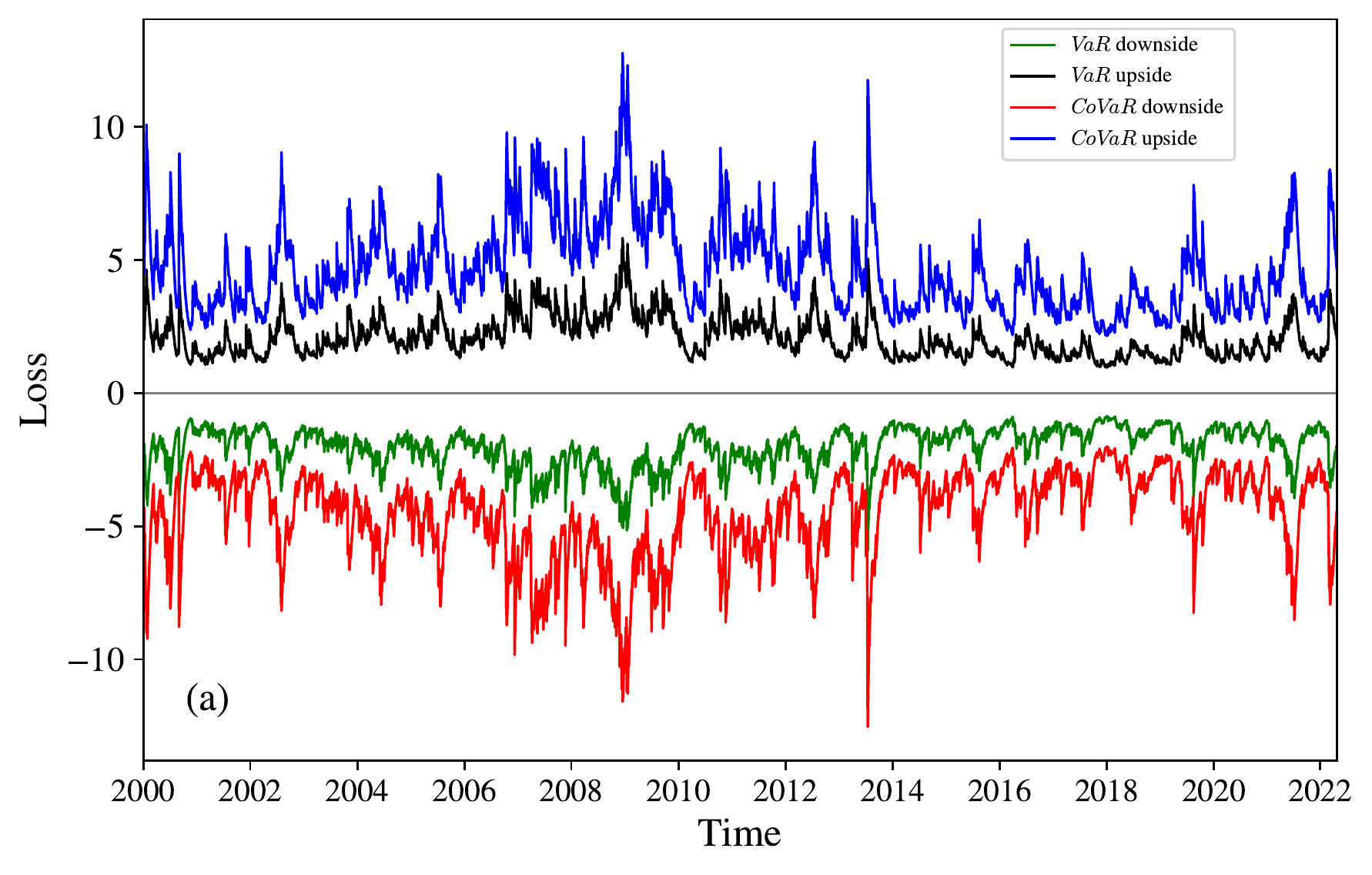}
\includegraphics[width=0.45\linewidth]{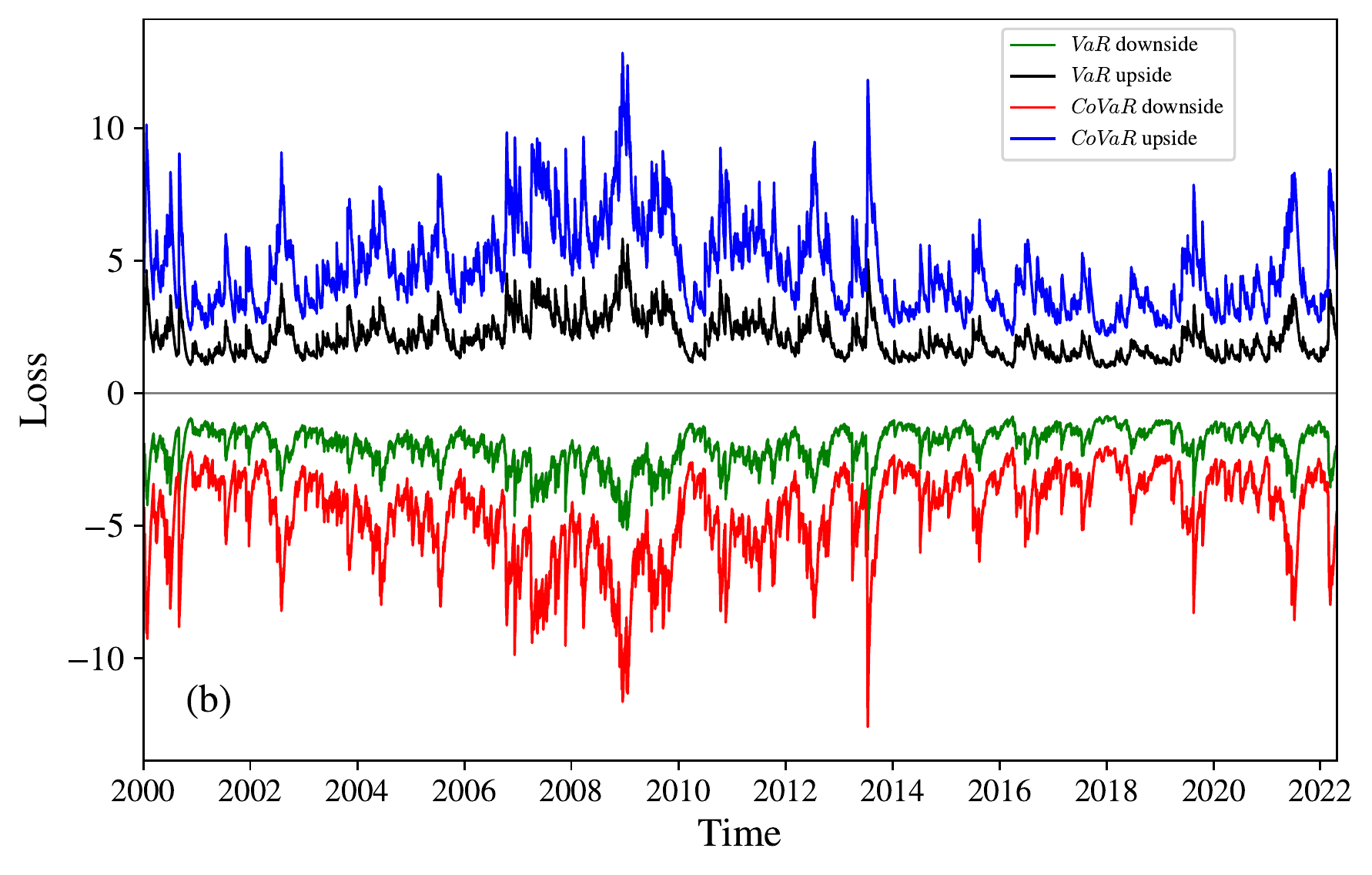}
\includegraphics[width=0.45\linewidth]{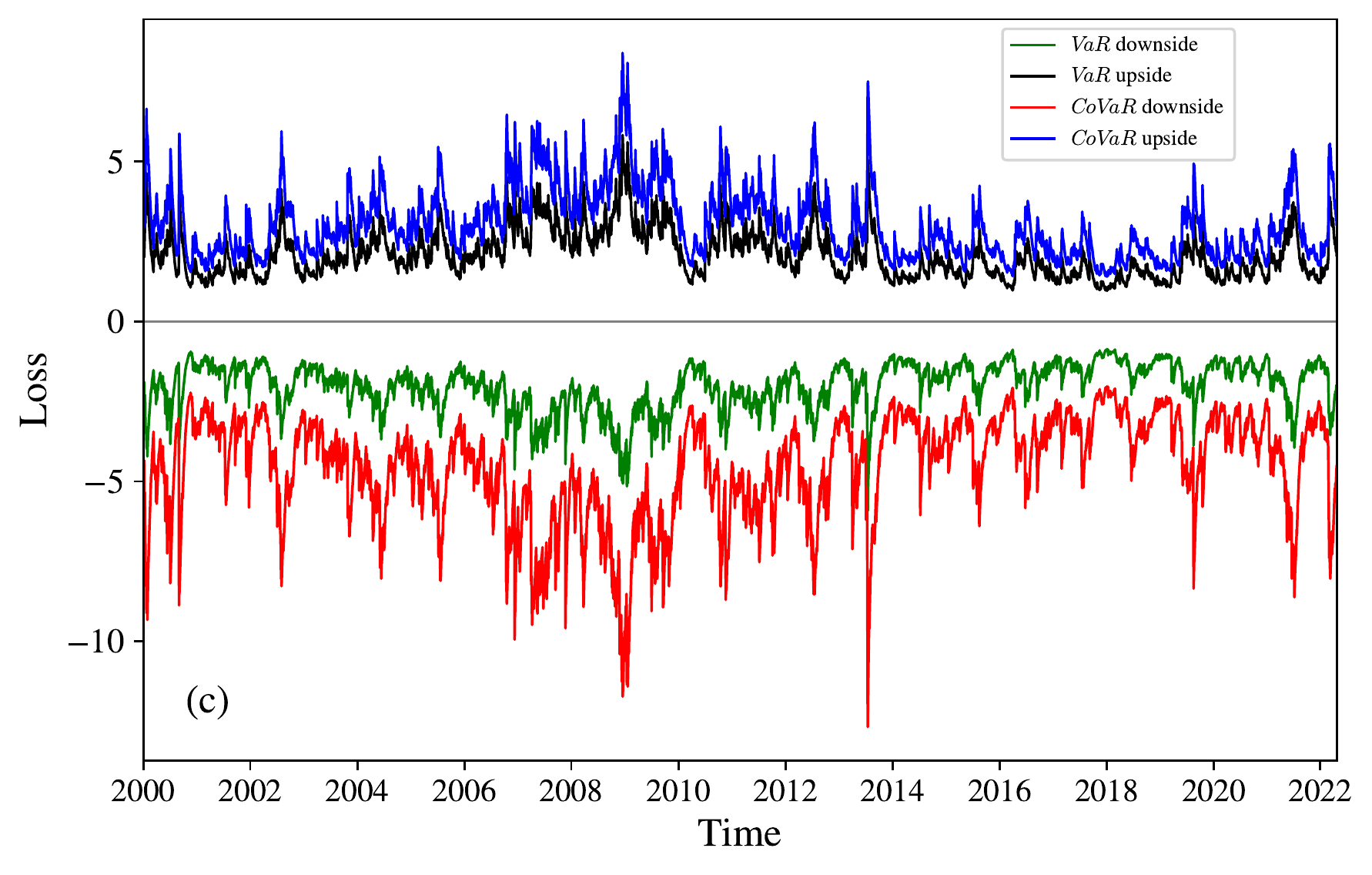}
\includegraphics[width=0.45\linewidth]{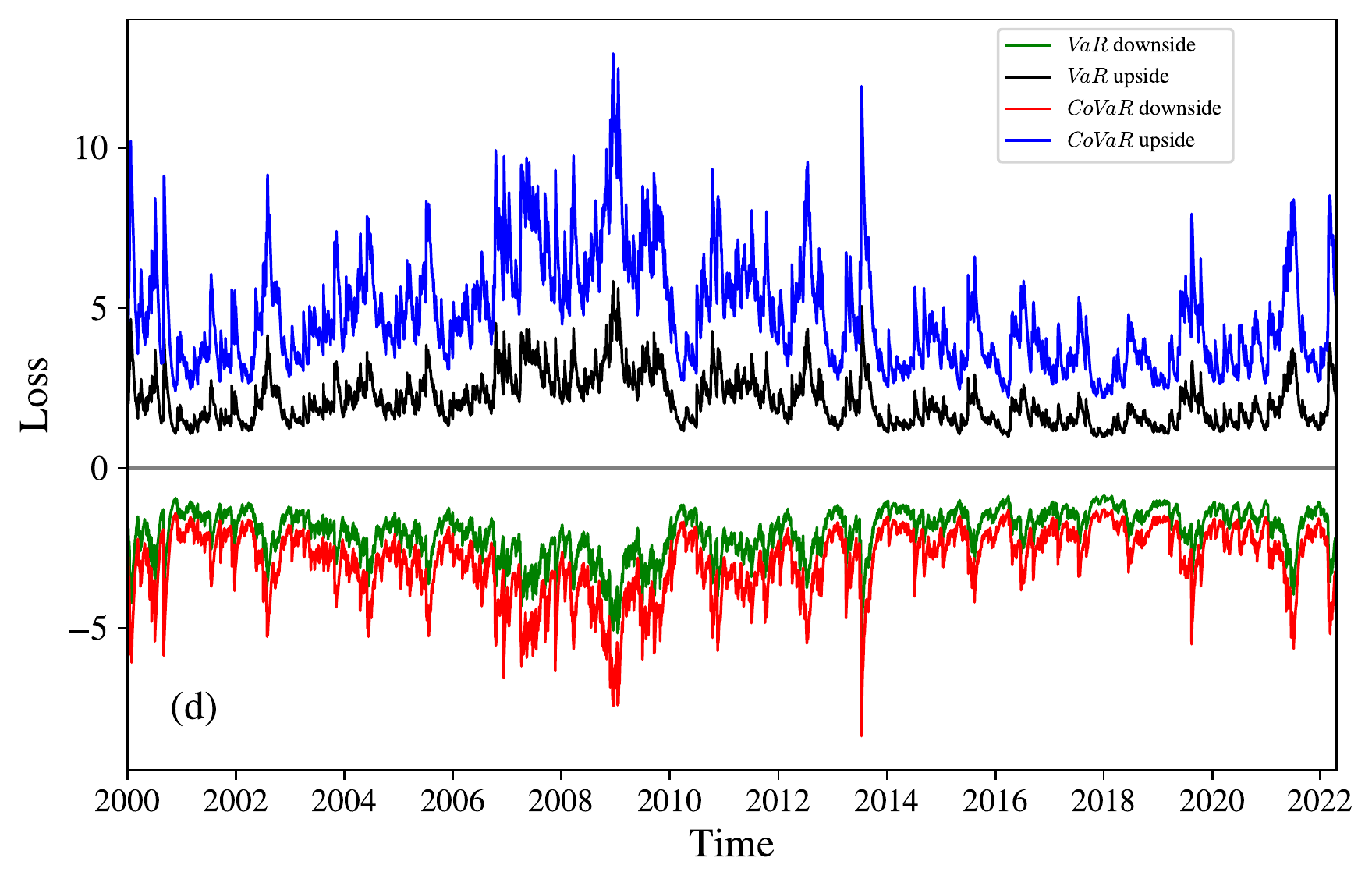}
\includegraphics[width=0.45\linewidth]{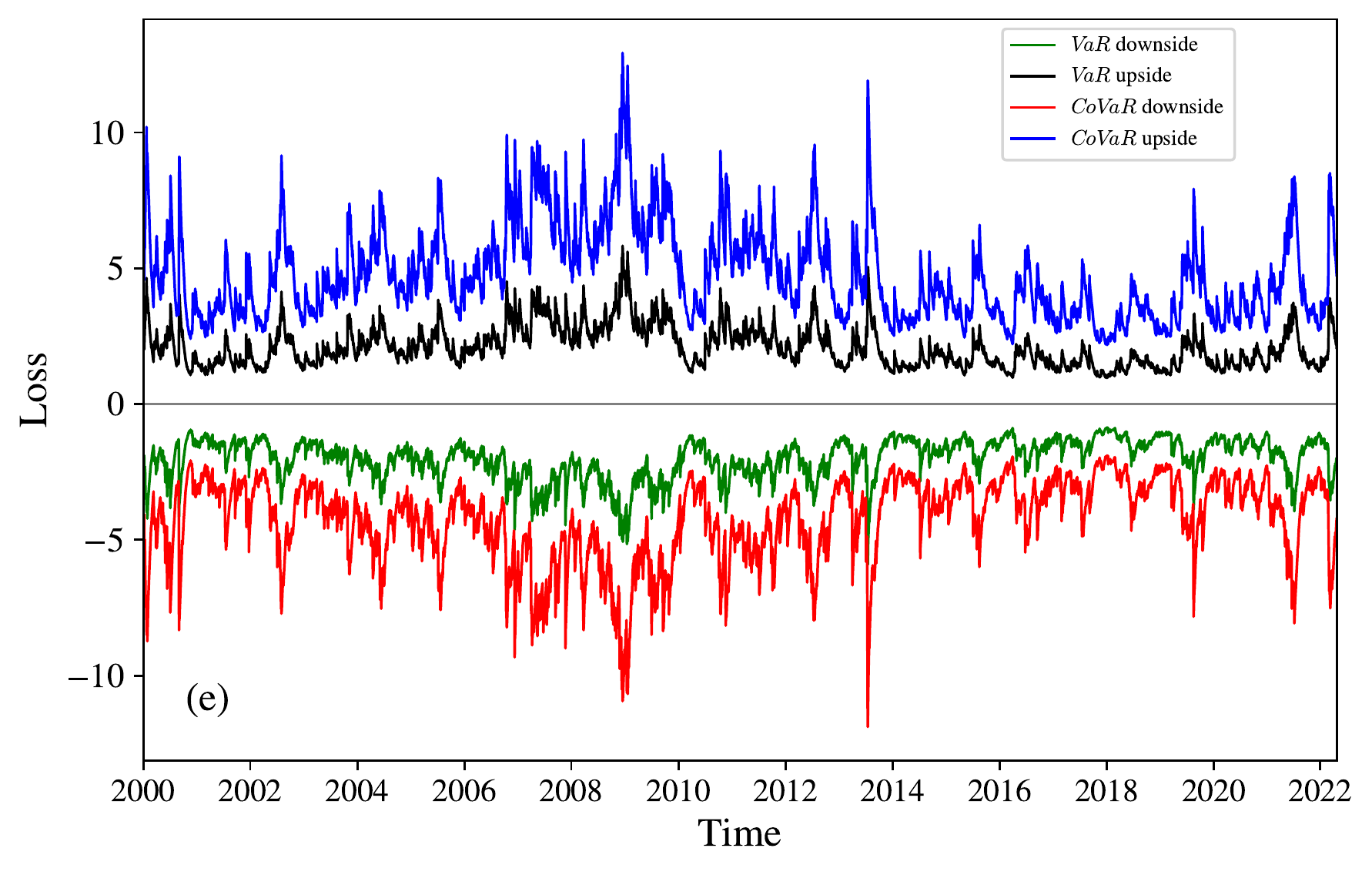}
\includegraphics[width=0.45\linewidth]{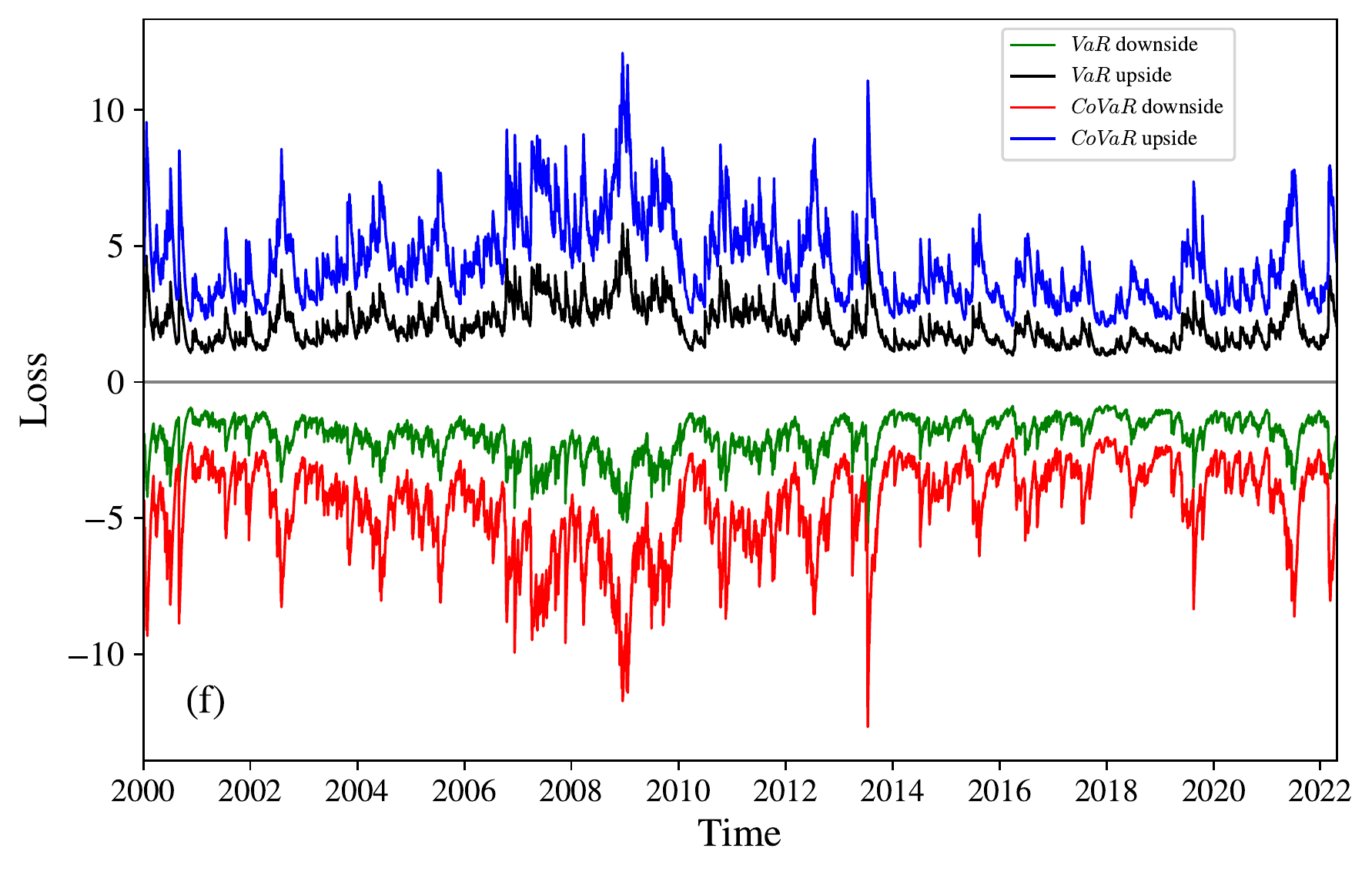}
\caption{$VaR$s and $CoVaR$s for maize based on Normal copula (a), Student-t copula (b), Clayton copula (c), survival Clayton copula (d), Gumbel copula (e), and survival Gumbel copula (f).}
\label{Fig:VaRs_Maize_singlecopula}
\end{figure}

\begin{figure}[!ht]
\centering
\includegraphics[width=0.45\linewidth]{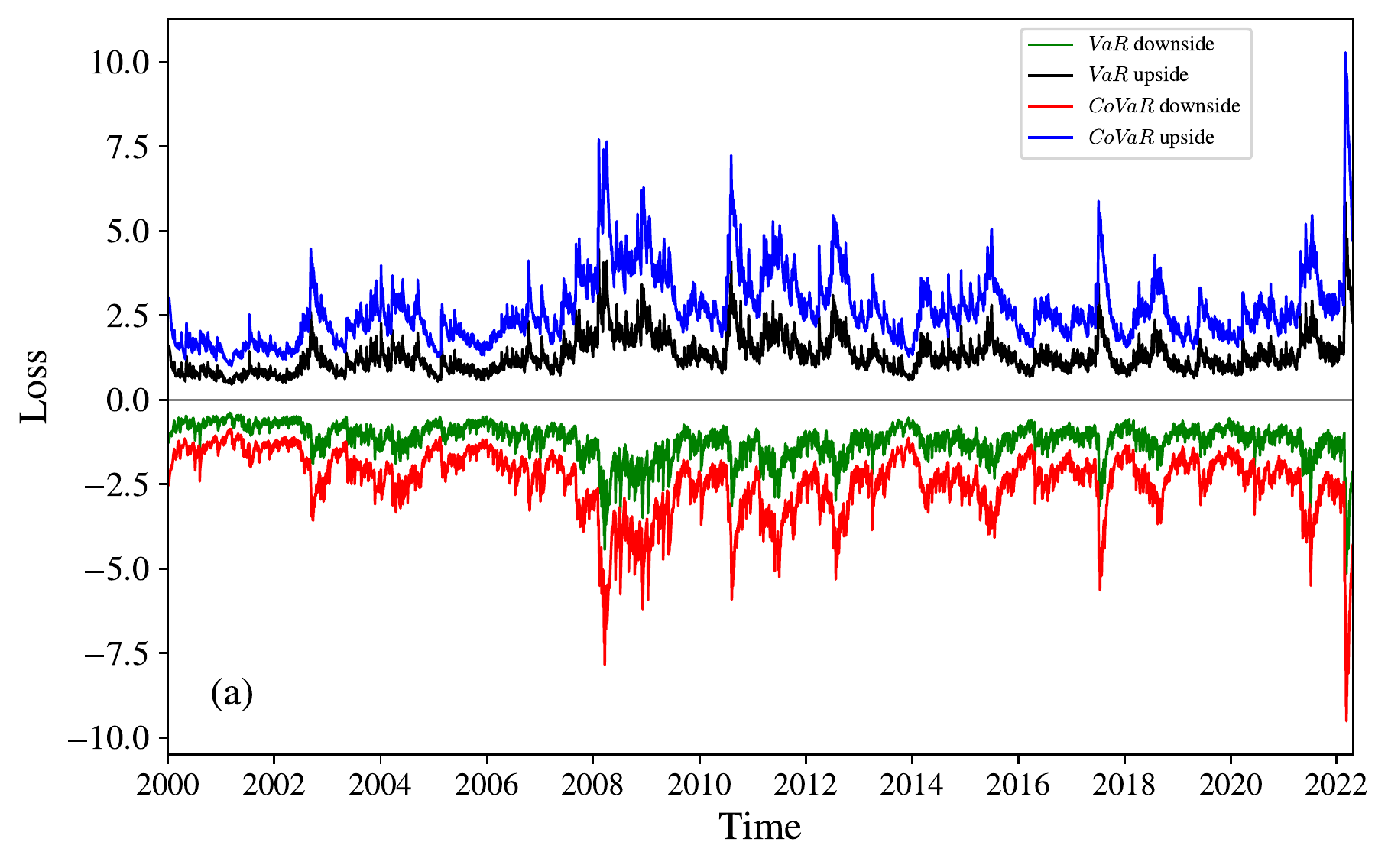}
\includegraphics[width=0.45\linewidth]{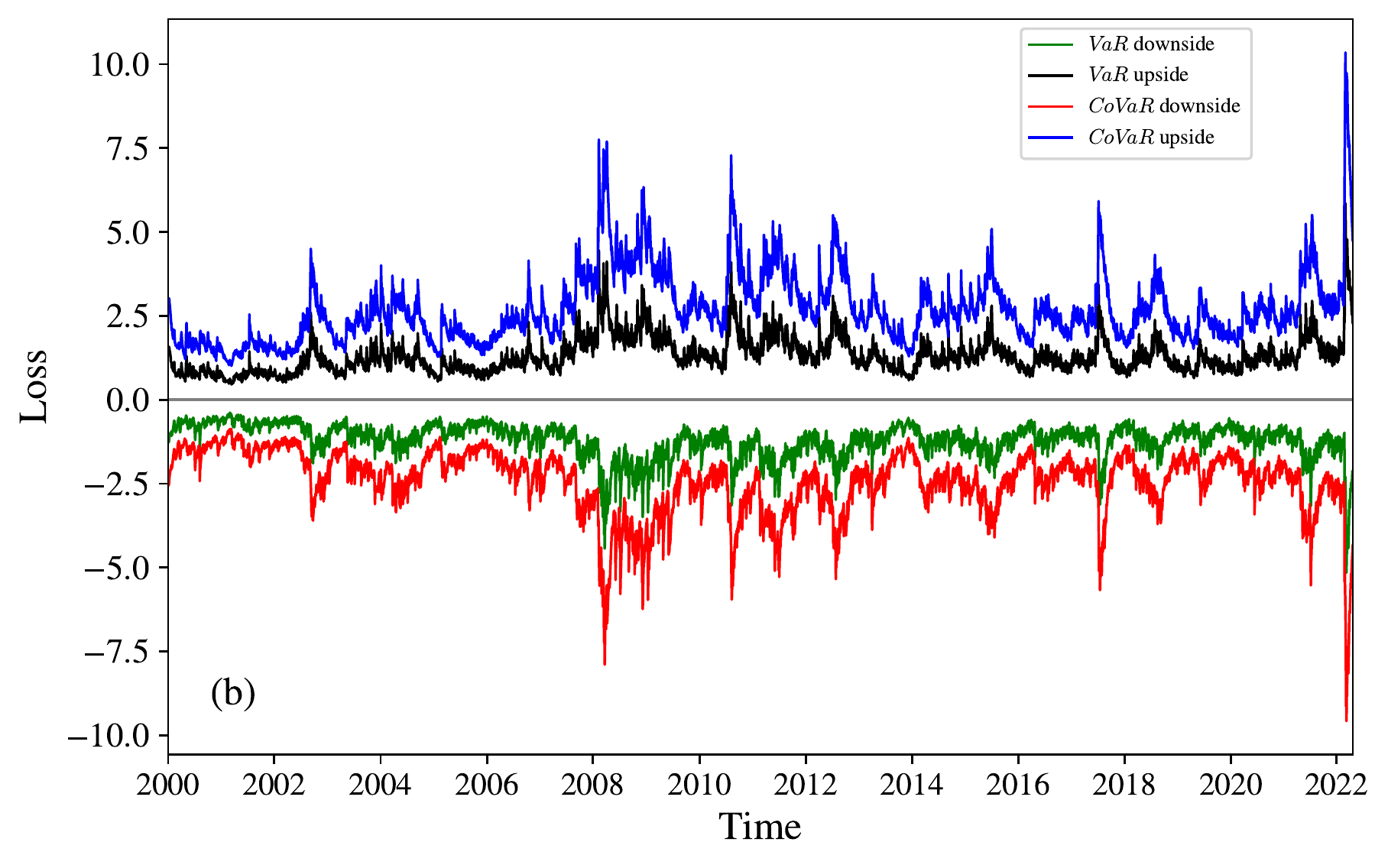}
\includegraphics[width=0.45\linewidth]{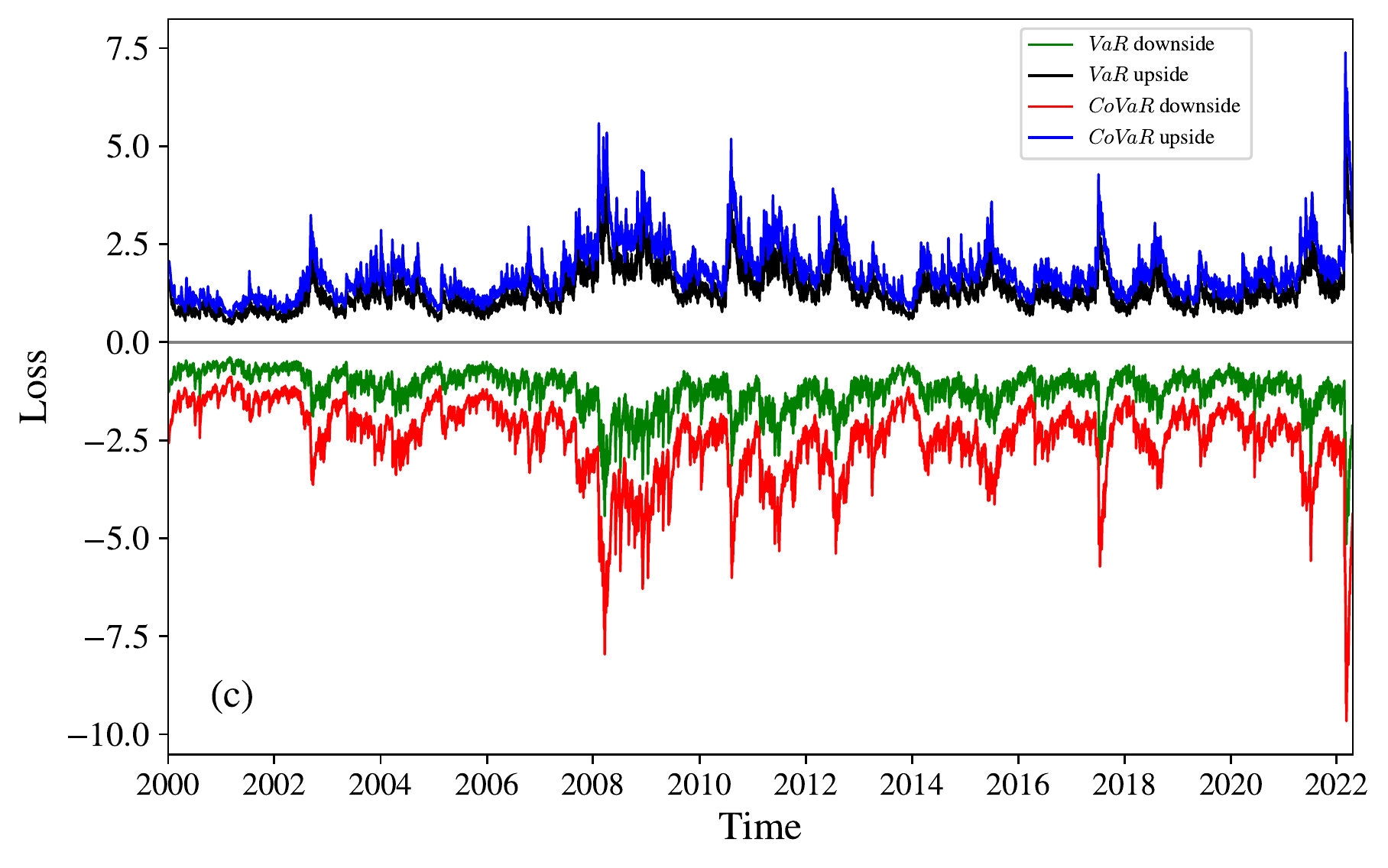}
\includegraphics[width=0.45\linewidth]{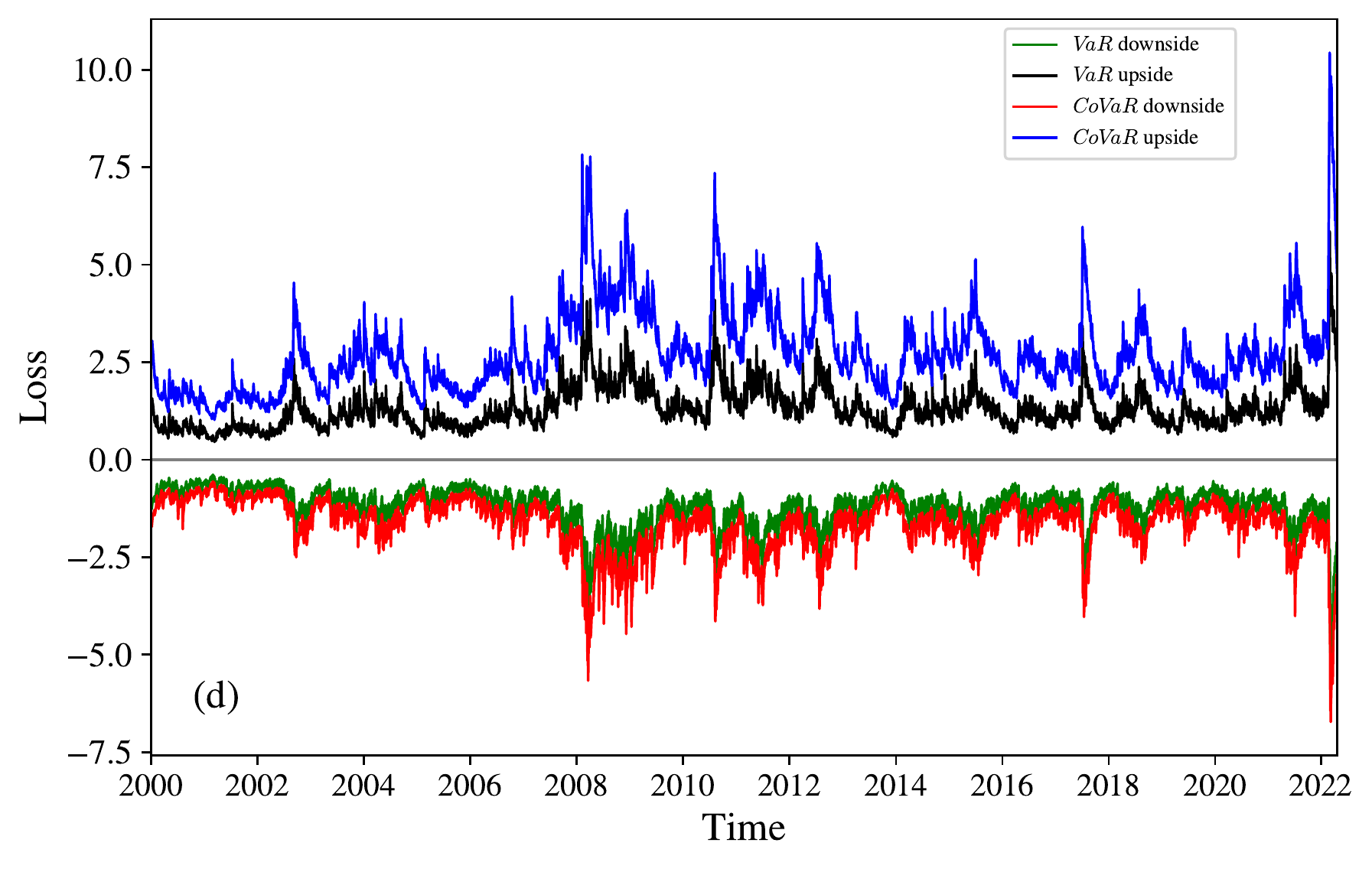}
\includegraphics[width=0.45\linewidth]{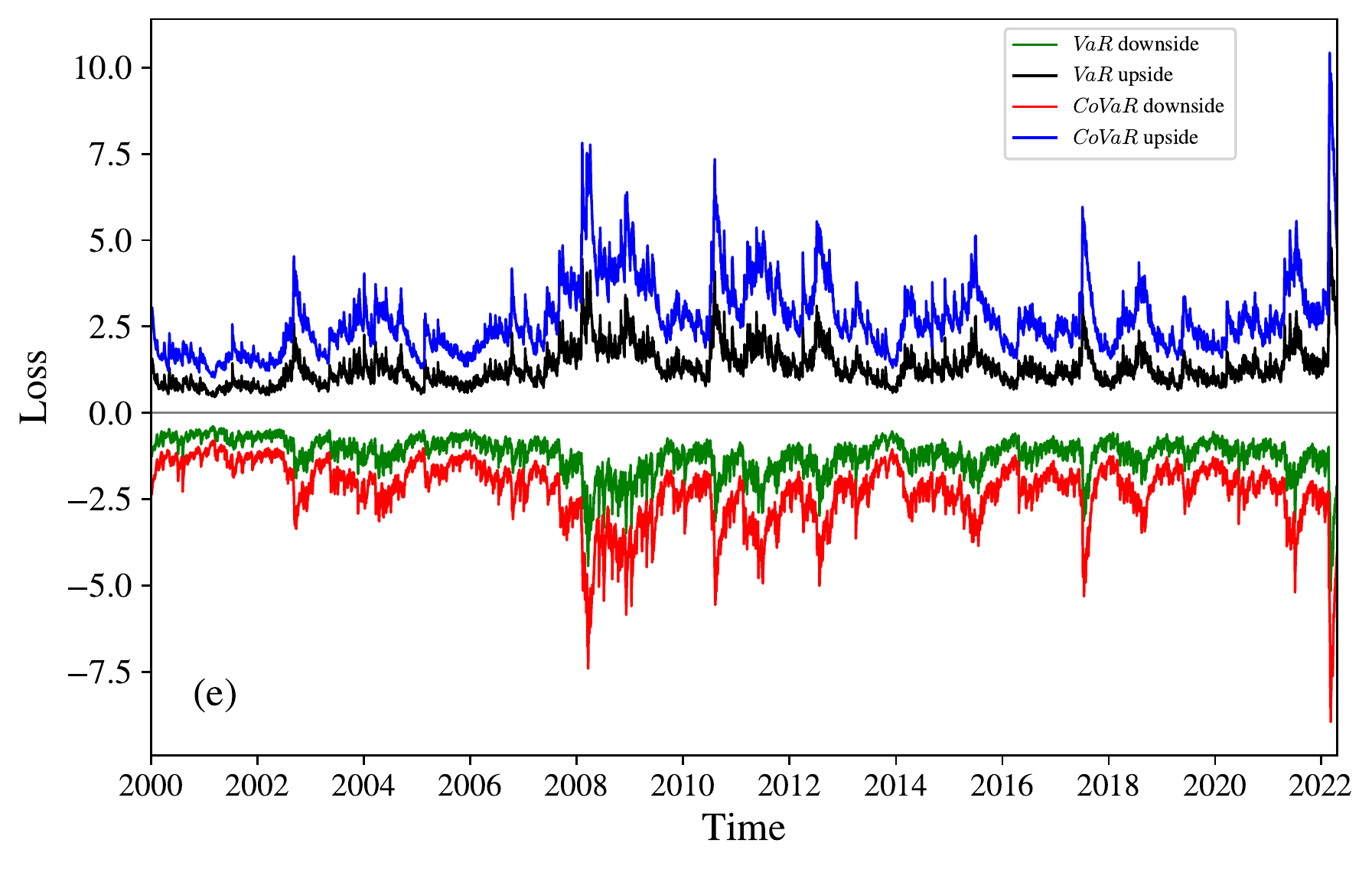}
\includegraphics[width=0.45\linewidth]{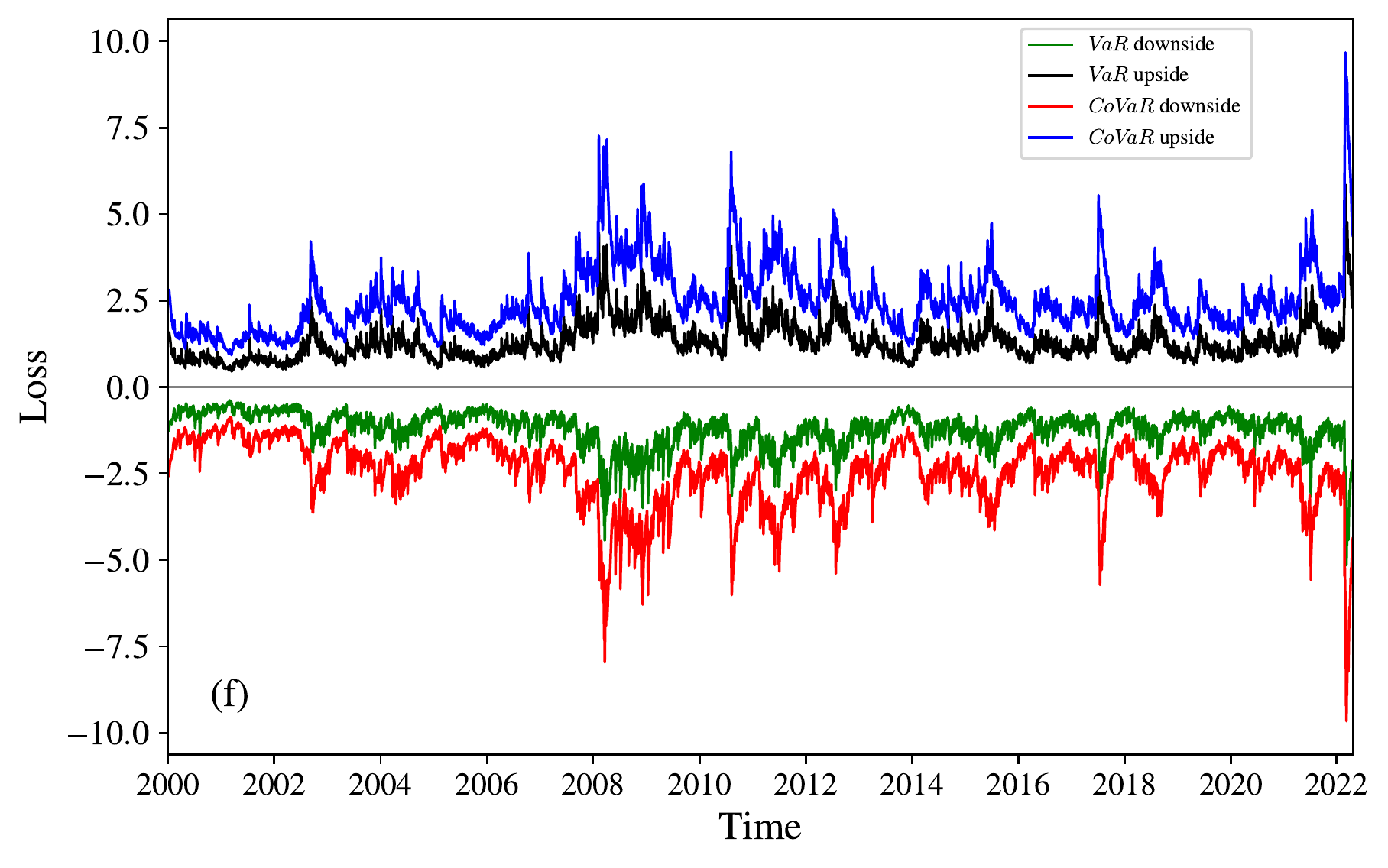}
\caption{$VaR$s and $CoVaR$s for wheat based on Normal copula (a), Student-t copula (b), Clayton copula (c), survival Clayton copula (d), Gumbel copula (e), and survival Gumbel copula (f).}
\label{Fig:VaRs_Wheat_singlecopula}
\end{figure}

\begin{figure}[!t]
\centering
\includegraphics[width=0.45\linewidth]{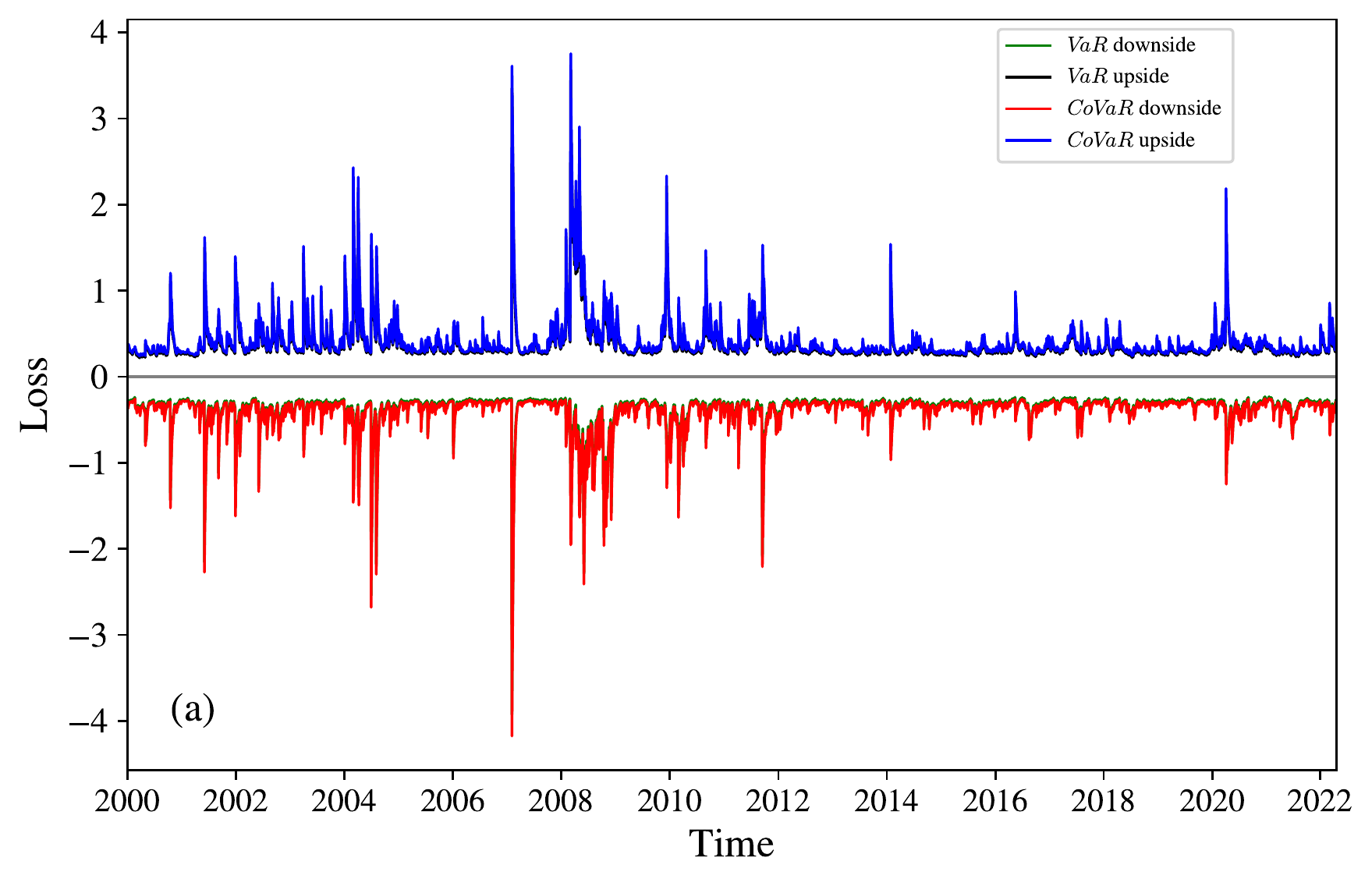}
\includegraphics[width=0.45\linewidth]{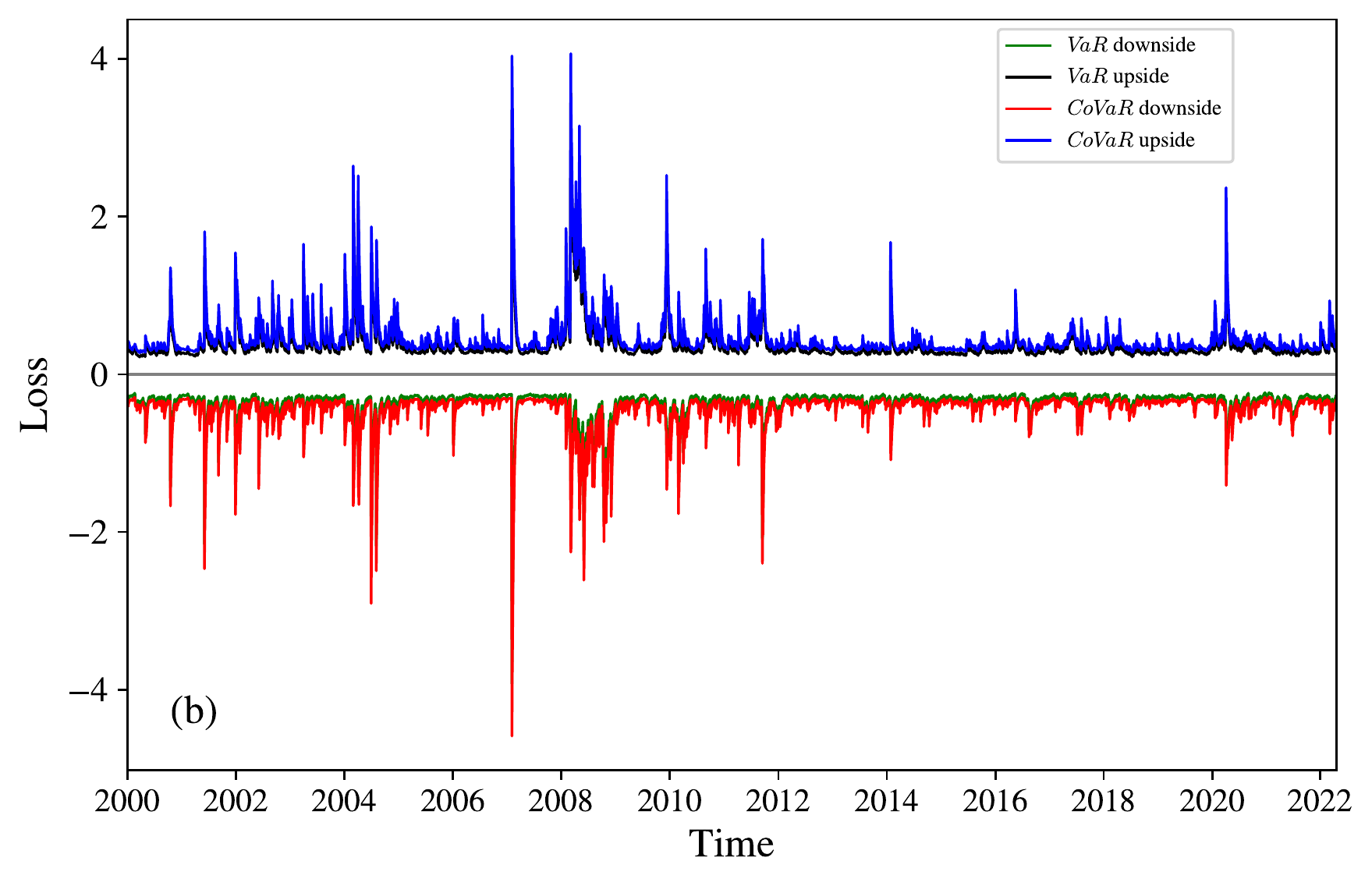}
\includegraphics[width=0.45\linewidth]{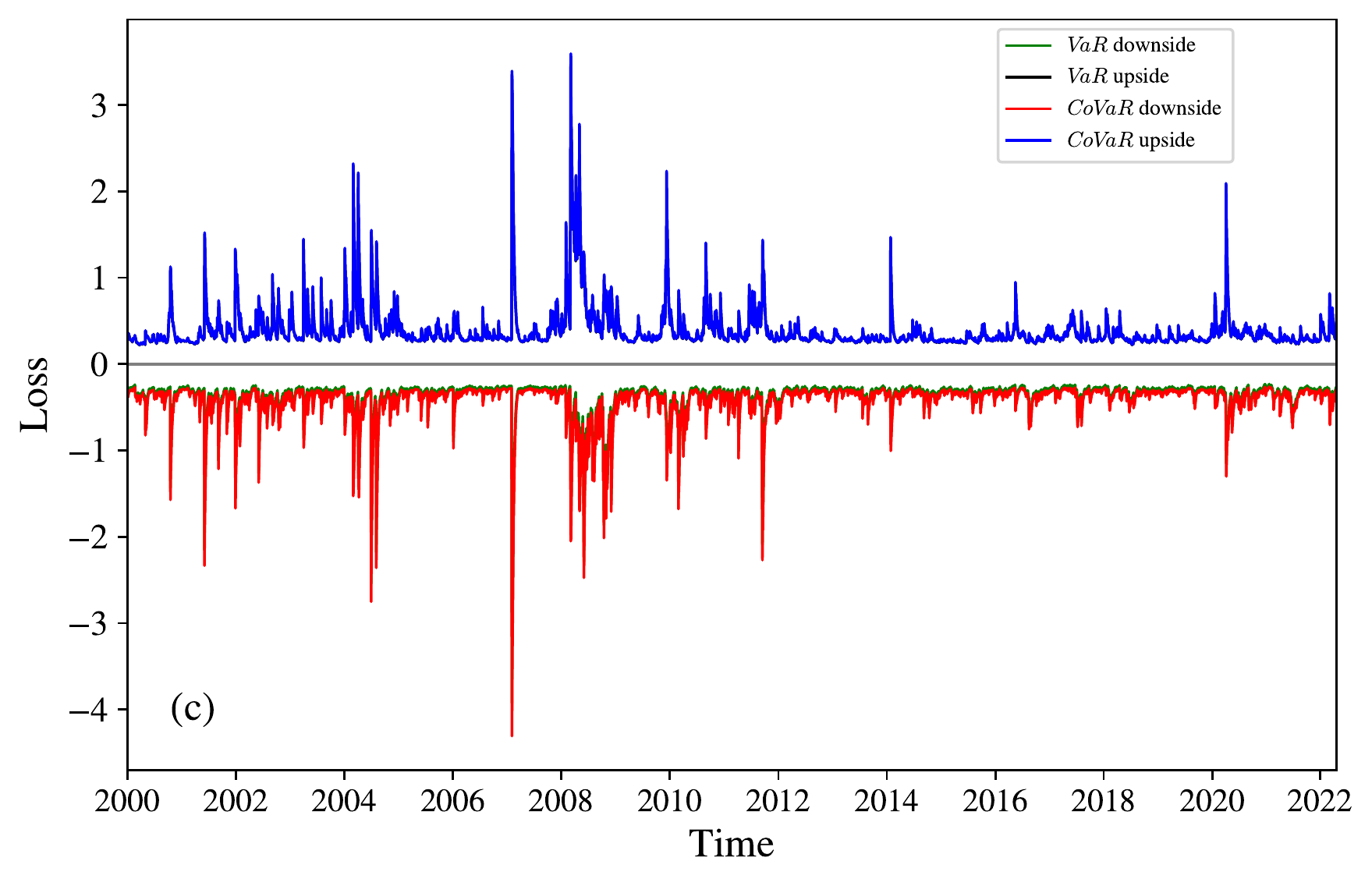}
\includegraphics[width=0.45\linewidth]{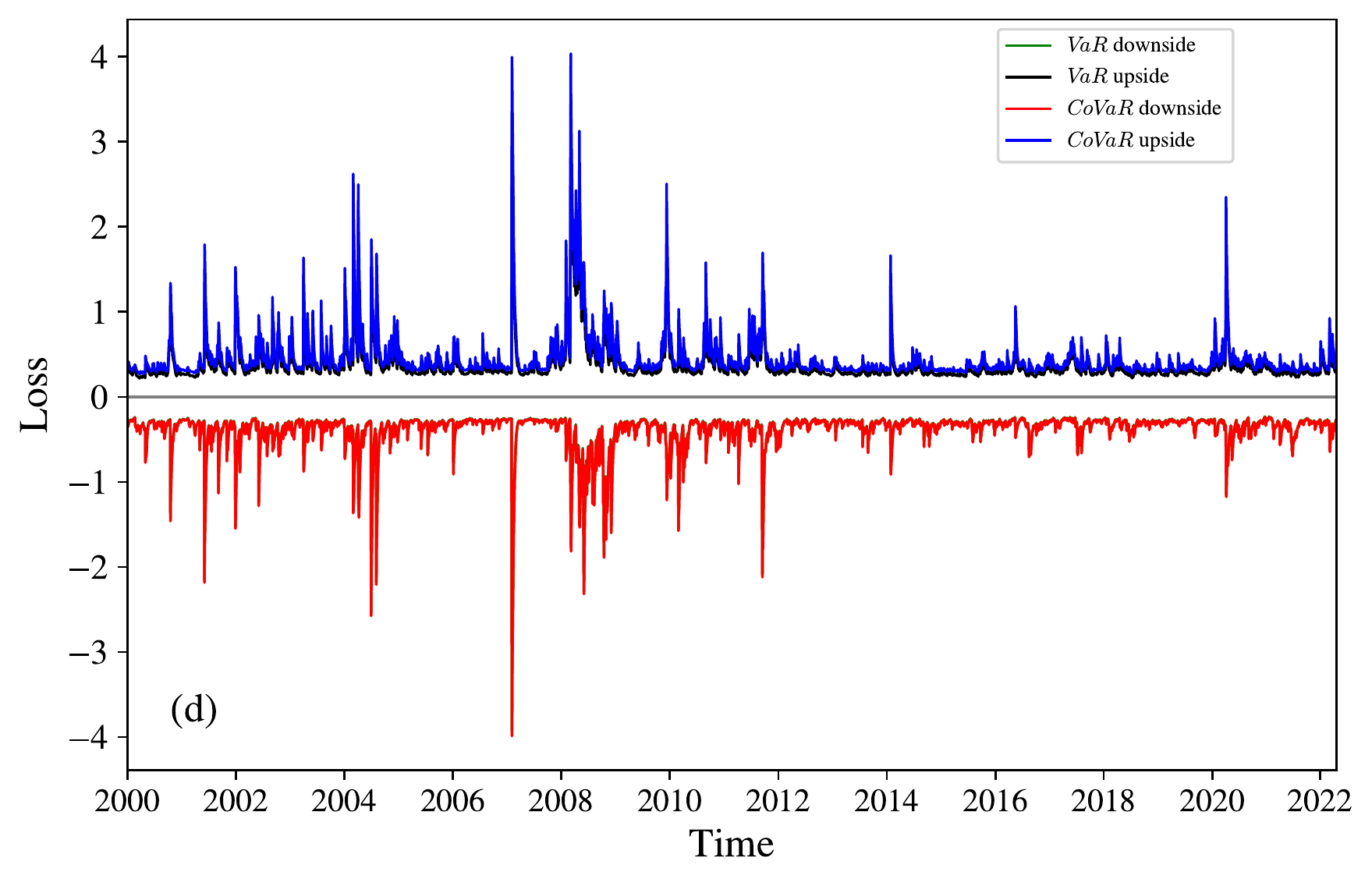}
\includegraphics[width=0.45\linewidth]{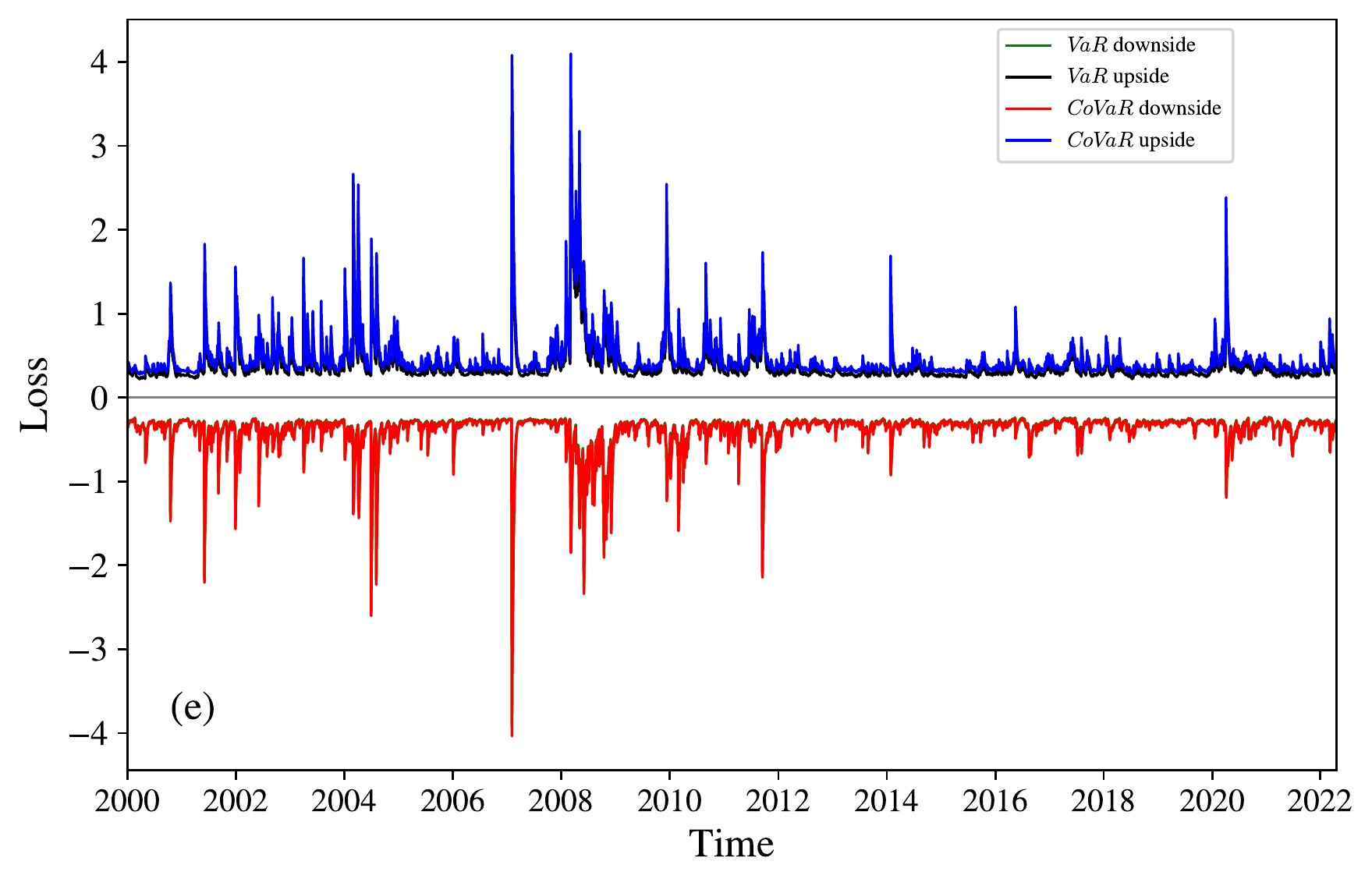}
\includegraphics[width=0.45\linewidth]{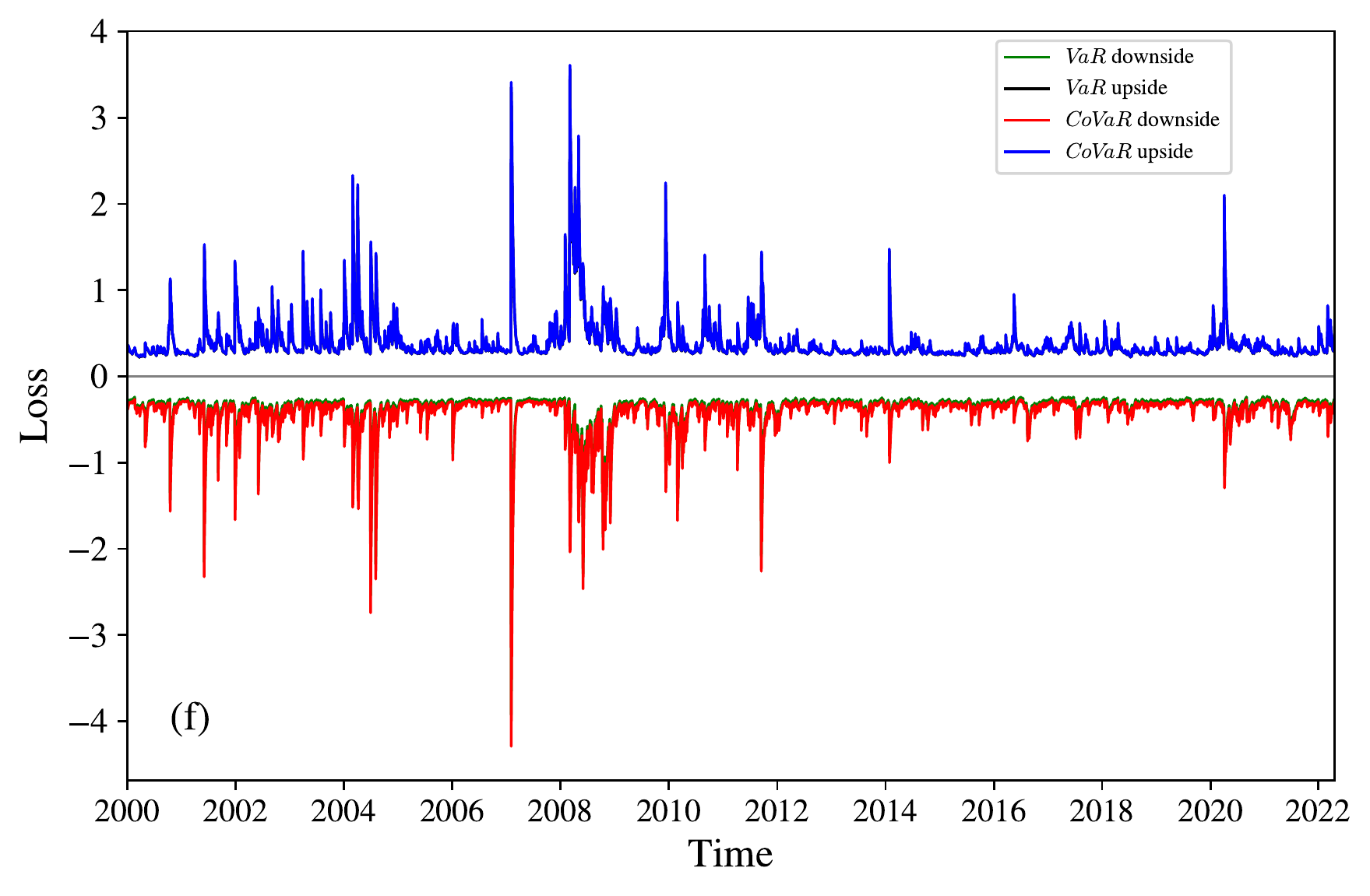}
\caption{$VaR$s and $CoVaR$s for rice based on Normal copula (a), Student-t copula (b), Clayton copula (c), survival Clayton copula (d), Gumbel copula (e), and survival Gumbel copula (f).}
\label{Fig:VaRs_Rice_singlecopula}
\end{figure}

Figures~\ref{Fig:VaRs_Soybean_singlecopula}--\ref{Fig:VaRs_Rice_singlecopula} depict the dynamic evolution of the downside and upside $VaR$s and $CoVaR$s for soybean, maize, wheat, and rice spot calculated by the estimated single copula models, respectively. It can be found that for the same agricultural commodity, the evolution processes of the downside and upside $VaR$s and $CoVaR$s exhibit similar trends in all cases, showing only slight magnitude differences. Furthermore, the downside $CoVaR$s for all agricultural commodities are smaller than the downside $VaR$s, and the upside $CoVaR$s are larger than the upside $VaR$s, indicating that for each agricultural commodity, the extreme downside and upside movements of the futures prices make a great impact on the spot prices, and thus the agricultural futures market has certain downside and upside risk spillover effects on the agricultural spot market.

From Figure~\ref{Fig:VaRs_Soybean_singlecopula}, we note that the $VaR$s and $CoVaR$s for soybean spot have mainly shown three large-scale fluctuations during 2000-2022, namely 2004-2006, 2008-2010 and 2020-2022. The world's leading producers and exporters of soybeans are the United States, Brazil, and Argentina. The supply-demand relationship was reversed due to the 2003-2004 drought in North America, which resulted in a significant decrease in soybean production. In addition, the continued easing of the dollar and the recovery of the global economy further boosted soybean prices. Then, the global soybean yields increased substantially, and the US dollar entered an interest rate hike cycle, bringing soybean prices down. This explains the first large-scale fluctuation of the risk spillovers from the soybean futures market to the soybean spot market. In 2007-2008, the extrusion of corn planting area and frequent floods affected the soybean output in the United States. From 2008 to 2009, Argentina raised export tariffs on several crops, including soybeans, because the main soybean-producing areas in Argentina were threatened by drought. The continuous falling output of soybeans in 2007-2009 resulted in the continuous rises in soybean prices. Commodity prices generally fell in the wake of the global financial crisis of 2008, and the global soybean production increased significantly in 2009-2010, which further drove soybean prices down. This explains the second large-scale fluctuation of the risk spillovers between soybean futures and spot markets. Since 2020, the prolonged La Nina phenomenon has led to frequent extreme weather events, including droughts and floods, which have greatly reduced soybean yields in the Americas. Moreover, the COVID-19 pandemic and the Russia-Ukraine conflict have further widened the food supply gap and thus driven food prices up. In this context, the third large-scale fluctuation of the risk spillovers measured by $VaR$s and $CoVaR$s reflects the increasing risks in the soybean futures and spot markets.

Compared with soybeans, the $VaR$s and $CoVaR$s for maize spot fluctuated more frequently from 2000 to 2022, especially during 2000-2001, 2008-2010, 2012-2014 and 2020-2022. As one of the most widely used grain varieties, maize can serve as feed and industrial raw materials, among other purposes. In 2000-2002, the global maize consumption demand increased steadily, but the maize supply was insufficient. Frequent episodes of extreme weather in 2006-2008, 2011-2012 and 2020-2022 led to widespread crop losses in the major maize-producing areas all over the world. Consequently, the tense between supply and demand, combined with the low stockpiles, pushed maize prices up. Moreover, with the vigorous promotion of new energy policies, the demand for biofuels such as fuel ethanol, most of which are processed from maize, continued to increase, further boosting the prices of maize. Similar to soybeans, the $VaR$s and $CoVaR$s for maize have fluctuated frequently since 2020, indicating that the risks in the maize futures and spot markets have increased.

As can be seen from Figure~\ref{Fig:VaRs_Wheat_singlecopula}, the $VaR$s and $CoVaR$s for wheat mainly experienced three periods of large fluctuations, including 2008-2009, 2010-2012 and 2021-2022. From 2006 to 2008, hurricanes, heat waves, floods and other meteorological disasters caused wheat production in the United States and Europe to decline year after year. In 2010-2012, the major wheat-producing areas including Russia, the Black Sea region and Australia suffered from a long spell of dry weather, which significantly cut the global wheat harvest. Meanwhile, a wide variety of trade protection policies were successively launched by several countries to restrict or prohibit the export of agricultural commodities and thus maintain their domestic food security, such as Argentina's increase in agricultural export taxes in 2007, Russia's ban on wheat export in 2010, and Ukraine's implementation of grain export quotas in 2010. These adjustments in agricultural trade policies exacerbated the shortage of wheat supply on the international market, further spurring price increases for wheat. Since August 2020, the widespread drought caused by the La Nina phenomenon has seriously affected the global grain yield, of which the impact on wheat is even worse because of the growing season. In addition, the Russian-Ukrainian conflict has a profound influence on the supply and demand situation of wheat because both Russia and Ukraine play a huge role in global wheat production and supply. Specifically, the Russian-Ukrainian conflict is generally thought to affect wheat prices from three aspects: first, it may derail the agricultural production process and then drag down the yield and export of wheat; second, it may lead to the cost increases for wheat by raising the prices of chemical fertilizers; third, it may push up international wheat prices by disrupting the grain supply chain. From Figure~\ref{Fig:VaRs_Wheat_singlecopula}, we note that the $VaR$s and $CoVaR$s for wheat have soared since 2022, implying a sharp increase in the risk of the global wheat market.

Combining with Figure~\ref{Fig:VaRs_Rice_singlecopula}, we can find that the evolution of the $VaR$s and $CoVaR$s for rice is quite different from that of the $VaR$s and $CoVaR$s for soybeans, maize and wheat. Specifically, the absolute values of the $VaR$s and $CoVaR$s for rice are relatively smaller than those of the $VaR$s and $CoVaR$s for the other agricultural commodities when the market is stable, but present jump characteristics when the market is at risk, such as 2004-2005, 2006-2010 and 2020-2021. India, the world's largest rice exporter and a dominant player in the international rice trade, was hit by devastating floods in 2005, 2006 and 2008, as well as a catastrophic drought in 2012, all of which caused rice harvests and exports to plunge and thus drove up rice prices. Since 2020, extreme weather events have reduced rice production in Pakistan and other countries, and increased the demand for imports, which further boosted the prices of rice. However, determined by dietary habits, the main rice producers and consumers are concentrated in Asia, and most of their harvests are kept for domestic consumption, which means that the relationship between rice production and consumption in world trade is relatively stable. In general, the risk of the international rice market is less than that of other agricultural markets.

Table~\ref{Tab:Statistics_VaR_Single} provides the summary statistics of $VaR$, $CoVaR$ and $\Delta CoVaR$ for the agricultural spot returns based on the estimated results of different single copula models, including mean values and standard deviations. The statistical results further corroborate the graphic evidence shown in Figures~\ref{Fig:VaRs_Soybean_singlecopula}--\ref{Fig:VaRs_Rice_singlecopula} that information from the agricultural futures market usually increase the risk exposure of the agricultural spot market. Specifically, the $CoVaR$s and $\Delta CoVaR$s calculated by different single copula models for the same agricultural commodity are different to some extent, but the mean values of the downside $CoVaR$s for all agricultural commodities are smaller than those of the downside $VaR$s, and the mean values of the upside $CoVaR$s are larger than those of the upside $VaR$s. Moreover, all the mean values of the downside $\Delta CoVaR$s are less than 0 while all the mean values of the upside $\Delta CoVaR$s are greater than 0, which implies that there exist both downside and upside risk spillover effects from the agricultural futures market to the agricultural spot market. Generally, the mean values of the downside $CoVaR$s for soybeans and maize are relatively small, and their mean values of the upside $CoVaR$s are relatively large, indicating that the spot markets for soybeans and maize are more sensitive to risk shocks from their corresponding futures markets. On the contrary, the rice spot has the largest mean value of the downside $CoVaR$s and the smallest mean value of the upside $CoVaR$s, suggesting that the extreme downside and upside risks of the international rice market are smaller than those of other agricultural commodities, which is also consistent with the above analysis of Figures~\ref{Fig:VaRs_Soybean_singlecopula}--\ref{Fig:VaRs_Rice_singlecopula}.

\begin{table}[!ht]
  \centering
  \setlength{\abovecaptionskip}{0pt}
  \setlength{\belowcaptionskip}{10pt}
  \caption{Summary statistics (mean and standard deviation) of $VaR$, $CoVaR$ and $\Delta CoVaR$ based on different single copula models}
  \setlength\tabcolsep{13.3pt}
    \resizebox{\textwidth}{!}{
    \begin{tabular}{l r@{.}l r@{.}l r@{.}l r@{.}l}
    \toprule
        & \multicolumn{2}{c}{Soybean} & \multicolumn{2}{c}{Maize} & \multicolumn{2}{c}{Wheat} & \multicolumn{2}{c}{Rice} \\
    \midrule
    \multicolumn{9}{l}{\textit{Panel A: VaR}} \\
    $VaR_{0.05}$ & $-$2&0431 (0.8697) & $-$1&9839 (0.7214) & $-$1&2016 (0.5033) & $-$0&3775 (0.2109) \\
    $VaR_{0.95}$ & 1&9855 (0.8141) & 2&0725 (0.7329) & 1&3037 (0.5283) & 0&3844 (0.2434) \vspace{1mm}\\
    \multicolumn{9}{l}{\textit{Panel B: Normal copula}} \\
    $CoVaR_{0.05}$ & $-$4&4510 (1.8585) & $-$4&4813 (1.6067) & $-$2&3987 (0.9462) & $-$0&4034 (0.2244) \\
    $CoVaR_{0.95}$ & 4&1702 (1.7105) & 4&6874 (1.6600) & 2&6633 (1.0338) & 0&4113 (0.2576) \\
    $\Delta CoVaR_{0.05}$ & $-$1&8796 (0.7760) & $-$1&9685 (0.7003) & $-$0&9159 (0.3519) & $-$0&0162 (0.0091) \\
    $\Delta CoVaR_{0.95}$ & 1&7031 (0.7031) & 2&0624 (0.7337) & 1&0421 (0.4004) & 0&0168 (0.0094) \vspace{1mm}\\
    \multicolumn{9}{l}{\textit{Panel C: Student-t copula}} \\
    $CoVaR_{0.05}$ & $-$4&4470 (1.8568) & $-$4&5011 (1.6137) & $-$2&4157 (0.9526) & $-$0&4467 (0.2472) \\
    $CoVaR_{0.95}$ & 4&1666 (1.7090) & 4&7082 (1.6674) & 2&6827 (1.0412) & 0&4560 (0.2815) \\
    $\Delta CoVaR_{0.05}$ & $-$1&8888 (0.7798) & $-$2&0002 (0.7116) & $-$0&9396 (0.3610) & $-$0&0603 (0.0337) \\
    $\Delta CoVaR_{0.95}$ & 1&7115 (0.7066) & 2&0956 (0.7455) & 1&0692 (0.4108) & 0&0624 (0.0348) \vspace{1mm}\\
    \multicolumn{9}{l}{\textit{Panel D: Clayton copula}}  \\
    $CoVaR_{0.05}$ & $-$4&4699 (1.8662) & $-$4&5358 (1.6260) & $-$2&4399 (0.9617) & $-$0&4173 (0.2317) \\
    $CoVaR_{0.95}$ & 2&9196 (1.1963) & 3&0393 (1.0750) & 1&7735 (0.7002) & 0&3886 (0.2456) \\
    $\Delta CoVaR_{0.05}$ & $-$1&8984 (0.7838) & $-$2&0228 (0.7196) & $-$0&9569 (0.3677) & $-$0&0311 (0.0174) \\
    $\Delta CoVaR_{0.95}$ & 0&4940 (0.2040) & 0&4884 (0.1737) & 0&2160 (0.0830) & 0&0012 (0.0007) \vspace{1mm}\\
    \multicolumn{9}{l}{\textit{Panel E: Survival Clayton copula}}  \\
    $CoVaR_{0.05}$ & $-$3&1158 (1.3092) & $-$2&9085 (1.0485) & $-$1&6334 (0.6604) & $-$0&3839 (0.2143) \\
    $CoVaR_{0.95}$ & 4&1873 (1.7175) & 4&7445 (1.6803) & 2&7106 (1.0517) & 0&4517 (0.2792) \\
    $\Delta CoVaR_{0.05}$ & $-$0&5818 (0.2402) & $-$0&4667 (0.1660) & $-$0&2017 (0.0775) & $-$0&0019 (0.0011) \\
    $\Delta CoVaR_{0.95}$ & 1&7202 (0.7102) & 2&1193 (0.7539) & 1&0891 (0.4185) & 0&0530 (0.0296) \vspace{1mm}\\
    \multicolumn{9}{l}{\textit{Panel F: Gumbel copula}}  \\
    $CoVaR_{0.05}$ & $-$4&3268 (1.8073) & $-$4&2374 (1.5201) & $-$2&2440 (0.8880) & $-$0&3889 (0.2168) \\
    $CoVaR_{0.95}$ & 4&1871 (1.7174) & 4&7435 (1.6799) & 2&7089 (1.0511) & 0&4607 (0.2840) \\
    $\Delta CoVaR_{0.05}$ & $-$1&7578 (0.7257) & $-$1&7327 (0.6164) & $-$0&7699 (0.2958) & $-$0&0066 (0.0037) \\
    $\Delta CoVaR_{0.95}$ & 1&7201 (0.7101) & 2&1184 (0.7536) & 1&0878 (0.4179) & 0&0665 (0.0372) \vspace{1mm}\\
    \multicolumn{9}{l}{\textit{Panel G: Survival Gumbel copula}}  \\
    $CoVaR_{0.05}$ & $-$4&4697 (1.8662) & $-$4&5347 (1.6257) & $-$2&4383 (0.9611) & $-$0&4156 (0.2308) \\
    $CoVaR_{0.95}$ & 4&0441 (1.6586) & 4&4305 (1.5688) & 2&4760 (0.9631) & 0&3906 (0.2467) \\
    $\Delta CoVaR_{0.05}$ & $-$1&8983 (0.7837) & $-$2&0219 (0.7193) & $-$0&9556 (0.3672) & $-$0&0331 (0.0185) \\
    $\Delta CoVaR_{0.95}$ & 1&5797 (0.6522) & 1&8139 (0.6453) & 0&8657 (0.3326) & 0&0036 (0.0020) \\
    \bottomrule
    \end{tabular}}%
  \begin{flushleft}
    \footnotesize
    \justifying Note: This table provides the summary statistics of downside and upside $VaR$, $CoVaR$ and $\Delta CoVaR$ for the spot returns of soybean, maize, wheat, and rice, based on the estimated results of six different single copula models, including mean values and standard deviations (in parentheses). The subscripts 0.05 and 0.95 refer to the downside and upside risk measures, respectively.
  \end{flushleft}
  \label{Tab:Statistics_VaR_Single}%
\end{table}%

In order to evaluate the robustness and significance of the downside and upside risk spillover effects, we adopt the K-S test to the $VaR$ and $CoVaR$ based on different single copula models, and the test results are presented in Table~\ref{Tab:Spillover_Test_Single}. For each spot return series, its downside $CoVaR$s conditional on the downside $VaR$s for futures returns are significantly smaller than its downside $VaR$s, and its upside $CoVaR$s conditional on the upside $VaR$s for futures returns are significantly larger than its upside $VaR$s, regardless of which single copula model is used for calculation. This proves that the futures market for each agricultural commodity has significant extreme downside and upside risk spillover effects on the corresponding spot market, and the two risk spillover effects are quite robust.

\begin{table}[!ht]
  \centering
  \setlength{\abovecaptionskip}{0pt}
  \setlength{\belowcaptionskip}{10pt}
  \caption{Hypothesis testing for downside and upside risk spillover effects based on different single copula models}
    \setlength\tabcolsep{10pt}
    \resizebox{\textwidth}{!}{
    \begin{tabular}{lcccc}
    \toprule
        & \multicolumn{1}{c}{Soybean} & \multicolumn{1}{c}{Maize} & \multicolumn{1}{c}{Wheat} & \multicolumn{1}{c}{Rice} \\
    \midrule
    \multicolumn{5}{l}{\textit{Panel A: Normal copula}} \\
    \makecell{$H_{0}: CoVaR_{0.05} = VaR_{0.05}$ \\ $H_{1}: CoVaR_{0.05} < VaR_{0.05}$} & 0.7129 [0.0000] & 0.7570 [0.0000] & 0.6642 [0.0000] & 0.1901 [0.0000] \\
    \makecell{$H_{0}: CoVaR_{0.95} = VaR_{0.95}$ \\ $H_{1}: CoVaR_{0.95} > VaR_{0.95}$} & 0.6917 [0.0000] & 0.7661 [0.0000] & 0.6894 [0.0000] & 0.1974 [0.0000] \vspace{1mm}\\
    \multicolumn{5}{l}{\textit{Panel B: Student-t copula}} \\
    \makecell{$H_{0}: CoVaR_{0.05} = VaR_{0.05}$ \\ $H_{1}: CoVaR_{0.05} < VaR_{0.05}$} & 0.7127 [0.0000] & 0.7604 [0.0000] & 0.6698 [0.0000] & 0.4478 [0.0000] \\
    \makecell{$H_{0}: CoVaR_{0.95} = VaR_{0.95}$ \\ $H_{1}: CoVaR_{0.95} > VaR_{0.95}$} & 0.6910 [0.0000] & 0.7692 [0.0000] & 0.6943 [0.0000] & 0.4320 [0.0000] \vspace{1mm}\\
    \multicolumn{5}{l}{\textit{Panel C: Clayton copula}} \\
    \makecell{$H_{0}: CoVaR_{0.05} = VaR_{0.05}$ \\ $H_{1}: CoVaR_{0.05} < VaR_{0.05}$} & 0.7145 [0.0000] & 0.7658 [0.0000] & 0.6762 [0.0000] & 0.2838 [0.0000] \\
    \makecell{$H_{0}: CoVaR_{0.95} = VaR_{0.95}$ \\ $H_{1}: CoVaR_{0.95} > VaR_{0.95}$} & 0.4314 [0.0000] & 0.3933 [0.0000] & 0.3643 [0.0000] & 0.0353 [0.0000] \vspace{1mm}\\
    \multicolumn{5}{l}{\textit{Panel D: Survival Clayton copula}} \\
    \makecell{$H_{0}: CoVaR_{0.05} = VaR_{0.05}$ \\ $H_{1}: CoVaR_{0.05} < VaR_{0.05}$} & 0.4557 [0.0000] & 0.3987 [0.0000] & 0.3522 [0.0000] & 0.0517 [0.0000] \\
    \makecell{$H_{0}: CoVaR_{0.95} = VaR_{0.95}$ \\ $H_{1}: CoVaR_{0.95} > VaR_{0.95}$} & 0.6938 [0.0000] & 0.7741 [0.0000] & 0.7004 [0.0000] & 0.4129 [0.0000] \vspace{1mm}\\
    \multicolumn{5}{l}{\textit{Panel E: Gumbel copula}}  \\
    \makecell{$H_{0}: CoVaR_{0.05} = VaR_{0.05}$ \\ $H_{1}: CoVaR_{0.05} < VaR_{0.05}$} & 0.6991 [0.0000] & 0.7214 [0.0000] & 0.6217 [0.0000] & 0.0873 [0.0000] \\
    \makecell{$H_{0}: CoVaR_{0.95} = VaR_{0.95}$ \\ $H_{1}: CoVaR_{0.95} > VaR_{0.95}$} & 0.6938 [0.0000] & 0.7741 [0.0000] & 0.7002 [0.0000] & 0.4514 [0.0000] \vspace{1mm}\\
    \multicolumn{5}{l}{\textit{Panel F: Survival Gumbel copula}} \\
    \makecell{$H_{0}: CoVaR_{0.05} = VaR_{0.05}$ \\ $H_{1}: CoVaR_{0.05} < VaR_{0.05}$} & 0.7145 [0.0000] & 0.7654 [0.0000] & 0.6758 [0.0000] & 0.2725 [0.0000] \\
    \makecell{$H_{0}: CoVaR_{0.95} = VaR_{0.95}$ \\ $H_{1}: CoVaR_{0.95} > VaR_{0.95}$} & 0.6760 [0.0000] & 0.7301 [0.0000] & 0.6388 [0.0000] & 0.0520 [0.0000] \\
    \bottomrule
    \end{tabular}}%
  \begin{flushleft}
    \footnotesize
    \justifying Note: This table reports the results (K-S statistic) of the K-S test for the significance of the downside and upside risk spillover effects based on various single copula models, where the $p$-values of test statistics are presented in square brackets. The rejection of the null hypotheses indicates the existence of significant risk spillover effects.
  \end{flushleft}
  \label{Tab:Spillover_Test_Single}%
\end{table}%

\subsection{Empirical results based on mixed copula models}

\subsubsection{Mixed copula estimation}

%\subsection{Mixed copula estimation between futures and spot agricultural commodities}

Based on the above analysis of the goodness-of-fit measures of different single copula models, we next construct mixed copula models combining the Gumbel copula and the survival Gumbel copula, which can take the possible asymmetric tail dependence into consideration, to further investigate the dependence structure between the agricultural futures and spot returns.

Table~\ref{Tab:Agro_Mixed_Copula_Estimation} reports the estimated results of the mixed copula models for soybeans, maize, wheat, and rice. A comparison of Table~\ref{Tab:Agro_Mixed_Copula_Estimation} and Table~\ref{Tab:Agro_Single_Copula_Estimation} shows that for soybeans, maize and wheat, their corresponding mixed copula models have larger LL and smaller AIC and BIC than the single copula models, indicating that the mixed copula models are capable of better describing the dependence structure between the agricultural futures and spot markets. For rice, however, the survival Clayton copula model is still the best choice. In addition, according to the parameter estimates in Table~\ref{Tab:Agro_Mixed_Copula_Estimation}, the weight of the Gumbel copula is not equal to that of the survival Gumbel copula in the corresponding mixed copula model for each pair of agricultural commodity, which verifies that the tail dependence structure between the agricultural futures and spot returns is asymmetric. More specifically, there exists strong lower tail dependence between the soybean futures and spot, while the upper tail dependence is greater than the lower tail dependence for each of the other three agricultural pairs, especially for rice, whose futures and spot markets have almost no lower tail dependence. The above findings are also supported by the lower and upper tail dependence estimates for each agricultural pair in Table~\ref{Tab:Agro_Mixed_Copula_Estimation}.

\begin{table}[!ht]
  \centering
  \setlength{\abovecaptionskip}{0pt}
  \setlength{\belowcaptionskip}{10pt}
  \caption{Estimation of mixed copula models between futures and spot return series of agricultural commodities}
  \setlength\tabcolsep{26pt}
  \begin{tabular}{lcccc}
    \toprule
        & \multicolumn{1}{c}{Soybean} & \multicolumn{1}{c}{Maize} & \multicolumn{1}{c}{Wheat} & \multicolumn{1}{c}{Rice} \\
    \midrule
    \multicolumn{5}{l}{\textit{Panel A: Parameter estimates}} \\
    $\theta_{c}^{1}$ & 2.0446 & 4.0731 & 2.5828 & 1.0131 \\
    $\theta_{c}^{2}$ & 5.7878 & 2.2462 & 2.7929 & 1.0000 \\
    $\omega_{1}$ & 0.2999 & 0.6152 & 0.5559 & 1.0000 \\
    $\omega_{2}$ & 0.7001 & 0.3848 & 0.4441 & 0.0000 \\
    $\lambda^{\mathrm{up}}$ & 0.1789 & 0.5010 & 0.3848 & 0.0178 \\
    $\lambda^{\mathrm{low}}$ & 0.6110 & 0.2457 & 0.3190 & 0.0000 \\
    $\tau$ & 0.7193 & 0.6711 & 0.6243 & 0.0129 \vspace{2mm}\\
    \multicolumn{5}{l}{\textit{Panel B: Goodness-of-fit measures}} \\
    LL & \multicolumn{1}{r}{5063.5010} & \multicolumn{1}{r}{3979.9250} & \multicolumn{1}{r}{3211.6890} & \multicolumn{1}{r}{1.9736} \\
    AIC & \multicolumn{1}{r}{$-$10121.0000} & \multicolumn{1}{r}{$-$7953.8510} & \multicolumn{1}{r}{$-$6417.3790} & \multicolumn{1}{r}{2.0528} \\
    BIC & \multicolumn{1}{r}{$-$10101.1000} & \multicolumn{1}{r}{$-$7933.9520} & \multicolumn{1}{r}{$-$6397.4800} & \multicolumn{1}{r}{21.9518} \\ 
    \bottomrule
    \end{tabular}%
  \begin{flushleft}
    \footnotesize
    \justifying Note: This table reports the parameter estimates and goodness-of-fit measures of the mixed copula models for soybeans, maize, wheat, and rice. $\theta_{c}^{1}$, $\theta_{c}^{2}$ denote the copula parameters, and $\omega_{1}$, $\omega_{2}$ denote the weight parameters of the Gumbel copula model and the survival Gumbel copula model. The lower tail dependence $\lambda^{\mathrm{low}}$ and the upper tail dependence $\lambda^{\mathrm{up}}$ for each agricultural pair are presented in the table, as well as the Kendall rank correlation coefficient $\tau$.
  \end{flushleft}
  \label{Tab:Agro_Mixed_Copula_Estimation}%
\end{table}%

\subsubsection{Risk spillover measure}

\begin{figure}[!t]
\centering
\includegraphics[width=0.45\linewidth]{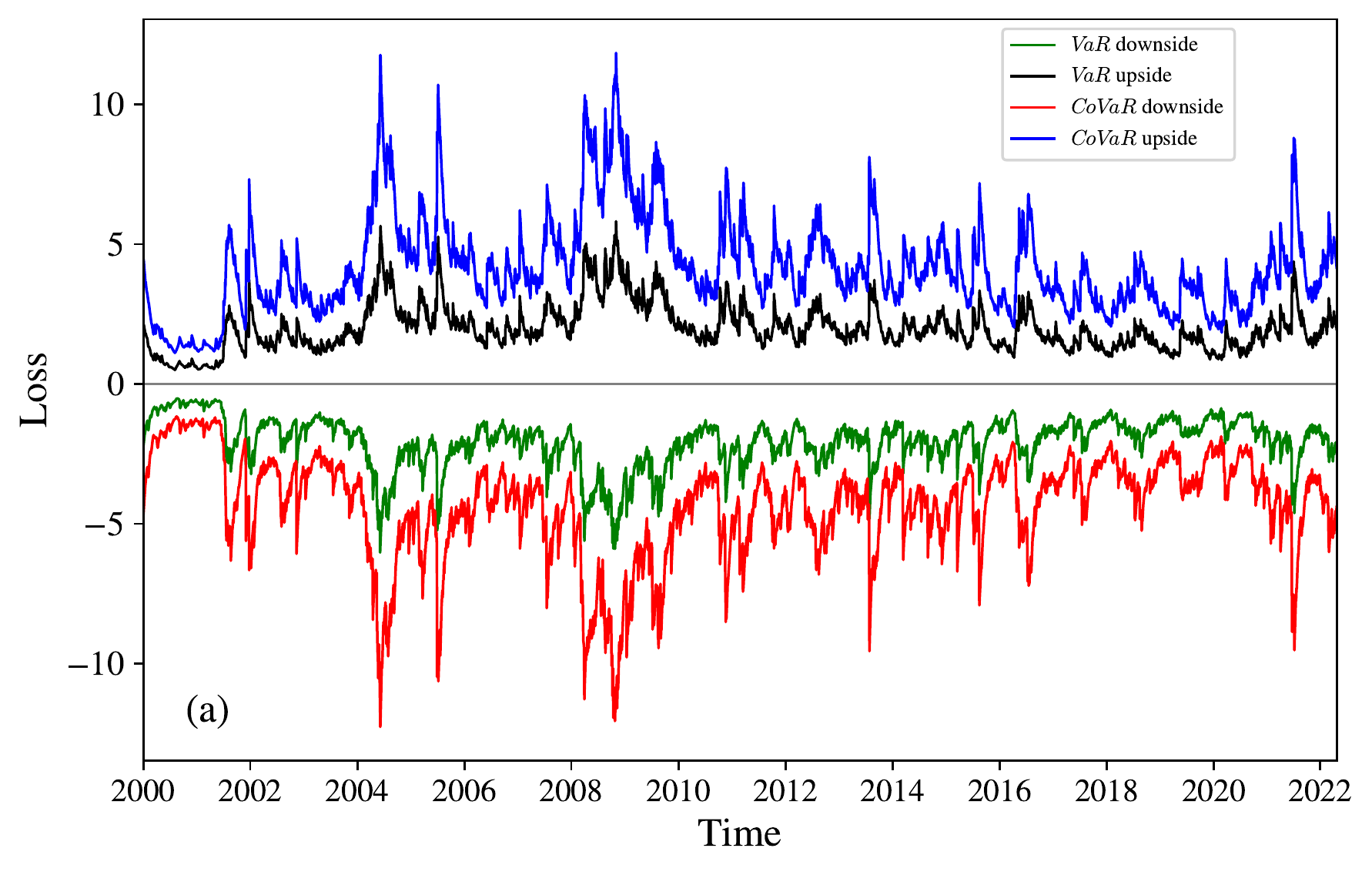}
\includegraphics[width=0.45\linewidth]{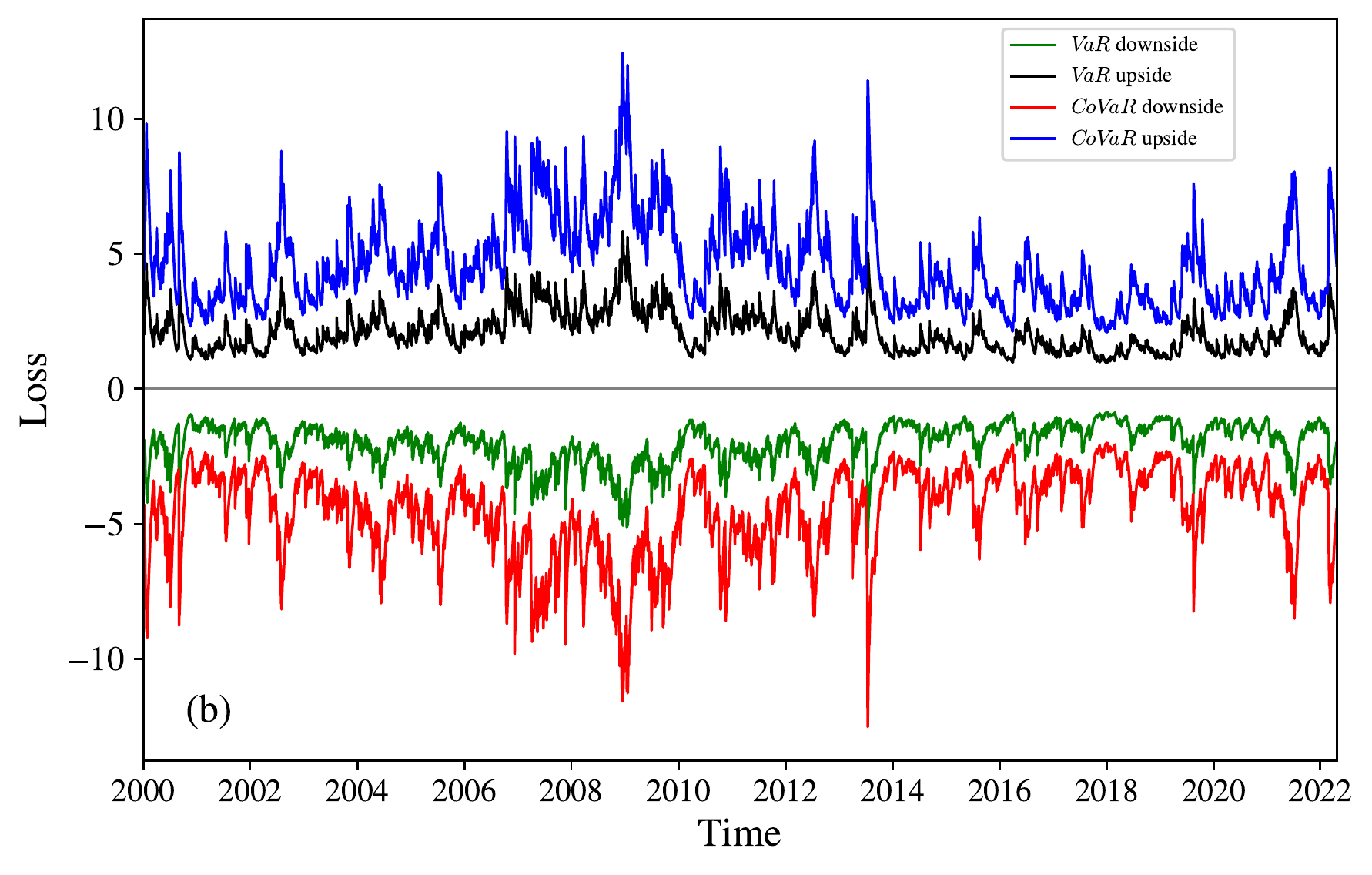}
\includegraphics[width=0.45\linewidth]{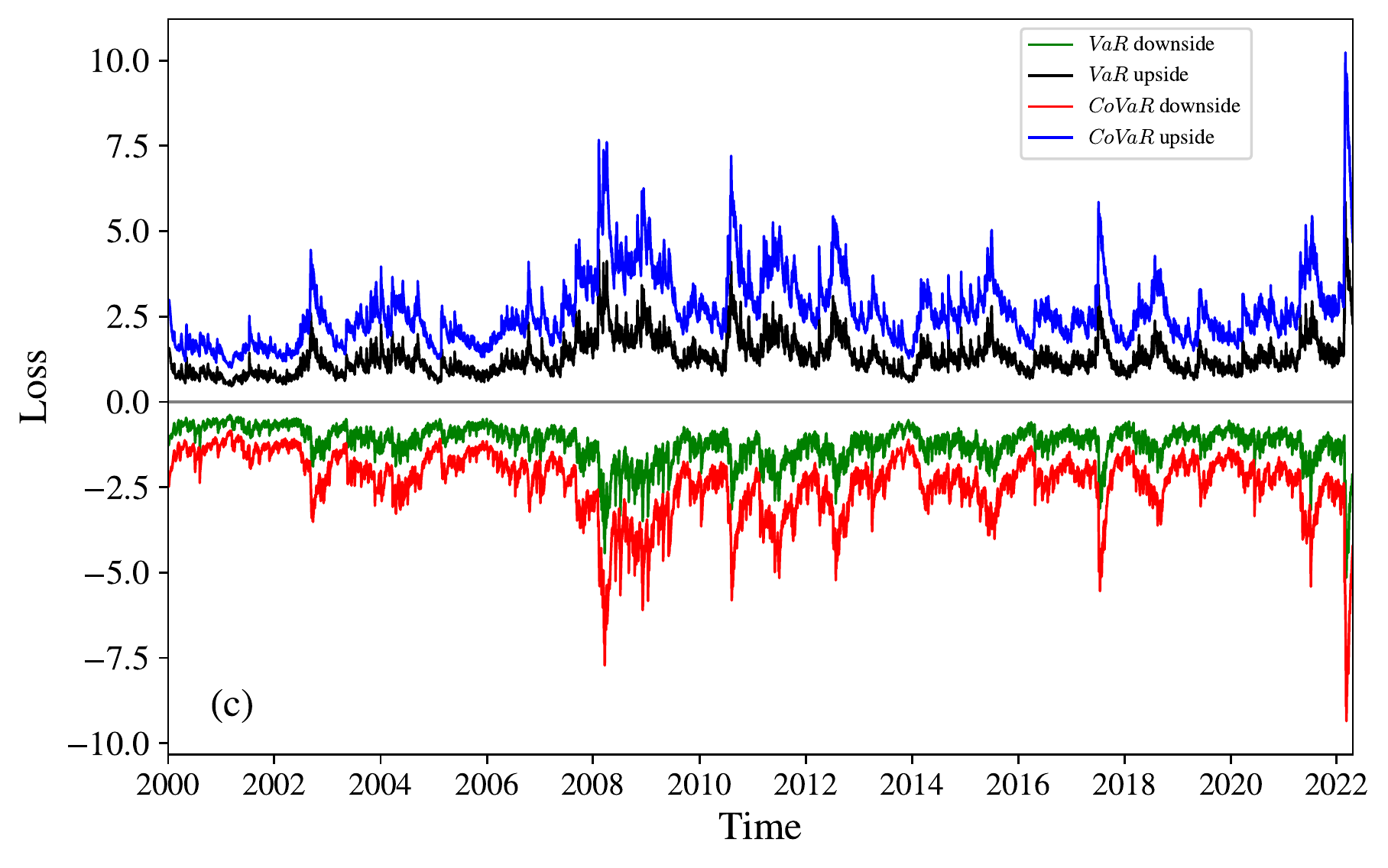}
\includegraphics[width=0.45\linewidth]{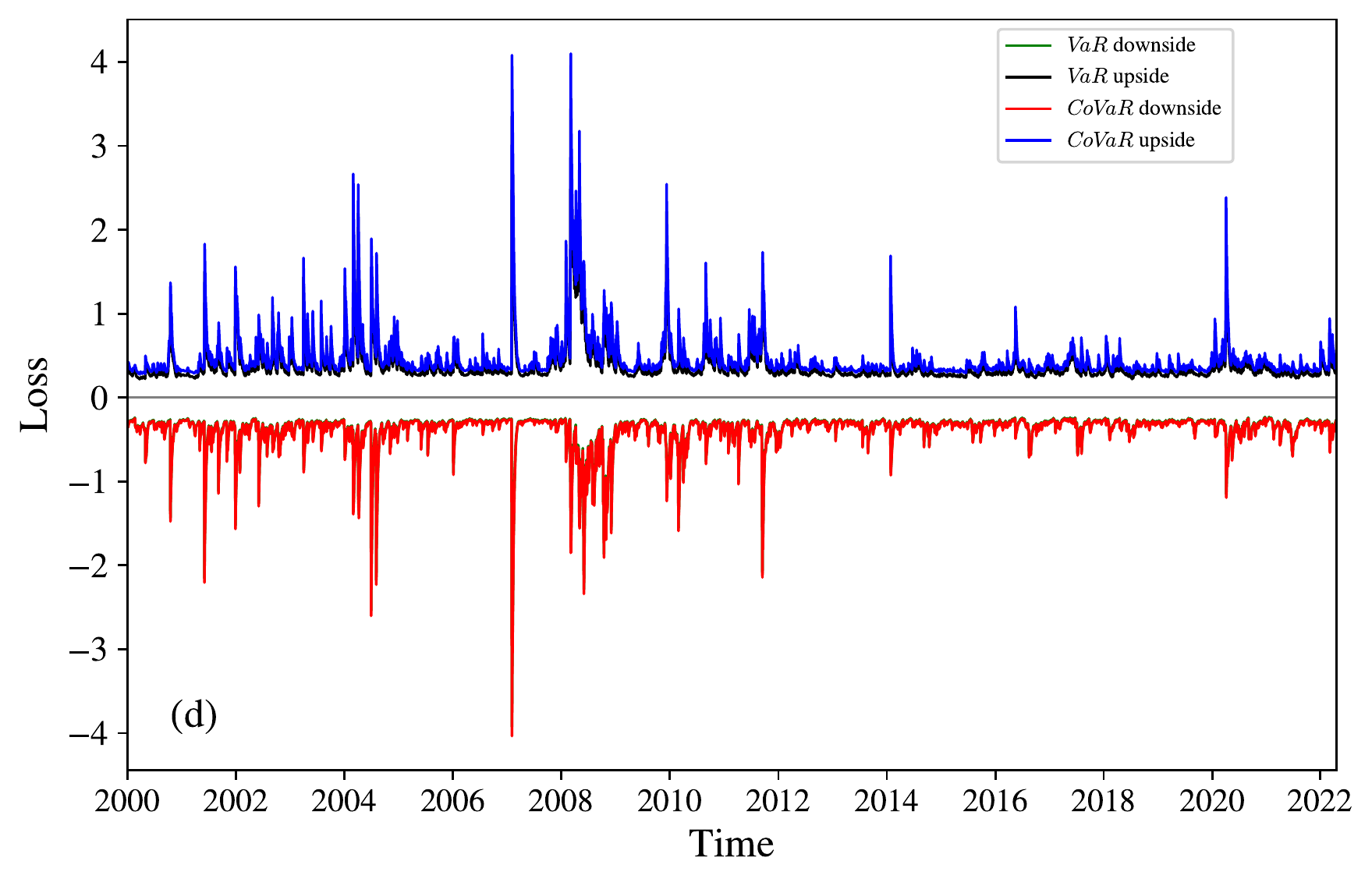}
\caption{$VaR$s and $CoVaR$s for soybean (a), maize (b), wheat (c), and rice (d) based on mixed copula models.}
\label{Fig:VaRs_mixedcopula}
\end{figure}

After analyzing the tail dependence structure between the agricultural futures and spot markets, we calculate the $CoVaR$s and $\Delta CoVaR$s for soybean, maize, wheat, and rice spot based on the estimated results of the mixed copula models to measure the extreme downside and upside risk spillovers from the agricultural futures returns to the agricultural spot returns. Figure~\ref{Fig:VaRs_mixedcopula} depicts the downside and upside $VaR$s and $CoVaR$s for the agricultural spot returns. We note that the shape of $CoVaR$s for each agricultural commodity in Figure~\ref{Fig:VaRs_mixedcopula}(a)--(d) is similar to that of $CoVaR$s in Figures~\ref{Fig:VaRs_Soybean_singlecopula}--\ref{Fig:VaRs_Rice_singlecopula}, with only slight difference in numerical values. Furthermore, the downside $CoVaR$s for each agricultural commodity are smaller than its downside $VaR$s, and the upside $CoVaR$s are larger than its upside $VaR$s, implying that the extreme downside and upside risks of the agricultural futures prices tend to spill over to the corresponding spot market, which is consistent with the results based on the single copula models.

Table~\ref{Tab:Statistics_VaR_Mix} provides the summary statistics of $VaR$, $CoVaR$ and $\Delta CoVaR$ calculated by the estimated results of the mixed copula models. It can be found that the maximum, minimum and mean values of the downside $CoVaR$s are smaller than the corresponding values of the downside $VaR$s for all agricultural commodities, and the maximum, minimum and mean values of the upside $CoVaR$s are larger than the corresponding values of the upside $VaR$s. Moreover, the maximum value of the downside $\Delta CoVaR$s for each agricultural commodity is less than 0, while the minimum value of the upside $\Delta CoVaR$s is greater than 0. These findings verify that the agricultural spot market is riskier when considering the extreme downside and upside comovement with its corresponding futures market. In addition, the soybean and maize spot markets are more sensitive to risk shocks from their futures markets, while the risk of the international rice market is minimal compared to other agricultural commodities.

\begin{table}[!ht]
  \centering
  \setlength{\abovecaptionskip}{0pt}
  \setlength{\belowcaptionskip}{10pt}
  \caption{Summary statistics of $VaR$, $CoVaR$ and $\Delta CoVaR$ based on mixed copula models}
  \setlength\tabcolsep{15pt}
  \resizebox{\textwidth}{!}{
    \begin{tabular}{l r@{.}l r@{.}l r@{.}l r@{.}l r@{.}l r@{.}l}
    \toprule
     & \multicolumn{2}{c}{Max} & \multicolumn{2}{c}{Min} & \multicolumn{2}{c}{Mean} & \multicolumn{2}{c}{Std. Dev.} & \multicolumn{2}{c}{Skewness} & \multicolumn{2}{c}{Kurtosis} \\
    \midrule
    \multicolumn{13}{l}{\textit{Panel A: Soybean}}  \\
    $VaR_{0.05}$ & $-$0&5145 & $-$6&0312 & $-$2&0431 & 0&8697 & $-$1&2161 & 1&9651 \\
    $VaR_{0.95}$ & 5&8200 & 0&4999 & 1&9855 & 0&8141 & 1&0965 & 1&7406 \\
    $CoVaR_{0.05}$ & $-$1&1657 & $-$12&2694 & $-$4&3244 & 1&8063 & $-$1&1717 & 1&8333 \\
    $CoVaR_{0.95}$ & 11&8414 & 1&1115 & 4&1714 & 1&7110 & 1&1133 & 1&7278 \\
    $\Delta CoVaR_{0.05}$ & $-$0&4892 & $-$4&9979 & $-$1&7662 & 0&7292 & $-$1&1387 & 1&7592 \\
    $\Delta CoVaR_{0.95}$ & 4&8275 & 0&4725 & 1&7060 & 0&7043 & 1&1387 & 1&7592 \vspace{1mm}\\
    \multicolumn{13}{l}{\textit{Panel B: Maize}}  \\
    $VaR_{0.05}$ & $-$0&8637 & $-$5&9173 & $-$1&9839 & 0&7214 & $-$1&0707 & 1&1446 \\
    $VaR_{0.95}$ & 5&8299 & 0&9593 & 2&0725 & 0&7329 & 1&0397 & 1&0413 \\
    $CoVaR_{0.05}$ & $-$2&0046 & $-$12&5275 & $-$4&4780 & 1&6055 & $-$1&0456 & 1&0330 \\
    $CoVaR_{0.95}$ & 12&4364 & 2&0909 & 4&5581 & 1&6141 & 1&0318 & 0&9899 \\
    $\Delta CoVaR_{0.05}$ & $-$0&8952 & $-$5&2265 & $-$1&9665 & 0&6996 & $-$1&0344 & 0&9904 \\
    $\Delta CoVaR_{0.95}$ & 5&1686 & 0&8853 & 1&9447 & 0&6918 & 1&0344 & 0&9904 \vspace{1mm}\\
    \multicolumn{13}{l}{\textit{Panel C: Wheat}}  \\
    $VaR_{0.05}$ & $-$0&3834 & $-$5&1504 & $-$1&2016 & 0&5033 & $-$1&7784 & 5&6294 \\
    $VaR_{0.95}$ & 5&8586 & 0&4543 & 1&3037 & 0&5283 & 1&7955 & 5&9415 \\
    $CoVaR_{0.05}$ & $-$0&8483 & $-$9&3514 & $-$2&3540 & 0&9293 & $-$1&7572 & 5&4891 \\
    $CoVaR_{0.95}$ & 10&2331 & 0&9941 & 2&6473 & 1&0278 & 1&7457 & 5&4641 \\
    $\Delta CoVaR_{0.05}$ & $-$0&3268 & $-$3&1896 & $-$0&8750 & 0&3362 & $-$1&7328 & 5&3327 \\
    $\Delta CoVaR_{0.95}$ & 3&7488 & 0&3841 & 1&0283 & 0&3951 & 1&7328 & 5&3327 \vspace{1mm}\\
    \multicolumn{13}{l}{\textit{Panel D: Rice}}  \\
    $VaR_{0.05}$ & $-$0&2308 & $-$3&9267 & $-$0&3775 & 0&2109 & $-$5&4178 & 45&9441 \\
    $VaR_{0.95}$ & 3&5655 & 0&2187 & 0&3844 & 0&2434 & 5&2593 & 39&4034 \\
    $CoVaR_{0.05}$ & $-$0&2397 & $-$4&0356 & $-$0&3889 & 0&2168 & $-$5&4156 & 45&8900 \\
    $CoVaR_{0.95}$ & 4&1002 & 0&2765 & 0&4607 & 0&2840 & 5&2510 & 39&4610 \\
    $\Delta CoVaR_{0.05}$ & $-$0&0046 & $-$0&0629 & $-$0&0066 & 0&0037 & $-$5&2576 & 41&3998 \\
    $\Delta CoVaR_{0.95}$ & 0&6358 & 0&0463 & 0&0665 & 0&0372 & 5&2576 & 41&3998 \\
    \bottomrule
    \end{tabular}}%
  \begin{flushleft}
    \footnotesize
    \justifying Note: This table provides the summary statistics of the downside and upside $VaR$, $CoVaR$ and $\Delta CoVaR$ for the spot returns of soybean, maize, wheat, and rice, based on the estimated results of the mixed copula models. The subscripts 0.05 and 0.95 refer to the downside and upside risk measures, respectively.
  \end{flushleft}
  \label{Tab:Statistics_VaR_Mix}%
\end{table}%

Table~\ref{Tab:Spillover_Test_Mix} further presents the test results for the extreme risk spillovers in the international soybean, maize, wheat, and rice markets, where the first two K-S tests are used to assess the significance of the downside and upside risk spillover effects, respectively. All the K-S statistics reject the null hypothesis, indicating that for each agricultural commodity, the extreme downside and upside risk spillover effects of the futures returns on the spot returns are both significant, that is, the information from the agricultural futures market significantly increases the risk exposure of the agricultural spot market. In addition, the third K-S test, which is employed to evaluate the possible asymmetry between the downside and upside risk spillover effects, shows that the downside risk spillover effects for both soybeans and maize are significantly stronger than their corresponding upside risk spillover effects, while there is no significant strength difference between the two risk spillover effects for wheat, and rice.

\begin{table}[!ht]
  \centering
  \setlength{\abovecaptionskip}{0pt}
  \setlength{\belowcaptionskip}{10pt}
  \caption{Hypothesis testing for downside and upside risk spillover effects based on mixed copula models}
  \setlength\tabcolsep{15.5pt}
    \resizebox{\textwidth}{!}{
    \begin{tabular}{lccc}
    \toprule
        &  \makecell{$H_{0}: CoVaR_{0.05} = VaR_{0.05}$ \\ $H_{1}: CoVaR_{0.05} < VaR_{0.05}$} & \makecell{$H_{0}: CoVaR_{0.95} = VaR_{0.95}$ \\ $H_{1}: CoVaR_{0.95} > VaR_{0.95}$} & \makecell{$H_{0}: \frac{CoVaR_{0.05}}{VaR_{0.05}} = \frac{CoVaR_{0.95}}{VaR_{0.95}}$ \\ $H_{1}: \frac{CoVaR_{0.05}}{VaR_{0.05}} > \frac{CoVaR_{0.95}}{VaR_{0.95}}$} \\
    \midrule
    Soybean & 0.6990 & 0.6918 & 0.0390 \\
        & [0.0000] & [0.0000] & [0.0002] \\
    Maize & 0.7567 & 0.7492 & 0.0416 \\
        & [0.0000] & [0.0000] & [0.0001] \\
    Wheat & 0.6523 & 0.6861 & 0.0228 \\
        & [0.0000] & [0.0000] & [0.0540] \\
    Rice & 0.0873 & 0.4517 & 0.0223 \\
        & [0.0000] & [0.0000] & [0.0618] \\
    \bottomrule
    \end{tabular}}%
  \begin{flushleft}
    \footnotesize
    \justifying Note: This table reports the results (K-S statistic) of the K-S test for the extreme risk spillovers based on the mixed copula models, where the $p$-values of test statistics are presented in square brackets. The first two K-S tests examine the significance of the downside and upside risk spillover effects, and the rejection of the null hypotheses indicates the existence of significant risk spillover effects. The third K-S test evaluates the possible asymmetry between the two risk spillover effects, and the rejection of the null hypotheses implies that the downside risk spillover effect is significantly stronger than the corresponding upside risk spillover effect.
  \end{flushleft}
  \label{Tab:Spillover_Test_Mix}%
\end{table}%

\section{Conclusions}
\label{S1:Conclude}

Based on the Copula-CoVaR method, we investigate the tail dependence structure and extreme risk spillovers between the global agricultural futures and spot markets, taking futures and spot of soybean, maize, wheat, and rice as examples. The ARMA-GARCH-skewed Student-t model is adopted to specify the marginal distribution for each agricultural return series, considering that the financial data tend to exhibit some stylized characteristics such as leptokurtosis, fat tail, and skewness, as well as series autocorrelation and conditional heteroscedasticity. We first analyze the tail dependence between the agricultural futures and spot returns using eight different single copula models, and find that the Student-t copula model performs best for soybeans, maize and wheat, while the survival Clayton copula is the best choice for rice.

However, with the acceleration of the process of global integration, the dependence structures among financial markets are becoming more and more complex, and gradually revealing the characteristics of nonlinearity and asymmetry, which means that the assumption of symmetric tail dependence in the Student-t copula may be too restrictive in empirical analysis. Hence, we further construct mixed copula models to explore the possible asymmetric tail dependence between the global agricultural futures and spot markets, and conclude that the tail dependence structures for the four futures-spot pairs are quite different, and each of them is asymmetric. Specifically, the lower tail dependence between the soybean futures and spot is far greater than the upper tail dependence, while the opposite is true for maize, wheat, and rice. In particular, there is basically no lower tail dependence between the rice futures and spot, indicating that the futures and spot markets for rice generally rise synchronously but fall asynchronously.

In addition, we utilize the estimates of the ARMA-GARCH-skewed Student-t model to calculate the $VaR$ measures, and quantify the $CoVaR$ and $\Delta CoVaR$ dynamics for the agricultural spot returns based on the estimated results of the single copula and mixed copula models, respectively, so as to assess the extreme risk spillovers from the agricultural futures markets to the spot markets. The empirical results show that the futures market for each agricultural commodity has significant and robust extreme downside and upside risk spillover effects on the spot market. Meanwhile, the spot markets for soybeans and maize are more sensitive to risk shocks from their corresponding futures markets, while the extreme risk of the international rice market is relatively small compared to other agricultural commodities. Particularly, the evolution of the $VaR$s and $CoVaR$s for rice is dissimilar to that of the $VaR$s and $CoVaR$s for soybeans, maize and wheat, which is reflected in the fact that the absolute values of the $VaR$s and $CoVaR$s for rice are much less than those of the $VaR$s and $CoVaR$s for the other agricultural commodities when the market is stable, but present jump characteristics when the market is at risk. Moreover, we apply the K-S test to further evaluate the possible asymmetry between the two risk spillover effects, and find that the downside risk spillover effects for both soybeans and maize are significantly stronger than their corresponding upside risk spillover effects, while there is no significant strength difference between the downside and upside risk spillover effects for wheat, and rice.

The agricultural futures market is a basic platform for producers or consumers to hedge price risks and investors to diversify asset allocation. However, with the continuous progress of commodity financialization, the major agricultural commodities have gradually become one of the objects of financial speculation. A flood of international speculative capital has flowed into the agricultural futures market, thus leading to more violent and frequent price fluctuations of agricultural futures. Excessive speculation, coupled with a more pessimistic outlook on the stability of the global agricultural system, has intensified the risks of the international agricultural futures market, which often significantly spill over to the agricultural spot market, distorting the spot prices and disturbing the smooth operation of the global agricultural market. As food is a basic and essential demand for human survival and development, countries or regions are supposed to strengthen cooperation to promote the formulation of multilateral rules for the supervision and governance on commodity transactions, and take effective measures to curb excessive speculation and facilitate the resolution of excess volatility in the agricultural futures market, so as to jointly maintain global food security.

For investors, agricultural commodities are an important tool for risk management and portfolio diversification. Our research on tail dependence and extreme risk spillovers inside the international agricultural market provides useful implications for financial asset pricing, risk management decisions, and investment strategy optimization. Specifically, when using agricultural futures for hedging, agricultural commodity dealers should not ignore the risk spillovers from the futures market, but pay close attention to different sorts of information, to make prudent trading decisions and avoid blindly following suit. Considering the tail dependence and extreme risk spillovers between the international agricultural futures and spot markets, investors and practitioners are expected to fully understand the degrees of financialization of various agricultural commodities, appropriately expand the scope of investment choices, such as energy, metal and other commodity futures, to seek diversified protection.

%\newpage
%
%\bibliography{E:/papers/Auxiliary/Bibliography}
%\bibliography{E:/Auxiliary/Bibliography}
\bibliography{BibRef_CoVaR}
%\bibliographystyle{plain}

% \end{CJK*}
\end{document}